\documentclass[pra,aps,amssymb,10pt,floatfix,showpacs,notitlepage, twocolumn, superscriptaddress]{revtex4-1}
\usepackage{bbm}
\usepackage{amsmath}
\usepackage{amssymb}
\usepackage{tensor}
\usepackage{color,graphicx}
\usepackage{hyperref}
\usepackage{interval}
\usepackage[disable]{todonotes}
\usepackage[acronym]{glossaries}
\usepackage{enumitem}

\newcommand{\ket}[1]{|#1\rangle}
\newcommand*{\bra}[1]{\langle#1|}

\newcommand*{\myexp}[1]{\left\langle#1\right\rangle}
\newcommand{\ie}{\textit{i.e.~}}
\newcommand{\wrt}{\textit{w.r.t.~}}
\DeclareMathOperator{\tr}{\mathrm{Tr}}
\newcommand{\cf}{{\it cf.~}}

\newcommand{\dvert}[1]{D_{y,#1}}
\newcommand{\dvertbare}{D_{y}}

\newcommand{\pmode}{\hat{a}}
\newcommand{\pmodenew}{a}
\newcommand{\acol}{A^{\text{col}}}
\newcommand{\bcol}{B^{\text{col}}}

\newcommand{\atwod}{A^{\mathrm{2D}}}
\newcommand{\btwod}{B^{\mathrm{2D}}}
\newcommand{\aoned}{\ket{\psi_{\text{1D}}(\acol) }}
\newcommand{\boned}{\ket{\psi_{\text{1D}}(\bcol) }}

\newcommand{\physdim}{d_p}
\newcommand{\deff}{\tilde{\physdim}}
\newcommand{\psioned}{\ket{\psi_{\text{1D}} }}
\newcommand{\parampump}{\phi_{\mathrm{pump}}}
\newcommand{\paramtriv}{\phi_{\mathrm{triv}}}
\newcommand{\rank}{r}
\newcommand{\MCMF}{\tilde{\Gamma}} 
\renewcommand{\Re}{\mathrm{Re}}
\renewcommand{\Im}{\mathrm{Im}}

\definecolor{comments}{RGB}{255,0,255}

\setacronymstyle{long-short}

\begin{document}
		
	\title{Fermionic tensor networks for higher order topological insulators from charge pumping}
	
	\author{Anna Hackenbroich}
	
	\affiliation{Max-Planck-Institute of Quantum Optics, Hans-Kopfermann-Str. 1, 85748 Garching, Germany}
	
	\affiliation{Munich Center for Quantum Science and Technology, Schellingstra{\ss}e 4, 80799 M{\"u}nchen, Germany}
	
	\author{B. Andrei Bernevig}
	
	\affiliation{Joseph Henry Laboratories and Department of Physics, Princeton University, Princeton, New Jersey 08544, USA}
	
	\affiliation{Max-Planck-Institute of Microstructure Physics, 06120 Halle, Germany}
	
	\author{Norbert Schuch}
	
	\affiliation{Max-Planck-Institute of Quantum Optics, Hans-Kopfermann-Str. 1, 85748 Garching, Germany}
	
	\affiliation{Munich Center for Quantum Science and Technology, Schellingstra{\ss}e 4, 80799 M{\"u}nchen, Germany}
	
	\author{Nicolas Regnault}
	
	\affiliation{Joseph Henry Laboratories and Department of Physics, Princeton University, Princeton, New Jersey 08544, USA}

	\affiliation{Laboratoire de Physique de l'\'Ecole normale sup\'erieure, ENS, Universit\'e PSL, CNRS, Sorbonne Universit\'e, Universit\'e Paris-Diderot, Sorbonne Paris Cit\'e, Paris, France.}
	
\newacronym{ti}{TI}{topological insulator}
\newacronym{tns}{TNS}{tensor network state}
\newacronym{mps}{MPS}{matrix product state}
\newacronym{peps}{PEPS}{projected entangled pair state}
\newacronym{ssh}{SSH}{Su-Schrieffer-Heeger}
\newacronym{oal}{OAI}{obstructed atomic insulator}
\newacronym[longplural={degrees of freedom}]{dof}{DOF}{degree of freedom}
	
		\begin{abstract}
			We apply the charge pumping argument to fermionic tensor network representations of $d$-dimensional \glspl*{ti} to obtain \glspl*{tns} for $(d+1)$-dimensional \glspl*{ti}. We exemplify the method by constructing a two-dimensional \gls*{peps} for a Chern insulator starting from a \gls*{mps} in $d=1$ describing pumping in the \gls*{ssh} model. In extending the argument to second-order \glspl*{ti}, we build a three-dimensional \gls*{tns} for a chiral hinge \gls*{ti} from a \gls*{peps} in $d=2$ for the \gls*{oal} of the quadrupole model. The $(d+1)$-dimensional \glspl*{tns} obtained in this way have a constant bond dimension inherited from the $d$-dimensional \glspl*{tns} in all but one spatial direction, making them candidates for numerical applications. From the $d$-dimensional models, we identify gapped next-nearest neighbour Hamiltonians interpolating between the trivial and \gls*{oal} phases of the fully dimerized \gls*{ssh} and quadrupole models, whose ground states are given by an \gls*{mps} and a \gls*{peps} with a constant bond dimension equal to $2$, respectively.

		\end{abstract}
	
	\maketitle
	
	\section{Introduction}
	
	Higher order \glspl*{ti}~\cite{benalcazar2017quantizedScience, schindler2018higher, benalcazar2017quantized, PhysRevLett.119.246401} have recently been introduced as a new class of symmetry protected topological systems generalizing the framework of TIs with surface states~\cite{RevModPhys.82.3045}. According to one definition, a TI of order $n$ in $d$ spatial dimensions has topological boundary modes at the $(d-n)$-dimensional intersection of $n$ crystal faces. In this terminology, conventional TIs such as Chern insulators~\cite{PhysRevLett.61.2015} are of order $n = 1$ with protected boundary modes at $(d-1)$-dimensional edges. On the other hand, second order TIs protected by mirror or rotation symmetries have zero-dimensional corner states in $d = 2$ dimensions (in which case they may exemplify obstructed atomic limits~\cite{bradlyn2017topological} with local Wannier states) and one-dimensional chiral or helical hinge states in $d = 3$ dimensions (and bulk bands without a local Wannier description). Prototypical examples of second order topological phases include the two-dimensional quadrupole model of Ref.~\cite{benalcazar2017quantizedScience}, a natural extension~\cite{2018arXiv181002373W} of the Su-Schrieffer-Heeger (SSH) model~\cite{PhysRevB.22.2099}, and the three-dimensional chiral hinge insulator of Ref.~\cite{schindler2018higher}, both of which have been experimentally observed in either materials~\cite{SchindlerBismuth} or mechanical~\cite{serra2018observation}, acoustic~\cite{HoaranAcousticHOTI, XiangAcoustic2Dchiral}, photonic~\cite{mittal2018photonic, hassan2018cornerphotonic, xie2018visualizationphotonic, yang2019gapped} and electrical~\cite{peterson2018quantized,imhof2018topolectrical,PhysRevB.99.020304} systems.

	The bulk-boundary correspondence states that the topological properties of a system are reflected in the excitations at its physical boundary. For instance, a two-dimensional Chern insulator is characterised by an integer number of gapless chiral edge modes~\cite{bernevig2013topological}. Similarly, second order TIs in two dimensions possess gapless corner modes at the intersection of two edges compatible with the crystal symmetry~\cite{benalcazar2017quantizedScience}. In three dimensions, one type of second order TI is characterised by the presence or absence of a chiral hinge mode~\cite{schindler2018higher}. For strong TIs~\cite{PhysRevLett.104.130502} as well as chiral topological phases, the universal features in the boundary energy spectrum are encoded in its entanglement spectrum (ES)~\cite{PhysRevLett.101.010504} characterizing the bulk entanglement properties. The convenience of this bulk characterization makes the ES an important tool for the numerical analysis of topological phases. Recently it was also observed, as expected, that the ES of higher order TIs displays characteristic $(d-n)$-dimensional boundary states as long as the entanglement cut preserves the crystal symmetries of the phase~\cite{wang2018entanglement, zhu2019identifying, takahashi2019bulkedge}.  
	
	A natural platform for the bulk-boundary correspondence is provided by tensor network states (TNSs)~\cite{Cirac_2009} in which the entanglement between physical particles is mediated through virtual particles hosting the lower-dimensional boundary theory~\cite{PhysRevB.83.245134}. Physical and virtual particles are related by a local tensor whose bond dimension determines the maximal amount of entanglement in the state~\cite{PhysRevB.73.085115, PhysRevLett.100.070502}. The structure of the local tensor encodes the topological properties of the quantum state, and an analysis of this relation has led to valuable insight for the systematic understanding of one-dimensional symmetry protected topological (SPT) phases~\cite{PhysRevB.84.165139} and intrinsically ordered topological phases~\cite{Schuch2010}. Free fermion systems can be described using Gaussian fermionic TNSs (GfTNSs) that are defined purely in terms of two-point correlation functions~\cite{Kraus}. Starting from such Gaussian TNSs, interacting states can be constructed, whose topological properties derive from the features of the initial Gaussian model~\cite{PhysRevLett.114.106803}.

	Many non-chiral topologically ordered phases in two dimensions~\cite{Kitaev2003, PhysRevB.71.045110} have a simple representation in terms of projected entangled pair states (PEPSs)~\cite{Verstraete:2004cf} with a finite and constant bond dimension~\cite{VerstraetePRL2006, gu2009tensor}. However, Gaussian PEPSs with chiral topological properties do not adequately describe the bulk of a gapped phase, since their correlation functions decay algebraically~\cite{PhysRevLett.111.236805,PhysRevB.92.205307}. Indeed, the existence of gapped PEPSs with finite bond dimension for SPT phases of free fermions is forbidden in dimensions $d > 1$~\cite{PhysRevB.92.205307, PhysRevB.95.115309}. The no-go theorem states that only Wannierizable phases corresponding to the product of one- and zero-dimensional systems can be exactly represented as a TNS.

	This no-go theorem does not prevent TNSs from being a useful numerical description at finite system size for chiral topological phases like the fractional quantum Hall effect~\cite{PhysRevB.86.245305}. Hence, numerically efficient TNSs for such phases are valuable, especially in two and three dimensions. One method for the construction of TIs in $d+1$ dimensions from $d$-dimensional TIs is given by charge pumping~\cite{PhysRevB.27.6083}. For instance, a two-dimensional Chern insulator is obtained from a charge pumping interpolation in the one-dimensional SSH model when the periodic time direction is identified with the momentum in the second spatial direction~\cite{karoly2016short}. For second order TIs, dipole pumping in the quadrupole model defines a three-dimensional chiral hinge insulator~\cite{benalcazar2017quantized, schindler2018higher}. In both cases, the zero-dimensional boundary modes of the $d$-dimensional model give rise to one-dimensional chiral boundary modes in the $(d+1)$-dimensional system.

	In this article, we exemplify how the charge pumping argument applied to $d$-dimensional GfTNSs with constant bond dimension yields gapped GfTNSs for TIs in $d + 1$ dimensions. By construction, the $(d + 1)$-dimensional TNS has a constant finite bond dimension in a hybrid coordinate system with $d$ real-space axes and one momentum axis in the additional dimension. In order to obtain a real-space tensor network for the $(d+1)$-dimensional state, we apply to the hybrid TNS an inverse Fourier transform (FT) in the direction $d + 1$. As a result, the bond dimension of the real-space local tensor in this direction generically grows with the system size due to the non-locality of the FT, whereas it is identical to the finite bond dimension of the $d$-dimensional TNS in the other $d$ directions. We apply this construction both to a matrix product state (MPS)~\cite{fannes1992finitely} for the SSH model in order to obtain a PEPS for a Chern insulator, and to a Gaussian PEPS with finite bond dimension for the topological quadrupole model in order to obtain a three-dimensional GfTNS for the second order chiral hinge state TI of Refs.~\cite{schindler2018higher, benalcazar2017quantized}. Therefore, our approach provides us with a gapped TNS with one-dimensional chiral boundary states and a constant finite bond dimension in all but one of the spatial directions. This representation is therefore potentially useful for tensor network algorithms.

	This article is structured as follows: In Sec.~\ref{sec:fMPS} we begin with a brief overview of charge pumping in the SSH model and fermionic TNSs, and continue by studying an MPS for the SSH ground state along the charge pumping interpolation. In the following Sec.~\ref{sec:ChernPEPS}, we employ this MPS to construct a two-dimensional PEPS describing a Chern insulator. In Sec.~\ref{sec:2DHOTI} we extend our method to the two-dimensional second order quadrupole insulator and the construction of a three-dimensional PEPS for the chiral hinge state higher order TI. Finally, we summarize our results and discuss remaining open questions in Sec.~\ref{sec:Conclusion}.

	\section{Fermionic MPS for charge pumping in the SSH model\label{sec:fMPS}}
	In this section we introduce a fermionic MPS that describes a charge pumping cycle in the SSH model corresponding to a Chern insulator with Chern number $\mathcal{C} = 1$ in two spatial dimensions. We begin by briefly reviewing the SSH model, the charge pumping argument and its relation to Chern insulators in Sec.~\ref{sec:SSHToChern}. We continue with a short introduction to fermionic MPSs in Sec.~\ref{sec:fMPS_fMPSAtHalfFilling}. In Sec.~\ref{sec:fMPS_Construction}, we construct the MPS for the SSH model and study its parent Hamiltonian.
	
	\subsection{Chern insulator from charge pumping in the SSH model\label{sec:SSHToChern}}

\begin{figure}
				\includegraphics[width = 0.99\linewidth]{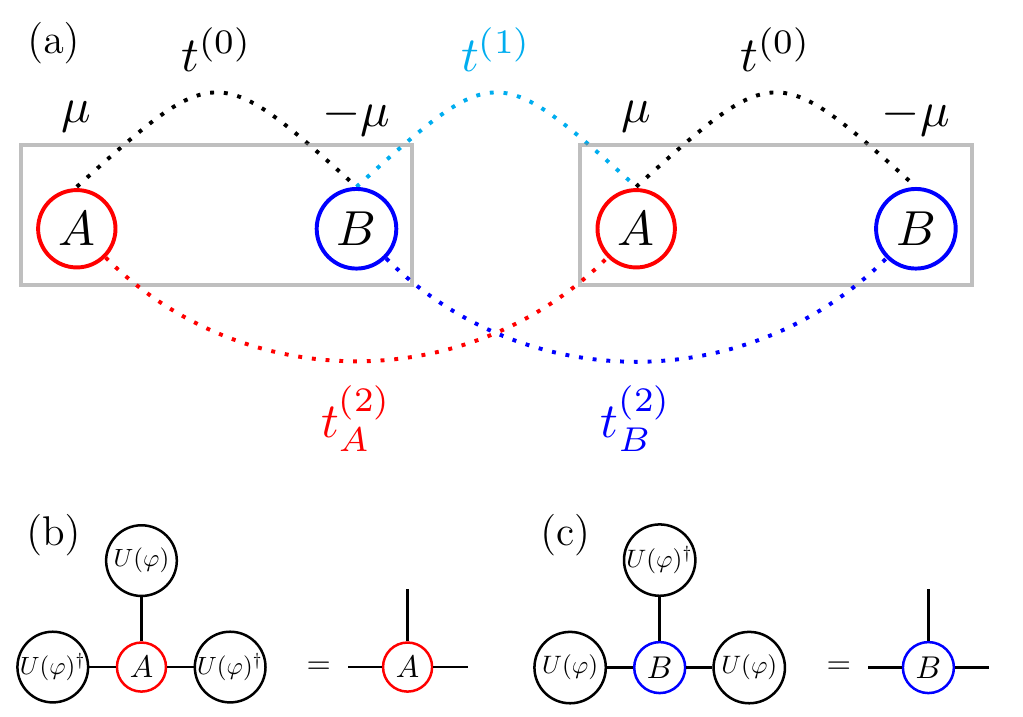}
			\caption{(a) Extended SSH model with a staggered chemical potential $\mu$, nearest-neighbour hoppings $t^{(0)}$ and $t^{(1)}$ within and between unit cells, and next-nearest neighbour hoppings $t_A^{(2)}$ and $t_B^{(2)}$ on the $A$ and $B$ sublattices, respectively. The unit cells consisting of one site of each sublattice are marked by gray rectangles. (b) Action on the local MPS tensors of the $\mathrm{U}(1)$ symmetry ensuring that the MPS of Eq.~\eqref{MPSMyInterpol} lies at half filling of the chain.\label{fig:SSH_Sketches}}
		\end{figure}

		The SSH model describes a one-dimensional chain of spinless fermions with two orbitals denoted $A$ and $B$ per unit cell~\cite{PhysRevB.22.2099}. We consider the model at half filling where the number of particles is equal to the number $N_x$ of unit cells. For open boundary conditions, the Hamiltonian reads		\begin{multline}\label{SSHHam}
		H_{\mathrm{SSH}} = t^{(0)} \sum_{x=0}^{N_x -1} (\pmode^{\dagger}_{A,x} \pmode_{B,x} + \mathrm{h.c.}) \\
		+ t^{(1)} \sum_{x=1}^{N_x-1} (\pmode^{\dagger}_{B,x-1} \pmode_{A,x} + \mathrm{h.c.}) 
		,
		\end{multline}
		where we use the notation $\pmode_{A,x}$ and $\pmode_{B,x}$ for the fermionic annihilation operators of the orbitals $A$ and $B$ in unit cell $x$ with $x = 0,\dotsc,N_x -1$, respectively. Here, $t^{(0)}$ denotes the hopping amplitude between $A$ and $B$ orbitals within the same unit cell, and $t^{(1)}$ the hopping amplitude between sites on neighboring unit cells (see Fig.~\ref{fig:SSH_Sketches}(a)). For $|t^{(1)}| > |t^{(0)}|$, the hopping between unit cells dominates the hopping within unit cells and the SSH model is in a phase topologically different from the case $|t^{(0)}| > |t^{(1)}|$. This phase is protected by spatial inversion symmetry and characterised by fermionic edge modes~\cite{karoly2016short} and a "filling anomaly"~\cite{2018arXiv181002373W}. It is called obstructed atomic insulator (OAI)~\cite{bradlyn2017topological}. When the intra-cell hopping $t^{(0)}$ vanishes in the deep OAI phase, the system is dimerized since its bulk splits into decoupled two-site blocks. In this case, the edge excitations are created by the mode operators $\pmode^{\dagger}_{A,0}$ and $\pmode^{\dagger}_{B,N_x -1}$ and have exactly zero energy in this specific model, but can generally be moved in energy. On the other hand, for $|t^{(0)}| > |t^{(1)}|$ the SSH model is trivial, with a dimerized point at $t^{(1)} = 0$ and no state at zero energy.
		
		A gapped interpolation between the trivial and OAI phases of the SSH model can be constructed by adding to the Hamiltonian $H_{\mathrm{SSH}}$ the term
		\begin{equation}
		\sum_{x=0}^{N_x -1} (\mu_A\pmode^{\dagger}_{A,x} \pmode_{A,x} + \mu_B\pmode^{\dagger}_{B,x} \pmode_{B,x})
		\end{equation}
		introducing a staggered chemical potential $\mu_A = - \mu_B = \mu$ which breaks the inversion symmetry as shown in Fig.~\ref{fig:SSH_Sketches}(a). In Ref.~\cite{benalcazar2017quantized}, the authors consider the periodic time-dependent Hamiltonian $H_{\mathrm{pump}}(t)$ with time $t \in (-\pi, \pi]$ defined by the couplings
		\begin{equation}\label{DimerizedInterpolationHamiltonian}
		(\mu,t^{(1)},t^{(0)}) = \begin{cases}
		(\cos (t), 0,-\sin(t)) & \quad -\pi < t \leq 0\\
		(\cos (t), \sin(t), 0) & \quad 0 < t \leq \pi\\
		\end{cases}.
		\end{equation}
		At $t = -\pi$ (equivalent to $t = \pi$), the system is in an atomic phase with only the $A$ orbitals occupied. For $ -\pi < t < 0$, the coupling between $A$ and $B$ sites in the same unit cell is non-zero and the charge is transferred from left to right by the changing chemical potential until only the $B$ orbitals are occupied at $t = 0$. At $t = - \pi/2$, the staggered chemical potential vanishes and the system is in the trivial dimerized phase of the SSH model. For $ 0 < t < \pi$, the intra-unit cell coupling vanishes whereas the hopping between different unit cells is non-zero and the charge is transferred from left to right such that at $ t = \pi$ the system returns to the state with all $A$ orbitals occupied. At $t = \pi /2$, the chemical potential vanishes and the system is in the OAI dimerized phase of the SSH model. The ground state of the pumping Hamiltonian $H_{\mathrm{pump}}(t)$ is continuous as a function of time $t$ if the chain has periodic and anti-periodic boundary conditions for $N_x$ odd and even, respectively.
		
		The charge pumping interpolation of Eq.~\eqref{DimerizedInterpolationHamiltonian} can be used to generate a lattice model in one dimension higher with the topology of a Chern insulator~\cite{benalcazar2017quantized}. Indeed, $H_{\mathrm{pump}}(t)$ corresponds to the time-dependent Bloch Hamiltonian 
		\begin{multline}\label{DimerizedInterpolationBlochHamiltonian}
		\mathcal{H}_{\mathrm{pump}} (k_x, t) = \left[t^{(0)}(t) + t^{(1)}(t) \cos k_x \right] \sigma_1 \\
		+ t^{(1)}(t) \sin k_x \times \sigma_2 + \mu (t) \,  \sigma_3,
 		\end{multline}
		where $\sigma_1 = (\begin{smallmatrix}
		0 & 1\\
		1 & 0
		\end{smallmatrix})$, $\sigma_2 = (\begin{smallmatrix}
		0 & -i\\
		i & 0
		\end{smallmatrix})$ and $\sigma_3 = (\begin{smallmatrix}
		1 & 0\\
		0 & -1
		\end{smallmatrix})$ are the Pauli matrices and $k_x = \frac{\pi}{N_x}(2j - N_x) \in [-\pi, \pi]$ for $0 \leq j \leq N_x -1$ the lattice momentum. Since the interpolation is cyclic, the time $t\in [-\pi, \pi]$ may be interpreted as the lattice momentum $k_y$ of a second spatial direction $y$ and Eq.~\eqref{DimerizedInterpolationBlochHamiltonian} as the Bloch Hamiltonian of a two-dimensional system with closed boundaries in both directions. This Hamiltonian has a Chern number $\mathcal{C} = 1$ due to the charge transport between unit cells induced by the interpolation for $t \in [0, \pi]$~\cite{benalcazar2017quantized}.
		
		We note that the cycle of Eq.~\eqref{DimerizedInterpolationHamiltonian} can be deformed to a smooth charge pumping interpolation, \emph{without changing the topology or breaking the dimerization}. In this cycle, the couplings $\mu(t)$, $t^{(1)}(t)$ and $t^{(0)}(t)$ and hence the Bloch Hamiltonian are smooth functions of the time $t$, \ie they are infinitely often continuously differentiable with respect to $t$. As explained in the previous paragraph, a two-dimensional Chern insulator model is obtained by identifying the time $t$ and the momentum $k_y$. Due to the smoothness of the cycle, the real-space representation of the Chern insulator has couplings which decay faster than any polynomial~\cite{PhysRevB.90.115133}. The smooth cycle is obtained by smoothing out the non-analyticities of Eq.~\eqref{DimerizedInterpolationHamiltonian} at $t = 0, \pi$ using smooth functions that interpolate between $0$ and $1$. This can be done in such a way that, like in Eq.~\eqref{DimerizedInterpolationHamiltonian}, at each time $t$ the smooth couplings satisfy $\mu(t) ^2 + t^{(1)}(t)^2 + t^{(0)}(t)^2 = 1$~\cite{deitmar2014principles}.

		\subsection{Fermionic MPSs for a half-filled lattice\label{sec:fMPS_fMPSAtHalfFilling}}
		Similarly to bosonic MPSs for spin chains, fermionic MPSs describing chains of electrons are obtained by associating virtual particles to each physical particle which mediate the entanglement between different physical constituents~\cite{Kraus}. In the case of fermionic tensor networks, the virtual particles obey fermionic statistics. A fermionic MPS with $f$ physical complex fermionic modes per site and $\xi$ virtual complex fermionic modes per lattice site and nearest-neighbour bond has physical dimension $\physdim=2^f$ and bond dimension $D=2^{\xi}$. The state is fully characterized by the set of local maps which relate the virtual and physical particles associated to one lattice site. With respect to orthonormal bases $\{\ket{i}\}$ with $i=0,\dotsc,\physdim-1$ for the physical Hilbert space and $\{\ket{l}\}$, $\{\ket{r}\}$ with $l,r=0,\dotsc,D -1$ for the left and right virtual spaces on one site, respectively, the local maps are represented by $\physdim$ local matrices $A^i_{lr}$ of dimension $D\times D$~\cite{bultinck2017fermionic}. 
		 
		In order to fix the global parity of the state, we restrict ourselves to parity-even local tensors which preserve the fermion parity between the physical and virtual layers~\cite{bultinck2017fermionic}. We choose the orthonormal basis for the physical Hilbert space such that all basis states $\ket{i}$ have either even fermion parity $|i| = 0$ or odd fermion parity $|i| = 1$, and similarly for the virtual Hilbert spaces. A local tensor is parity-even if for all non-vanishing elements $A^i_{lr} \neq 0$, the total fermion parity $|i| + |l| + |r| = 0$ is even, such that
		\begin{equation}\label{Parity}
		(-1)^{|i|} A^i_{lr} = (-1)^{|l|} A^i_{lr} (-1)^{|r|}
		\end{equation}
		for all configurations of $i$, $l$ and $r$ (where no summation is implied). If the MPS is constructed from tensors $A[j]$ at each site $j$ with this property, the state on a closed chain with $N_x$ sites is given by
		\begin{multline}\label{fMPSBasis}
		\ket{\psi} = \sum_{i_0,\dotsc,i_{N_x-1}} \sum_{l} (-1)^{\epsilon |l|} \times \\
		\left(A[0]^{i_0}\dotsm A[N_x-1]^{i_{N_x-1}}\right)_{ll}  \ket{i_0,\dotsc,i_{N_x-1}}.
		\end{multline} 
		Here, $|l|$ is the parity of the virtual basis state on the link between the first and last site, and $\epsilon = 1$ or $\epsilon = 0$ corresponding to periodic and anti-periodic boundary conditions for the many-body state, respectively~\cite{bultinck2017fermionic}. 
		
		Below, we construct a fermionic MPS with parity-even local tensors for the ground state of the SSH model along the dimerized charge pumping interpolation. Due to the parity symmetry of Eq.~\eqref{Parity}, this MPS necessarily has an even number of physical particles on a closed chain where all virtual bonds are contracted. On the other hand, the ground state of the SSH model is half-filled such that the number of particles is odd if the number of unit cells is odd. Therefore, to construct the parity-even SSH MPS we use a many-body basis built by acting with creation operators on the state $\ket{\Omega}$ that contains $N_x$ physical particles. To that end, we define new mode operators $\pmodenew_{A,x}$ and $\pmodenew_{B,x}$ for the physical fermions by performing a particle-hole transformation on all $B$ orbitals of the SSH chain while leaving the $A$ orbitals unaltered,
		\begin{subequations}\label{PHTrafo}
		\begin{gather}
		\pmodenew_{A,x} \equiv \pmode_{A,x},\\
		\pmodenew_{B,x} \equiv \pmode^{\dagger}_{B,x}.
		\end{gather}
		\end{subequations}
		The state $\ket{\Omega}$ is the vacuum of the new operators, given in terms of the original vacuum state $\ket{\hat{\Omega}}$ with $\pmode_{A,x}\ket{\hat{\Omega}} = \pmode_{B,x}\ket{\hat{\Omega}} = 0$ as
		\begin{equation}\label{PHVacuum}
		\ket{\Omega} = \prod_{x} \pmode^{\dagger}_{B,x} \ket{\hat{\Omega}}.
		\end{equation}
		It satisfies $\pmodenew_{A,x}\ket{\Omega} = \pmodenew_{B,x}\ket{\Omega} = 0$. Therefore, the new vacuum contains $N_x$ of the original fermions and thus is half-filled. The MPS of the SSH model along the interpolation is then defined with respect to the Fock states constructed from the modes $\pmodenew^{\dagger}_{A,x}$ and $\pmodenew^{\dagger}_{B,x}$ acting on the new vacuum $\ket{\Omega}$. This particle-hole transformation is motived by our desire to write the MPS in terms of parity-even local tensors, which will in turn enable us to express the state as a GfTNS and thereby compute a Bloch parent Hamiltonian analytically.
		
		\subsection{MPS for the SSH model\label{sec:fMPS_Construction}}
		In this Section, we study a fermionic MPS for the SSH model. In Sec.~\ref{sec:fMPS_Construction_Def}, we define the state in terms of its local tensors and identify the $\mathrm{U}(1)$ symmetry leading to a conserved number of particles. In Sec.~\ref{sec:fMPS_PH}, we derive a parent Hamiltonian for the MPS, allowing us to conclude in Sec.~\ref{sec:fMPS_Construction_ChargePumping} that the MPS describes the SSH model along the charge pumping cycle of Eq.~\eqref{DimerizedInterpolationHamiltonian}.

		\subsubsection{Local tensors and $\mathrm{U}(1)$ symmetry\label{sec:fMPS_Construction_Def}}
		In order to describe the ground state of the SSH model along the dimerized charge pumping interpolation, we consider a fermionic MPS with physical dimension $\physdim = 2$ corresponding to one fermionic mode per site and bond dimension $D_{\mathrm{SSH}} = 2$. This is the minimal bond dimension for an MPS along the charge pumping cycle, since the ES of an open SSH chain in the OAI and trivial dimerized phases has two degenerate levels \wrt all cuts between and within unit cells, respectively. 
		
		The fermionic MPS is translation invariant with a unit cell of two sites. It is therefore fully specified by the local matrices $A^i_{lr}$ and $B^i_{lr}$ with $i, l, r \in \{0, 1\}$ for sites on the $A$ and $B$ sublattices, respectively. In terms of the local tensors the physical state on a closed chain is given as
		\begin{multline}\label{SSHMPSState}
		\ket{\psi} = \sum_{\substack{i_{A,0},\dotsc,i_{A,N_x-1}\\i_{B,0},\dotsc,i_{B,N_x-1}}} \ket{i_{A,0},i_{B,0},i_{A,1},\dotsc,i_{B,N_x-1}}  \\
		\sum_{l} (-1)^{\epsilon |l|} \times \left(A^{i_{A,0}}B^{i_{B,0}}A^{i_{A,1}}\dotsm B^{i_{B,N_x-1}}\right)_{ll},
		\end{multline}
		where $\epsilon = 1$ or $\epsilon = 0$ for periodic and anti-periodic boundary conditions. The physical many-body basis state is
		\begin{multline}
		\ket{i_{A,0},i_{B,0},i_{A,1},\dotsc,i_{B,N_x-1}} = \\
		(\pmodenew_{A,0}^{\dagger})^{i_{A,0}} (\pmodenew_{B,0}^{\dagger})^{i_{B,0}} (\pmodenew_{A,1}^{\dagger})^{i_{A,1}} \dotsm (\pmodenew_{B,N_x-1}^{\dagger})^{i_{B,N_x-1}}\ket{\Omega}
		\end{multline}
		with the vacuum $\ket{\Omega}$ from Eq.~\eqref{PHVacuum}. Guided by the dimerized limits to be discussed below, for the local MPS matrices we make the ansatz
		\begin{subequations}\label{MPSMyInterpol}
			\begin{gather}
			A^0 = \begin{pmatrix}
			\gamma & 0\\
			0 & 0
			\end{pmatrix},\,
			A^1 = \begin{pmatrix}
			0 &  \beta\\
			-\alpha & 0
			\end{pmatrix},\\
			B^0  = \begin{pmatrix}
			\gamma & 0\\
			0 & 0
			\end{pmatrix},\,
			B^1  = \begin{pmatrix}
			0 & -\alpha\\
			-\beta & 0
			\end{pmatrix},
			\end{gather}
		\end{subequations}
		depending on parameters $\alpha, \beta, \gamma \in \mathbb{R}$. Note that the normalised quantum state defined by these local MPS matrices depends only on the two quotients $\alpha/\gamma$ and $\beta/\gamma$. Nonetheless, we choose to work with the parametrisation of Eq.~\eqref{MPSMyInterpol} since the case $\gamma = 0$ can be treated more conveniently without divergences.
		
		In order to motivate the ansatz of Eq.~\eqref{MPSMyInterpol} for the local matrices, let us see how the MPS from Eq.~\eqref{SSHMPSState} can describe the ground state of the SSH model both in the trivial and the OAI dimerized phases with appropriate choices for the parameters $\alpha$, $\beta$ and $\gamma$. In the trivial dimerized phase of the SSH model, each unit cell decouples from the rest of the system and is in the state $(\pmodenew_A^{\dagger}\pmodenew_B^{\dagger} - 1)\ket{\Omega}$. In order to ensure the absence of entanglement on the bonds between unit cells, the blocked MPS matrices $(A^{i_A}B^{i_B})_{lr}$ for one unit cell should be non-zero only if $l = r = 0$. Then, the restriction of the MPS to the unit cell, $\sum_{i_A, i_B}(A^{i_A}B^{i_B})_{00}\ket{i_A,i_B}$, is proportional to $(\pmodenew_A^{\dagger}\pmodenew_B^{\dagger} - 1)\ket{\Omega}$ if the blocked MPS matrices have only two non-zero entries $(A^0B^0)_{00} = - (A^1B^1)_{00}$. Therefore, the MPS represents the trivial dimerized phase of the SSH model if the parameters are chosen as $\alpha = 0$ and $|\beta| = |\gamma| \neq 0$. Similarly, in the OAI dimerized phase the chain splits into decoupled plaquettes composed of two adjacent sites from different unit cells. The state on each of the plaquettes is given by the superposition $(\pmodenew_B^{\dagger}\pmodenew_A^{\dagger} + 1)\ket{\Omega}$ such that the blocked MPS matrices for the plaquette should have only two non-zero entries $(B^0A^0)_{00} = (B^1A^1)_{00}$. This is achieved if $\beta = 0$ and $|\alpha| = |\gamma| \neq 0$. The ansatz of Eq.~\eqref{MPSMyInterpol} is chosen such as to allow an interpolation between these two cases.		
		
		The MPS of Eq.~\eqref{SSHMPSState} with the parametrisation of Eq.~\eqref{MPSMyInterpol} has a $\mathrm{U}(1)$ symmetry which ensures that the physical state lies at half filling of the chain. The $\mathrm{U}(1)$ rotation acts on a single complex fermion with the matrix $U (\varphi)= \big(\begin{smallmatrix}
		1 & 0 \\
		0 & e^{i\varphi}
		\end{smallmatrix}\big)$. The local tensors for the two sublattices are invariant under a combination of $\mathrm{U}(1)$ rotations of the physical and virtual legs as sketched in Fig.~\ref{fig:SSH_Sketches}(b) and (c) (see for instance Ref.~\cite{PhysRevLett.100.167202} for MPSs with physical symmetries), 
		\begin{subequations}\label{LocalU1}
			\begin{gather}
			\label{LocalU1A} A^i_{lr} = \sum_j\sum_{l'r'} U(\varphi)_{ij}U(\varphi)_{ll'}^{\dagger} A^j_{l'r'} U(\varphi)_{r'r}^{\dagger},\\
			B^i_{lr} = \sum_j\sum_{l'r'} U(\varphi)^{\dagger}_{ij} U(\varphi)_{ll'} B^j_{l'r'} U(\varphi)_{r'r}.
			\end{gather}
		\end{subequations}
		If we choose $\varphi = \pi$, these identities correspond to the parity symmetry of Eq.~\eqref{Parity}, showing that the local MPS tensors are parity-even. For each nearest-neighbour bond, one virtual leg transforms with $U(\varphi)$ and the other with $U(\varphi)^{\dagger}$ in Eq.~\eqref{LocalU1}, such that the two $\mathrm{U}(1)$ rotations cancel if the bond is contracted. Therefore, the state on a chain with closed boundaries after the contraction of all virtual bonds is invariant under the physical part of the $\mathrm{U}(1)$ rotations of Eq.~\eqref{LocalU1}, given by staggered transformations $U(\varphi)$ and $U(\varphi)^{\dagger}$ on $A$ and $B$ orbitals, respectively. Invariance under this global symmetry implies that
		\begin{multline}\label{GeneratorU1MPS}
		 0 = \myexp{\sum _{x} \left[ \pmodenew^{\dagger}_{A,x}\pmodenew_{A,x} - \pmodenew^{\dagger}_{B,x}\pmodenew_{B,x} \right]} =  \\
		\myexp{\sum _{x } \left[ \pmode^{\dagger}_{A,x}\pmode_{A,x} + \pmode^{\dagger}_{B,x}\pmode_{B,x}\right]} - N_x, 
		\end{multline}
		forcing the number of particles measured in terms of the original physical modes to be equal to the number of unit cells as required for the SSH ground state. In Eq.~\eqref{LocalU1}, the physical legs on the $A$ and $B$ sublattices transform as particles and holes, respectively, as expected due to the particle-hole transformation of Eq.~\eqref{PHTrafo}. In addition, Eq.~\eqref{LocalU1} implies that the virtual legs on the $A$ and $B$ sublattices transform as hole-like and particle-like degrees of freedom (DOFs), respectively.

		In order to gain a better understanding of the parameters $\alpha$, $\beta$ and $\gamma$ as well as of the systems described by the MPS from Eq.~\eqref{SSHMPSState}, it is helpful to consider a parent Hamiltonian for which it is the exact ground state. This Hamiltonian can be computed directly in terms of its Bloch representation once the MPS of Eq.~\eqref{SSHMPSState} is expressed as a Gaussian fermionic TNS.
		 
		 \subsubsection{Parent Hamiltonian\label{sec:fMPS_PH}}	
		 Since the charge pumping Hamiltonian of Eq.~\eqref{DimerizedInterpolationHamiltonian} describes non-interacting fermions, its ground state is a fermionic Gaussian state. It is thus fully characterised by its covariance matrix (CM) (see Appendix~\ref{sec:AppendixDefCM} for a summary of our conventions)~\cite{bravyi2005lagrangian}. As we review in Appendix~\ref{sec:AppendixGfPEPSConstruction}, the tensor network formalism may be used to construct Gaussian fermionic TNSs which are ground states of free fermion Hamiltonians~\cite{Kraus}. The CM in Fourier space of a translationally invariant Gaussian TNS is given by a simple expression which can often be evaluated analytically. Then, any positive function $\epsilon (\mathbf{k}) > 0$ on the Brillouin zone gives rise to a parent Hamiltonian with dispersion relation $\epsilon(\mathbf{k})$, whose Bloch representation is obtained by multiplying the CM by $\epsilon(\mathbf{k})$.
		 
		 For all values of the parameters $\alpha$, $\beta$ and $\gamma$, the MPS of Eq.~\eqref{MPSMyInterpol} corresponds to a Gaussian fermionic TNS whose Fourier CM is computed analytically in Appendix~\ref{sec:AppendixSSHMPS}. We show that this MPS with bond dimension $D = 2$ is the unique ground state of a longer-range SSH-like model with a staggered chemical potential $\mu$ and next-nearest neighbour hoppings $t^{(2)}_{A}$ and $t^{(2)}_{B}$ on the $A$ and $B$ sublattices, respectively, as sketched in Fig.~\ref{fig:SSH_Sketches}(a). The coupling constants of the parent Hamiltonian $H_{\mathrm{MPS}}$ depend on the parameters of the MPS as
		 \begin{subequations}\label{CouplingsMPSPH}
		 	\begin{gather}
		 	\mu  = \frac{\gamma^4 - \alpha^4-\beta^4}{\alpha^4 + \beta^4 + \gamma^4},\\
		 	t^{(0)} = \frac{2\beta^2 \gamma^2}{\alpha^4 + \beta^4 + \gamma^4},\\
		 	t^{(1)} = \frac{2\alpha^2 \gamma^2}{\alpha^4 + \beta^4 + \gamma^4},\\
		 	t^{(2)}_{A} = - t^{(2)}_{B} = \frac{-2\alpha^2 \beta^2 }{\alpha^4 + \beta^4 + \gamma^4},
		 	\end{gather}
		 \end{subequations}
	 	 where $t^{(0)}$ and $t^{(1)}$ denote the hopping amplitudes between nearest-neighbour sites on the same and adjacent unit cells, respectively. 
	 	 
	 	 Depending on the parameter values, $H_{\mathrm{MPS}}$ describes different phases of the fermionic chain. If any two out of the three parameters $\alpha$, $\beta$, $\gamma$ vanish, the system is in an atomic state without entanglement between different sites. Indeed, if only $\gamma$ is non-zero, all hopping constants vanish and the chemical potential is $\mu = +1$ such that we obtain the state with all $B$ sites occupied. On the other hand, if either only $\alpha$ or only $\beta$ is non-zero, all hopping constants vanish and the chemical potential is $\mu = -1$ such that the MPS is equal to the state with all $A$ sites occupied. This implies that the atomic state with occupied $A$ orbitals is obtained from two distinct parameter configurations $(\alpha, \beta, \gamma) = (1, 0, 0)$ and $(\alpha, \beta, \gamma) = (0, 1, 0)$ whose corresponding local MPS matrices are related by a virtual unitary gauge transformation with representation matrix $\sigma_1$.
	 	 
	 	 On the other hand, if only $\gamma$ and $\beta$ are non-zero, the system is in a dimerized phase where the next-nearest neighbour hopping as well as the nearest-neighbour hopping between unit cells vanishes, whereas the nearest-neighbour hopping within unit cells is finite. Unless $|\gamma| = |\beta|$, inversion symmetry is broken by the non-zero chemical potential. For $|\gamma| = |\beta|$, we obtain the trivial dimerized phase of the SSH model. Similarly, if only $\gamma$ and $\alpha$ are non-zero, all couplings vanish except for the nearest-neighbour hopping between unit cells and the chemical potential. For $|\gamma| = |\alpha|$, inversion symmetry is restored and we obtain the OAI dimerized phase of the SSH model. Finally, if all three parameters are non-zero, both nearest-neighbour hopping constants $t^{(0)}$ and $t^{(1)}$ are non-zero and there is an additional next-nearest neighbour hopping $t^{(2)}_{A} = - t^{(2)}_{B}$ which is odd under inversion.
	 	 
	 	 \begin{figure}[t]
	 	 		\includegraphics[width = 0.99\linewidth]{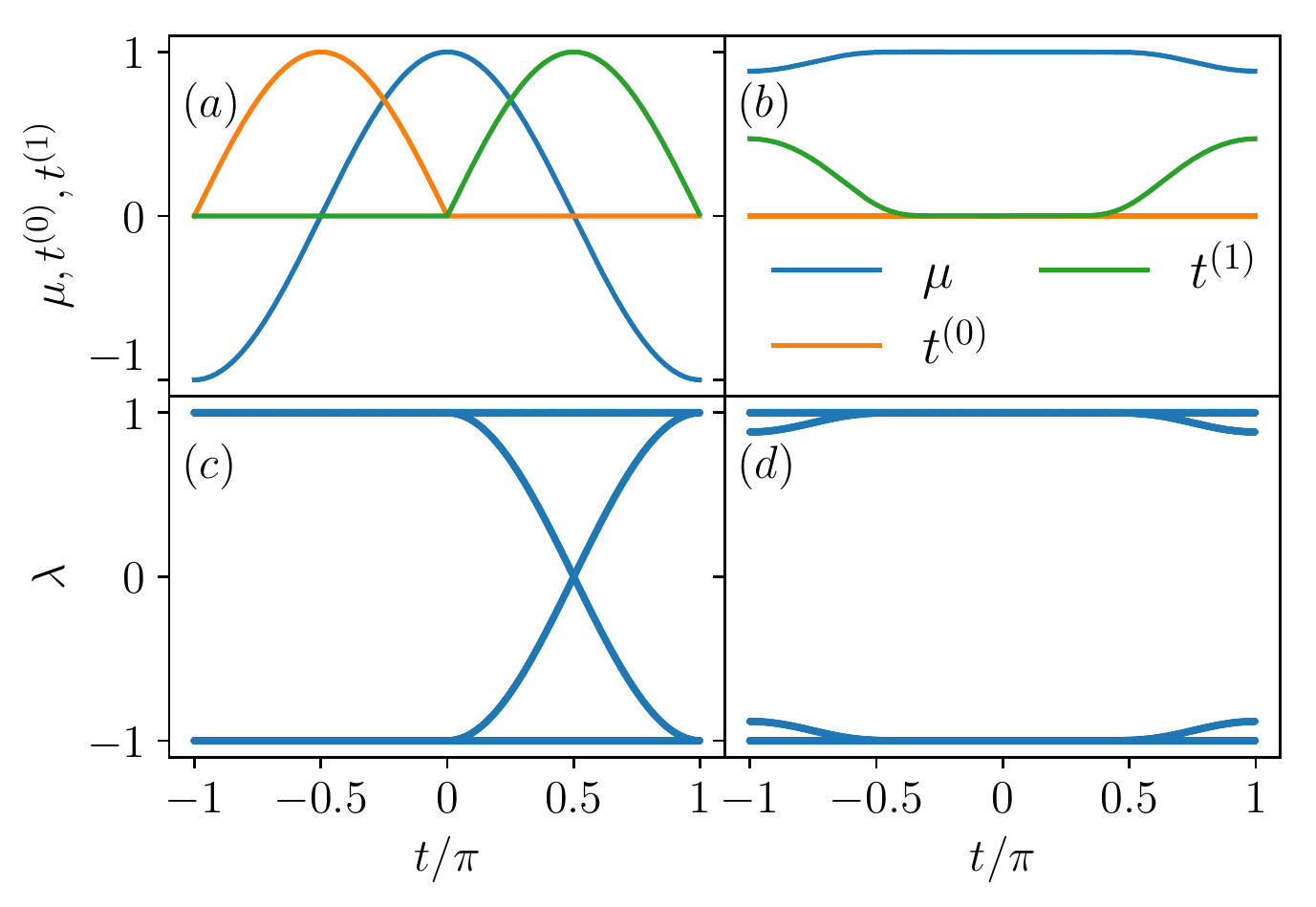}
	 	 \caption{(a), (b) Coupling constants of $H_{\mathrm{MPS}}$ and (c), (d) single-particle ES for the MPS of Eq.~\eqref{MPSMyInterpol} along topologically non-trivial and trivial interpolations. In (a) and (c), along the charge pumping interpolation of Eq.~\eqref{ParametrisationDimerized} corresponding to a Chern insulator with Chern number $\mathcal{C} = 1$. In (b) and (d), along the topologically trivial interpolation of Eq.~\eqref{MPSInterpolationTrivial}. In both (a) and (b), the vanishing next-nearest neighbour coupling $t^{(2)}_{A} = t^{(2)}_{B} = 0$ of $H_{\mathrm{MPS}}$ is not shown. The ES in (c) and (d) was computed on half of a periodic chain with $N_x = 10$ unit cells.\label{fig:SSH_ESAndPH}}
	 	 \end{figure}
	 	 
	 	 \subsubsection{Charge pumping interpolation\label{sec:fMPS_Construction_ChargePumping}}
	 	 
	 	 We now use the MPS of Eq.~\eqref{MPSMyInterpol} to describe charge pumping by considering the evolving state obtained from time-dependent parameters $\alpha(t)$, $\beta(t)$ and $\gamma(t)$ with time $t \in (- \pi, \pi]$. In Eq.~\eqref{CouplingsMPSPH} we saw that the MPS can represent charge pumping with a dimerized nearest-neighbour as well as a longer-range non-dimerized Hamiltonian. For simplicity, we focus on the dimerized case. For the MPS given by the parametrisation
	 	 \begin{multline}\label{ParametrisationDimerized}
	 	 \parampump: \quad (\alpha(t), \beta (t), \gamma (t)) = \\
	 	 \begin{cases}
	 	 (0, \sqrt {|\sin t/2|}, \sqrt {|\cos t/2|}) & -\pi < t \leq 0\\
	 	 (\sqrt {|\sin t/2|}, 0, \sqrt {|\cos t/2|}) & 0 < t \leq \pi
	 	 \end{cases},
	 	 \end{multline} 
	 	 the parent Hamiltonian $H_{\mathrm{MPS}}$ is exactly equal to the time-dependent Hamiltonian $H_{\mathrm{pump}}(t)$ of the dimerized charge pumping cycle of Eq.~\eqref{DimerizedInterpolationHamiltonian}. The evolution of the couplings in $H_{\mathrm{MPS}}$ along the parametrisation is shown in Fig.~\ref{fig:SSH_ESAndPH}(a). Hence, the MPS of Eq.~\eqref{MPSMyInterpol} can describe the ground state of the pumping model of Eq.~\eqref{DimerizedInterpolationHamiltonian} with the topology of a Chern insulator for all times $t\in (-\pi, \pi]$. 
		 
		 The ground state of the charge pumping cycle $H_{\mathrm{pump}}(t)$ and its correlation functions are continuous along the entire interpolation. However, the local MPS tensors parametrised by $\parampump$ are discontinuous at the point $t = \pm \pi$ along the cycle: $\alpha \rightarrow 0$ and $\beta \rightarrow 1$ as $t \rightarrow -\pi$ from above, whereas $\alpha \rightarrow 1$ and $\beta \rightarrow 0$ as $t \rightarrow \pi$ from below. In fact, at $t = \pm \pi$ the system is in the atomic state with all $A$ orbitals occupied. As discussed in the paragraph beneath Eq.~\eqref{CouplingsMPSPH}, there are two distinct configurations for the MPS parameters corresponding to this state which are attained for $t = \pm \pi$. 
		 
		 For the MPS we are considering, the discontinuity in the parametrisation $\parampump$ is related to the chiral edge mode of the Chern insulator defined by the interpolation. Indeed, charge pumping in the SSH model with open boundaries corresponds to a Chern insulator on a cylinder with periodic and open boundaries in the vertical and horizontal directions, respectively. In the topological phase with Chern number $\mathcal{C} = 1$, each edge of the cylinder hosts a one-dimensional chiral mode~\cite{bernevig2013topological}. These modes are reflected in the ES of the SSH chain along the charge pumping cycle. For instance, the single-particle ES~\cite{peschel2003calculation} of half the chain with periodic boundaries computed from the representation of the state as a Gaussian fermionic TNS along $\parampump$ is shown in Fig.~\ref{fig:SSH_ESAndPH}(c). It has both a left- and a right-moving mode which are localized at the two virtual edges and which are degenerate in the SSH topological phase at $t = \pi/2$. 
		 
		 Using the representation of the MPS in terms of local tensors, we can also compute the many-body ES of half of an infinite chain with open boundaries. Indeed, the many-body ES is isometric to the spectrum of the logarithm of the normalised and symmetrised left- and right fixed point of the MPS transfer matrix~\cite{PhysRevB.83.245134}. In the case of the MPS from Eq.~\eqref{MPSMyInterpol}, the fixed point is a matrix of dimension $2 \times 2$ with eigenvalues
		 \begin{equation}\label{ManyBodyESMPS}
		 \frac{1}{2} \pm  \frac{-\alpha ^4 + \beta ^4 + \gamma^4}{2\sqrt{\left(-\alpha ^4 + \beta ^4 + \gamma^4\right)^2 + 4 \alpha^2 \beta^2 \gamma ^4}}.
		 \end{equation}   
		 Therefore, the many-body ES has two non-trivial levels that are related by normalisation and which describe the spinless fermion at the single virtual boundary of the half-infinite chain. If and only if $\alpha ^4 = \beta ^4 + \gamma^4$, the two eigenvalues from Eq.~\eqref{ManyBodyESMPS} are degenerate. This corresponds to a crossing of the Fermi level by the edge fermion. If the interpolation describes a Chern insulator with Chern number $\mathcal{C} = 1$, a degeneracy of the eigenvalues from Eq.~\eqref{ManyBodyESMPS} should therefore occur exactly once along the cycle. For the parametrisation $\parampump$ this happens in the SSH topological phase at $t = \pi/2$ with $\alpha = \gamma$ and $\beta = 0$. Indeed, $\alpha ^4 < \beta ^4 + \gamma^4$ for $t < \pi /2$ and $\alpha ^4 > \beta ^4 + \gamma^4$ for $t> \pi /2$. Hence, the non-trivial Chern number implies that for MPS of the form of Eq.~\eqref{MPSMyInterpol} the parametrisation has to be discontinuous along the cycle in order to combine the two parts of the interpolation where $\alpha ^4 \gtrless \beta ^4 + \gamma^4$ with only a single degenerate point.
		 
		 We emphasize that this discontinuity in the MPS interpolation is required by the topology of the charge pumping cycle, even if the pumping cycle itself is infinitely often continuously differentiable. For example, the smooth deformation of the topological pumping cycle, whose existence is discussed in Sec.~\ref{sec:SSHToChern}, has an MPS ground state of the form of Eq.~\eqref{MPSMyInterpol}. Due to the non-trivial topology of the cycle, this MPS has a discontinuous interpolation despite the smoothness of its Bloch Hamiltonian.   
		
		 The MPS of Eq.~\eqref{MPSMyInterpol} can also describe cyclic interpolations corresponding to topologically trivial two-dimensional models. For instance, the MPS with parametrisation
		 \begin{equation}\label{MPSInterpolationTrivial}
		 \paramtriv : \quad (\alpha(t), \beta (t), \gamma (t)) = \left(\frac{1}{2} e^{-\left(\tan \frac{|t|}{2}\right)^{-2}}, 0, 1 \right)
		 \end{equation} 
		 for $t \in (-\pi, \pi]$ corresponds to the ground state of a model with finite chemical potential and nearest-neighbour hopping $t^{(1)}$ between unit cells, whereas the hopping within unit cells and the next-nearest neighbour hopping vanish (see Fig.~\ref{fig:SSH_ESAndPH}(b)). From the single-particle ES in Fig.~\ref{fig:SSH_ESAndPH}(d) we see that the conduction and valence bands in this system are not connected by the edge modes such that the Chern number is zero. Indeed, both the MPS parametrisation $\paramtriv$ and the corresponding Bloch parent Hamiltonian are smooth along the cycle.

	\section{Chern insulator PEPS from SSH charge pumping\label{sec:ChernPEPS}}
	 
	 	In this section we discuss how charge pumping can be used to construct tensor networks in $d+1$ dimensions starting from $d$-dimensional TNSs. Specifically, we show how the MPS from Sec.~\ref{sec:fMPS} for the SSH model along a charge pumping cycle leads to a two-dimensional PEPS for a Chern insulator. In Sec.~\ref{sec:HybridTNS} we define a hybrid real-momentum space PEPS with finite bond dimension for the Chern insulator. In Sec.~\ref{sec:PEPSFromMPS}, we perform an inverse FT in the direction $d+1$ required to transform the state to a fully real-space representation. In Sec.~\ref{sec:ChernPEPSRealSpace} we study how the resulting state can be expressed as a real-space PEPS, whose bond dimension is investigated in Sec.~\ref{sec:ChernPEPSBondDim}.
	 	
	 	\subsection{$(d+1)$-dimensional TIs from charge pumping\label{sec:ChargePumpIntro}}
	 			
		Charge pumping provides a systematic method to obtain a TI in $d+1$ spatial dimensions from a $d$-dimensional TI. Indeed, if the Bloch Hamiltonian of the $d$-dimensional model is smooth along the charge pumping interpolation as a function of the cyclic time $t \in (-\pi, \pi]$, $t$ can be identified with the momentum $k_{d+1}$ in the $(d+1)^{\text{st}}$ direction of the $(d+1)$-dimensional system. The time-dependent Hamiltonian of the $d$-dimensional model then gives the Bloch Hamiltonian of the $(d+1)$-dimensional system as a function of $k_{d+1}$. For instance, charge pumping in the SSH model with $d = 1$ defines a two-dimensional Chern insulator with Chern number $\mathcal{C} =1$ (\cf Sec.~\ref{sec:SSHToChern}).
		
		We discretize the $(d+1)^{\text{st}}$ dimension with a finite number $N_{d+1}$ of lattice sites. The discrete momentum values in $(- \pi, \pi]$ are $k_{d+1}^{(j)} = \frac{\pi}{N_{d+1}}(2j - N_{d+1})$ for even $N_{d+1}$, and $k_{d+1}^{(j)} = \frac{\pi}{N_{d+1}}(2j - N_{d+1} + 1)$ for odd $N_{d+1}$. They are identified with discrete times $t^{(j)} =  k_{d+1}^{(j)}$ for $ j = 0, \dotsc, N_{d+1} -1$. 
		
		Let us express the identification of time $t$ and momentum $k_{d+1}$ as a relation between the mode operators of the hybrid and real-space systems. The $d$-dimensional system has annihilation operators $\pmode_{\tau, \mathbf{x}}(t^{(j)})$ for the physical fermion on the orbital $\tau = 1, \dotsc, f$ on the unit cell $\mathbf{x} \in \mathbb{Z}^d$, where $f$ is the number of orbitals per unit cell. They depend on the discretized time value $t^{(j)}$ along the pumping interpolation at which the $d$-dimensional model is evaluated. For example, in the SSH model we have mode operators $\pmode_{A, x}(t^{(j)}), \pmode_{B, x}(t^{(j)})$ with $x = 0, \dotsc, N_x -1$.
		
		The $(d+1)$-dimensional model obtained from charge pumping has the same number of orbitals as the $d$-dimensional model it is derived from. For instance, the Chern insulator constructed from the SSH model also has sublattices $A$ and $B$. Therefore, the physical creation operators of the $(d+1)$-dimensional system in real space are $\pmode_{\tau, (\mathbf{x}, x_{d+1})}$, where $x_{d+1} = 0, \dotsc, N_{d+1} -1$ is the real-space coordinate in the $(d+1)^{\text{st}}$ direction and $\tau = 1, \dotsc, f$. The charge pumping construction of the $(d+1)$-dimensional model requires periodic boundary conditions in the direction $d+1$. We may therefore consider the FT of the mode operators in the direction $d+1$, while keeping the real-space coordinate $\mathbf{x}$ in the other $d$ dimensions, given by
		\begin{subequations}\label{FTDPlus1}
			\begin{equation}
			\pmode_{\tau, (\mathbf{x}, k_{d+1}^{(j)})} = \sum_{x_{d+1} = 0}^{N_{d+1} - 1} \mathcal{F}_{k_{d+1}^{(j)}x_{d+1}} \pmode_{\tau, (\mathbf{x},x_{d+1})}
			\end{equation}
			with
			\begin{equation}
			\mathcal{F}_{k_{d+1}^{(j)}x_{d+1}} = \frac{e^{-i k_{d+1}^{(j)}x_{d+1}}}{\sqrt{N_{d+1}}}.
			\end{equation}
		\end{subequations}
				
		In terms of these mode operators, the identification of time $t$ and momentum $k_{d+1}$ is the expressed by the identity
		\begin{equation}\label{IdentificationModes}
		\pmode_{\tau, \mathbf{x}}(t^{(j)}) = \pmode_{\tau, (\mathbf{x}, k_{d+1}^{(j)})}
		\end{equation}				 
		for all d-dimensional unit cells $\mathbf{x}$ and sublattices $\tau = 1, \dotsc, f$. For example, in the case of the SSH charge pumping we have $\pmode_{\tau, x} (t^{(j)}) = \pmode_{\tau, (x, k_y^{(j)})}$ for $\tau = A, B$ and $x = 0, \dotsc, N_x -1$.
		
		The Bloch Hamiltonian of the $(d+1)$-dimensional model is the time-dependent Hamiltonian of the $d$-dimensional system. Hence, with the identification of Eq.~\eqref{IdentificationModes}, the $(d+1)$-dimensional ground state is given by the direct product 
		\begin{equation}\label{GSFromPumpingGeneral}
		\ket{\psi_{d+1}} = \bigotimes _{j =0}^{N_{d+1}-1} \ket{\psi_{d}(t^{(j)})}
		\end{equation}		
		of the many-body ground states $\ket{\psi_{d}(t)}$ of the $d$-dimensional model evaluated at the $N_{d+1}$ discrete times along the interpolation. From Eq.~\eqref{IdentificationModes} it is clear that this defines the ground state \wrt hybrid $(d+1)$-dimensional real-momentum space coordinates $(\mathbf{x}, k_{d+1}^{(j)})$. 
		
		In order to obtain the state in terms of $(d+1)$-dimensional real-space coordinates $(\mathbf{x},x_{d+1})$, we need to apply the inverse of the FT of Eq.~\eqref{FTDPlus1} to the hybrid state of Eq.~\eqref{GSFromPumpingGeneral}. If the $d$-dimensional pumping Bloch Hamiltonian is smooth as a function of time, the real-space correlation functions are guaranteed to decay faster than any polynomial~\cite{PhysRevB.90.115133}.

		\subsection{Hybrid real-momentum space Chern PEPS\label{sec:HybridTNS}}
		
		We now specialize this construction, which we explained above in terms of generic free-fermionic TIs, to $d$-dimensional TIs which are described by a GfTNS at all times along their charge pumping cycle. We will thereby obtain a GfTNS for the $(d+1)$-dimensional TI.
		
		For the remainder of this section and for pedagogical purposes, we will focus on the one-dimensional case (\ie $d=1$) and the corresponding notation. We will mostly rely on charge pumping in the SSH model which is described by the MPS with bond dimension $D_{\mathrm{SSH}}= 2$ from Eq.~\eqref{SSHMPSState}. Hence, we will obtain a Gaussian fermionic PEPS for the Chern insulator with Chern number $\mathcal{C} =1$. Extensions to other models and higher dimensions are straightforward. 
		
		In case the $d$-dimensional ground state along the pumping cycle is given as a GfTNS, Eq.~\eqref{GSFromPumpingGeneral} allows to define the $(d+1)$-dimensional ground state as a hybrid TNS, where the first $d$ dimensions correspond to real space and the $(d+1)^{\text{st}}$ dimension is expressed in momentum space. Indeed, the local tensor of the hybrid TNS at the position $(\mathbf{x}, k_{d+1}^{(j)})$ is given by the local tensor of the $d$-dimensional model at the $d$-dimensional real-space position $\mathbf{x}$ and time $t^{(j)}$.
		
		Thus, the virtual DOFs of the hybrid TNS in the first $d$ directions are identical to those of the $d$-dimensional model. In other words, the identification of time and momentum from Eq.~\eqref{IdentificationModes} holds not just for the physical mode operators, but also for the virtual mode operators in the first $d$ dimensions.
		
		On the other hand, due to the direct product in the direction of $k_{d+1}$ in Eq.~\eqref{GSFromPumpingGeneral}, the hybrid TNS does not need virtual legs in the direction $d+1$, and we say that the bond dimension in this direction is equal to one (implying that the contraction of this direction corresponds trivially to a product).

				\begin{figure*}[th]
						\includegraphics[width = 0.99\linewidth]{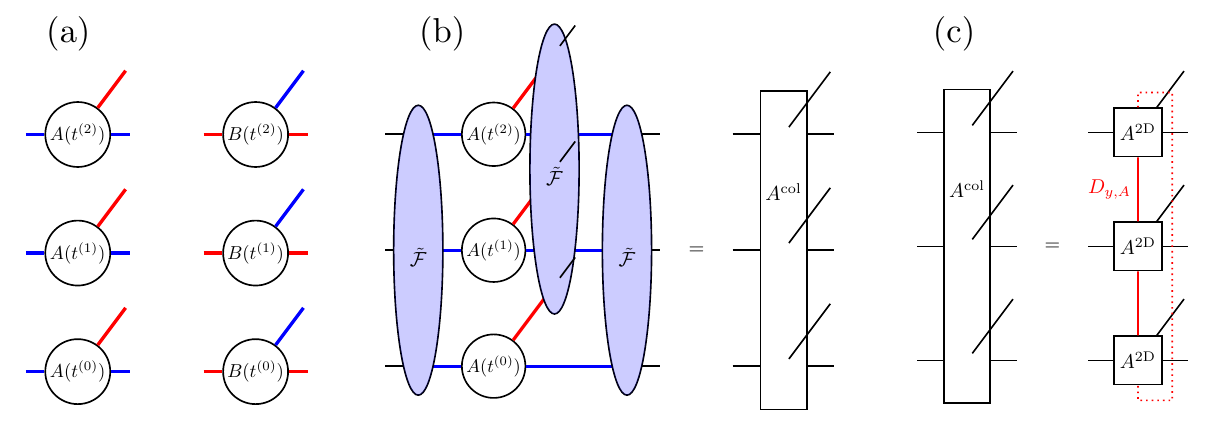}	
					\caption{(a) Hybrid real-momentum space PEPS for $N_y = 3$ sites in the vertical direction obtained by stacking rows of the MPS of Eq.~\eqref{MPSMyInterpol} evaluated at different times $t^{(j)}$ along the charge pumping interpolation. Physical and virtual legs transforming as particles and holes under the $\mathrm{U}(1)$ symmetry of Eq.~\eqref{LocalU1} are marked in red and blue, respectively. Since the vertical bond dimension is trivial, \ie equal to one, it is omitted in this sketch. (b)~Inverse FT $\tilde{\mathcal{F}}$ acting separately on the physical and horizontal virtual legs of one column of $A$ sites of the hybrid PEPS. The result defines the column tensor $\acol$ for the real-space PEPS. (c)~Decomposition of the column tensor $\acol$ into the contraction of local PEPS tensors $\atwod$ for the two-dimensional state. The vertical virtual legs of $\atwod$ marked in red have a bond dimension $\dvert{A}$ which generally grows with the system size $N_y$ due to the non-locality of the inverse FT. The inverse FT and decomposition of a column of $B$ sites is analogous.\label{fig:SketchChernPEPSfromMPS}}
				\end{figure*}

		The hybrid Chern PEPS obtained from the SSH charge pumping MPS of Eq.~\eqref{SSHMPSState} is sketched in Fig.~\ref{fig:SketchChernPEPSfromMPS}(a) in the hybrid coordinate system where the horizontal axis describes $x$ and the vertical axis corresponds to $k_y \equiv k_{d+1}$. The local tensor for a site on the $A$ sublattice at the position $(x, k_y^{(j)})$ is given by the SSH local tensor $A^{i}_{lr}(t^{(j)})$ at the time $t^{(j)}$ along the cycle, and similarly for the $B$ sublattice. In the horizontal direction, the two-dimensional hybrid state hence inherits both the translation invariance and the constant finite bond dimension $D_{\mathrm{SSH}}$ of the MPS. As discussed in the previous paragraph, virtual legs in the vertical direction are not needed for the hybrid Chern PEPS and are hence not shown in Fig.~\ref{fig:SketchChernPEPSfromMPS}(a).

		\subsection{Inverse Fourier transform\label{sec:PEPSFromMPS}} 		
		
		As discussed below Eq.~\eqref{GSFromPumpingGeneral}, the hybrid state obtained from charge pumping is mapped to a $(d+1)$-dimensional real space coordinate system by applying an inverse FT in the direction $d+1$ to the fermionic mode operators. For GfTNSs -- which have virtual in addition to physical DOFs -- the inverse FT should be applied to both the physical mode operators and the virtual fermionic mode operators in the first $d$ directions. For the hybrid Chern PEPS, we therefore apply the inverse FT in the vertical direction to the physical and the horizontal virtual legs. The extension of the FT to the horizontal virtual modes amounts to a virtual basis change and does not alter the physical state, but ensures that the real-space PEPS is invariant under vertical translations $y \mapsto y + 1$ of its physical and virtual legs (see also Appendix~\ref{sec:AppendixColumnCM}). 
		
		For the hybrid Chern PEPS, the correct definition of the inverse FT in the vertical direction $y$ entails a subtlety. Indeed, recall that the SSH charge pumping MPS from Eq.~\eqref{SSHMPSState} is defined \wrt mode operators $\pmodenew_{\tau, x}(t^{(j)})$ related to the $\pmode_{\tau, x}(t^{(j)})$ used in Eq.~\eqref{IdentificationModes} by the particle-hole transformation of Eq.~\eqref{PHTrafo}. For the Chern PEPS, we perform an analogous particle-hole transformation in two-dimensional real space and define new mode operators
		\begin{subequations}\label{PHTrafoChern}
		\begin{gather}
		\pmodenew_{A, (x,y)} = \pmode_{A, (x,y)},\\
		\pmodenew_{B, (x,y)} = \pmode^{\dagger}_{B, (x,y)}.
		\end{gather}
		\end{subequations}
		The operators $\pmodenew_{A, (x,y)}, \pmodenew^{\dagger}_{A, (x,y)}, \pmodenew_{B, (x,y)}, \pmodenew^{\dagger}_{B, (x,y)}$ span the basis in which we want to express the real-space Chern PEPS.
		
		The SSH particle-hole transformation of Eq.~\eqref{PHTrafo} acts within the set of operators at time $t^{(j)}$. Since we identify time and momentum, this is equivalent to a particle-hole transformation acting on the modes of the Chern PEPS in momentum space rather than real space like in Eq.~\eqref{PHTrafoChern}. Due to the anti-unitarity of the particle-hole transformation, the inverse FT relating the modes $\pmodenew_{B, x} (t^{(j)})$ and $\pmodenew_{B, (x,y)}$ on the $B$ sublattice therefore contains an additional complex conjugation. Hence, the physical mode operators $\pmodenew_{\tau, x} (t^{(j)})$ of the SSH models, providing the basis for the hybrid Chern PEPS, and the mode operators $\pmodenew_{\tau, (x,y)}$ for the real-space Chern PEPS are related as
		\begin{subequations}\label{InverseFTPHComplete}
		\begin{equation}
		\pmodenew_{\tau, (x,y)} = \sum_{j = 0}^{N_y-1}  \tilde{\mathcal{F}}_{\tau, \tau; y,t^{(j)}} \pmodenew_{\tau, x}(t^{(j)})
		\end{equation}
		with the inverse vertical FT
		\begin{equation}\label{InverseFTPH}
		\tilde{\mathcal{F}}_{\tau, \tau'; y,t^{(j)}} = \frac{\delta_{\tau, \tau' }}{\sqrt{N_y}} e^{i \eta_{\tau}  y t^{(j)}}.
		\end{equation}
		\end{subequations}
		Here, $\eta_A = 1$ and $\eta_B = -1$ for the physical modes which are particle-like on the $A$ sublattice and hole-like on the $B$ sublattice.

		The inverse FT of the horizontal virtual legs of the hybrid PEPS is analogous to Eq.~\eqref{InverseFTPHComplete} for the physical modes. Here, the particle- or hole-like character of the virtual modes is determined by their transformation under the $\mathrm{U}(1)$ symmetry of Eq.~\eqref{LocalU1}. Specifically, the left and right virtual legs on the $A$ sublattice transform as holes, \ie $\eta_{L, A} = \eta_{R, A} = -1$, whereas the left and right virtual legs on the $B$ sublattice transform as particles such that $\eta_{L, B} = \eta_{R, B} = 1$. 
		
		Let us now study how the inverse vertical FT $\tilde{\mathcal{F}}$ acts on the hybrid Chern PEPS. Due to the translation invariance in the horizontal direction, it is enough to consider one column of the hybrid state given by the sites $(x, k_y^{(j)})$ for a fixed horizontal position $x$ and $j = 0, \dotsc, N_y -1$ (see Fig.~\ref{fig:SketchChernPEPSfromMPS}(a)). Under $\tilde{\mathcal{F}}$, the hybrid column is mapped to one column of the two-dimensional real-space state, given by the sites $(x, y)$ with $y = 0, \dotsc, N_y -1$. 
		
		Since the inverse FT $\tilde{\mathcal{F}}$ is non-local, for a generic interpolation there are long-range correlations in the real-space column $\{(x, y)\}_{0 \leq y \leq N_y -1}$ of the PEPS. In the tensor network language, the real-space column is therefore described by a single tensor obtained from the application of $\tilde{\mathcal{F}}$ to one column of the hybrid PEPS (in contrast to the hybrid column $\{(x, k_y^{(j)})\}_{0 \leq j \leq N_y -1}$ which is described by a product of individual tensors for each $k_y^{(j)}$, signifying the absence of correlations in the vertical direction). For a real-space column of $A$ sites, we denote this tensor by $\acol$ as shown in Fig.~\ref{fig:SketchChernPEPSfromMPS}(b). Similarly, we define a tensor $\bcol$ for a real-space column of $B$ sites. The tensors $\acol$ and $\bcol$ have $N_y$ physical legs of dimension $\physdim = 2$ and $N_y$ left and right virtual legs of dimension $D_{\mathrm{SSH}} = 2$, which describe the physical and horizontal virtual DOFs of the real-space PEPS.

		The SSH pumping MPS is a Gaussian MPS for free fermions. Thus, the column tensors $\acol$ and $\bcol$ are also Gaussian states, which can be described by their covariance matrices (CMs). These CMs are computed in App.~\ref{sec:AppendixColumnCM} for the general $d$-dimensional case. There we show that the CMs of the real-space columns $\acol$ and $\bcol$, defined by the application of the inverse FT of Eq.~\eqref{InverseFTPHComplete} to the physical and horizontal virtual legs of the hybrid columns, are given by the inverse FT of the time-dependent CMs describing the local SSH tensors $A^i_{lr} (t^{(j)})$ and $B^i_{lr} (t^{(j)})$ along the pumping cycle. This result fully characterizes the column tensors $\acol$ and $\bcol$.

		\subsection{Translation invariant real-space PEPS\label{sec:ChernPEPSRealSpace}}

		From the previous subsections, we are now ready to discuss how the two-dimensional real-space state can be expressed as a TNS with local tensors for each site. For that purpose, the column tensors $\acol$ and $\bcol$ have to be decomposed into a column of local PEPS tensors $\atwod$ and $\btwod$ with both horizontal virtual legs of dimension $D_{\mathrm{SSH}}$ and vertical virtual legs with dimension $\dvert{A}$ and $\dvert{B}$, respectively. We require the local PEPS tensors to be identical on all sites of the same sublattice as shown in Fig.~\ref{fig:SketchChernPEPSfromMPS}(c) in order to make the invariance under real-space translations explicit.
		
		The decomposition of the column into PEPS tensors can be achieved using tools developed for one-dimensional MPSs~\cite{2019arXiv190610144S, perez2006matrix}. To that end, we define one-dimensional pure states $\aoned$ and $\boned$, whose many-body wave functions are the tensor elements of $\acol$ and $\bcol$, respectively. Therefore, $\aoned$ effectively describes a chain of length $N_y$, where the local Hilbert space at each site of the fictitious chain is the tensor product of the physical and horizontal virtual Hilbert spaces of $\acol$ at the corresponding physical site (and similarly for $\boned$ and $\bcol$). Pictorially, $\aoned$ and $\boned$ are obtained by merging at each site the physical and horizontal virtual legs of $\acol$ and $\bcol$ into an effective physical index of dimension $\deff = \physdim D_{\mathrm{SSH}}^2$ per site.
		
		Our goal is now to express $\aoned$ and $\boned$ as translation invariant MPSs with periodic boundary conditions. Then, after unfolding the effective physical index into the physical and horizontal virtual indices of the PEPS, the local tensors of these MPSs will define the PEPS tensors $\atwod$ and $\btwod$, respectively. Similarly, the MPS bond dimensions $\dvert{A}$ and $\dvert{B}$ are equal to the vertical bond dimensions of the PEPS local tensors $\atwod$ and $\btwod$, respectively.
		
		Numerically, the translation invariant MPSs for $\aoned$ and $\boned$ can be obtained as follows. Multiple steps are necessary since the common and stable MPS algorithms rely on the presence of open boundary conditions and do not directly find a periodic translation invariant MPS for a translation invariant state. Instead, in a first step the Gaussian states $\aoned$ and $\boned$ can be decomposed into non-translation invariant Gaussian MPSs with open boundary conditions by performing successive Schmidt decompositions of the system~\cite{PhysRevB.92.075132, 2019arXiv190610144S}. We are not aware of any method to transform a Gaussian open-boundary MPS into a Gaussian translation invariant MPS. We therefore interpret the Gaussian MPSs for $\aoned$ and $\boned$ as regular fermionic open-boundary MPSs by choosing physical and virtual basis states and computing the local MPS tensors for each site. Finally, these open-boundary MPSs are transformed into translation invariant and generically non-Gaussian MPSs with periodic boundary conditions following the standard procedure described in Ref.~\onlinecite{perez2006matrix}. This approach leads to the bond dimension of the translation invariant MPS being given by the sum of the bond dimensions of the open-boundary MPS for each site. Therefore, the bond dimension grows at least linearly with the system size~\cite{perez2006matrix}.

		\subsection{Vertical PEPS bond dimension\label{sec:ChernPEPSBondDim}}
		
		Above, we showed that the inverse FT of the hybrid PEPS can be decomposed into a $(d+1)$-dimensional real-space TNS, where the bond dimension in the first $d$ directions is equal to that of the original $d$-dimensional state. The core question is the scaling of the bond dimension in the $(d+1)^{\mathrm{st}}$ real space dimension \wrt the system size in this direction.

		\subsubsection{Lower bound from ES}
		
		The bond dimension of a tensor network is intimately related to the entanglement between its physical particles: For a TNS describing a physical system with a subsystem $\mathcal{A}$, the total dimension of all virtual legs at the boundary $\partial \mathcal{A}$ can be no smaller than the rank of the state's Schmidt decomposition into the DOFs of $\mathcal{A}$ and its complement. This constraint allows us to infer a lower bound for the vertical bond dimension of the real-space PEPS from the ES of the one-dimensional column states. 

			\begin{figure}[t]		
				\includegraphics[width = 0.5\linewidth]{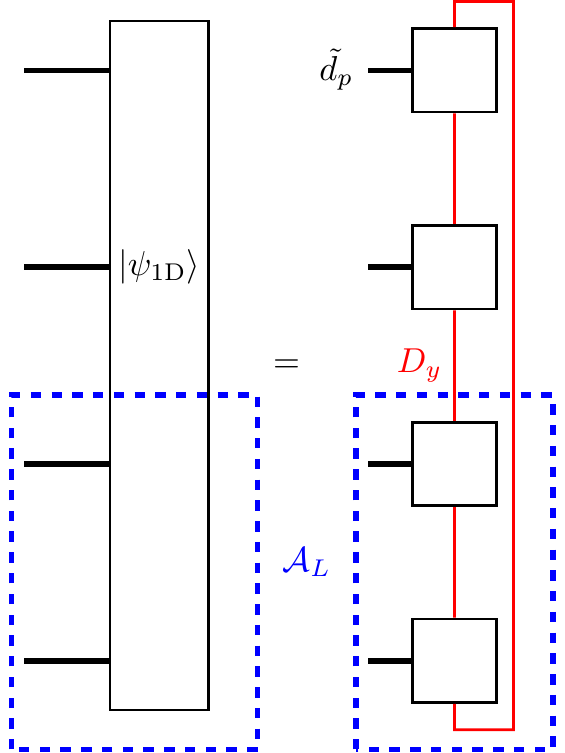}	
				\caption{Schmidt decomposition of the one-dimensional effective state $\psioned$ for the PEPS column with physical dimension $\deff$ into the subsystem $\mathcal{A}_L$ of the first $L$ sites and its complement. When $\psioned$ is written as a translation invariant MPS with bond dimension $\dvertbare$, the virtual boundary of the subsystem $\mathcal{A}_L$, marked in blue, crosses two bonds of the MPS.\label{fig:ColumnDecomposition}}
			\end{figure}		
		
		We consider a generic one-dimensional effective column state denoted $\psioned$ with an effective physical dimension $\deff$. Below, we will choose either $\psioned = \aoned$ or $\psioned = \boned$ with $\deff= \physdim D_{\mathrm{SSH}}^2$. We assume that $\psioned$ is expressed as a translation invariant MPS with a bond dimension $\dvertbare$ corresponding to the vertical PEPS bond dimension (see Fig.~\ref{fig:ColumnDecomposition}). Such a representation can be obtained for instance using the procedure described in the previous subsection. We now perform a Schmidt decomposition of the pure state $\psioned$ \wrt the subsystem $\mathcal{A}_L$ of the first $L$ sites $0 \leq y \leq L-1$, and denote the rank of the decomposition by $\rank_L(\psioned)$. As sketched in Fig.~\ref{fig:ColumnDecomposition}, the cut between $\mathcal{A}_L$ and its complement crosses exactly two virtual bonds of the MPS, namely those between sites $(L-1, L)$ and $(N_y-1, 0)$, each of dimension $\dvertbare$. Since the virtual dimension can be no smaller than the Schmidt rank for any subsystem size $L$, we obtain the bound
		\begin{equation}\label{BondDimensionOBCES}
		\dvertbare ^2 \geq \max_{1 \leq L \leq N_y -1} \rank_L(\psioned).
		\end{equation}

		The Schmidt rank $\rank_L(\psioned)$ may be obtained by counting the non-trivial levels in the single-particle ES of the Gaussian column state $\psioned$. This ES can be computed numerically based on the explicit result for the CM of $\psioned$ given in Appendix~\ref{sec:AppendixColumnCM}. Since the state $\psioned$ with physical dimension $\deff$ has $\log_2 (\deff)$ free fermionic modes per lattice site, its single-particle ES \wrt the subsystem $\mathcal{A}_L$ consists of levels $0 \leq |\lambda_i(L)| \leq 1$ with $ 1 \leq i \leq L \log_2 (\deff)$ (see Eq.~\eqref{DefSPES} for a definition of the single particle ES and its levels $|\lambda_j|$). For an improved numerical stability, we actually compute $\mu_i(L) = \sqrt{1 - \lambda_i(L)^2}$ that may be conveniently extracted from the CM (see Eq.~\eqref{OffDiagonalSPES}). Here, values $\mu_i(L) = 1$ and $\mu_i(L) = 0$ correspond to maximally entangled and non-entangled modes, respectively. The many-body Schmidt rank of the state $\psioned$ is therefore given by the exponential
		\begin{equation}
		\rank_L(\psioned) = 2^{\# \{\mu_i( L) \, > \, 0\}}
		\end{equation}
		of the number of entangled modes in the single-particle ES with $\mu_i(L)$ bigger than zero.

				\begin{figure*}[t]		
					\includegraphics[width = 0.99\linewidth]{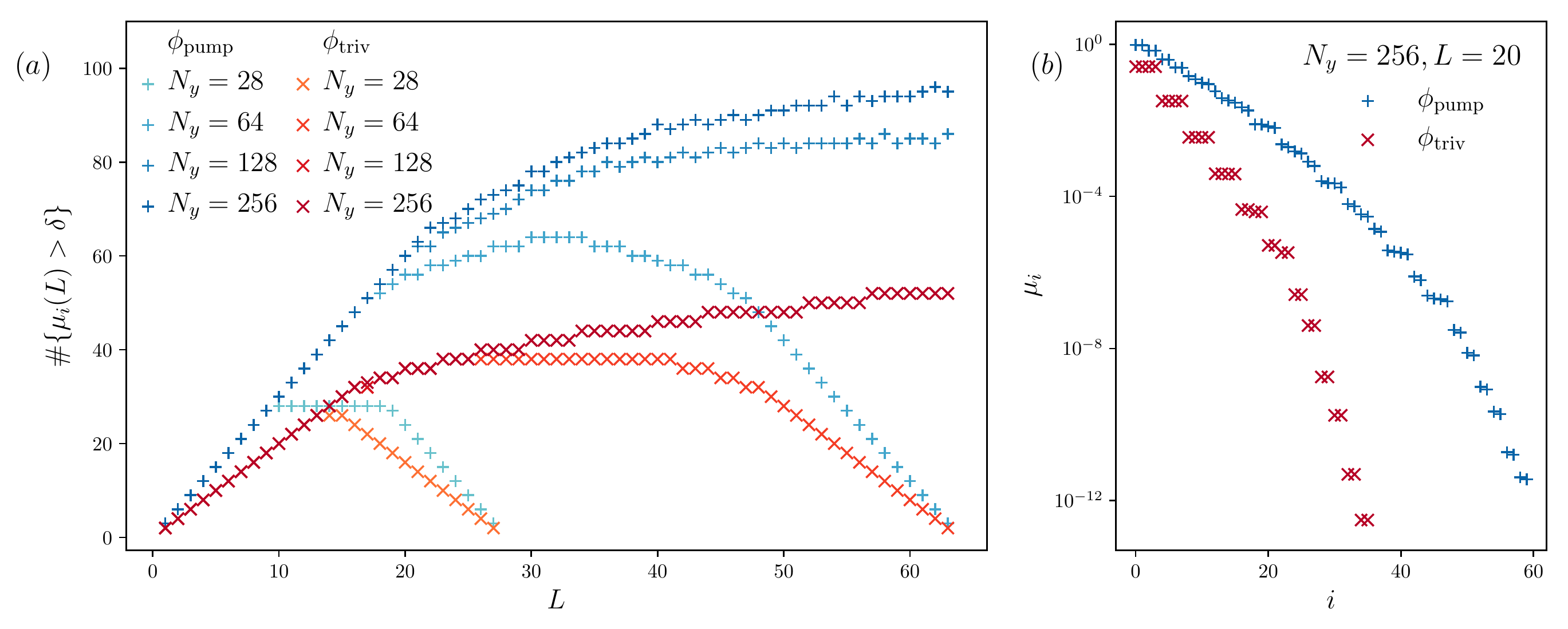}	
					\caption{Single-particle ES of a column $\acol$ of the two-dimensional real-space TNSs defined by two different parametrisations for the MPS of Eq.~\eqref{MPSMyInterpol}: the charge pumping interpolation $\parampump$ of Eq.~\eqref{ParametrisationDimerized} leading to a Chern insulator, and the parametrisation $\paramtriv$ of Eq.~\eqref{MPSInterpolationTrivial} leading to a topologically trivial two-dimensional state. (a)~Number of modes in the single-particle ES with a finite entanglement corresponding to a value $\mu_i > \delta$ with $\delta = 10^{-13}$ the numerical accuracy for different system sizes $N_y$ as a function of the subsystem size $L$. (b)~Single-particle ES for a column with $N_y = 256$ sites \wrt the subsystem $\mathcal{A}_{L}$ with $L = 20$. The double degeneracy in the ES corresponding to $\paramtriv$ is due to the decoupling of all right virtual legs since $\beta(t) = 0$ along $\paramtriv$.\label{fig:ESPumpingPEPS}}
				\end{figure*}	

		\subsubsection{Exponential growth}
		 
		The number of entangled modes in the single-particle ES of a column $\acol$ of unit cells of the real-space PEPS is displayed in Fig.~\ref{fig:ESPumpingPEPS}(a) for two different cyclic interpolations of the MPS from Eq.~\eqref{MPSMyInterpol}: on one hand, the parametrisation $\parampump$ of Eq.~\eqref{ParametrisationDimerized} giving rise to a two-dimensional Chern insulator, and on the other hand the parametrisation $\paramtriv$ of Eq.~\eqref{MPSInterpolationTrivial} corresponding to a topologically trivial two-dimensional state. The corresponding data of the column $\bcol$ is identical.
		
		For $\parampump$ the number of entangled modes is equal to
		\begin{equation}\label{NumberModesESExact}
		\# \{\mu_i( L) \, > \, 0\} = \min \{ 3L, \, N_y\}
		\end{equation}
		when $L \leq N_y /2$ (the spectra for $L$ and $N_y - L$ are identical). In Appendix~\ref{sec:AppendixESUOne} we show that this is the maximal number of entangled modes which is compatible with the global $\mathrm{U}(1)$ symmetry of $\acol$ inherited from the SSH model MPS. Here, the factor $3$ in Eq.~\eqref{NumberModesESExact} is a consequence of the column tensor having three fermions (one physical fermion and two virtual fermions) per site. The validity of Eq.~\eqref{NumberModesESExact} can be seen in Fig.~\ref{fig:ESPumpingPEPS}(a) for the smallest system size $N_y = 28$. For larger system sizes, the number of entangled modes shown in Fig.~\ref{fig:ESPumpingPEPS}(a) is lower than Eq.~\eqref{NumberModesESExact} because of the finite numerical resolution, as shown by the ES in Fig.~\ref{fig:ESPumpingPEPS}(b). We have checked that for the Chern PEPS obtained from the MPS corresponding to the smooth charge pumping cycle described in Sec.~\ref{sec:SSHToChern}, the number of entangled modes is also given by Eq.~\eqref{NumberModesESExact}.
		
		For $\paramtriv$, the number of entangled modes is given by $2L$ when $L \leq N_y /2$. This is lower than Eq.~\eqref{NumberModesESExact} for $\parampump$ except at $L = N_y /2$, where both numbers agree. As discussed in Appendix~\ref{sec:AppendixESUOne}, this reduction of entangled modes is due to the decoupling of all right virtual legs and is not related to topology. Indeed, for the trivial cycle, the parameter $\beta (t) = 0$ vanishes for the entire duration of the interpolation $\paramtriv$. Any small perturbation of $\paramtriv$ by a non-zero $\beta$ increases the number of entangled modes to Eq.~\eqref{NumberModesESExact}. Moreover, these additional decoupled modes disappear when blocking the two columns $\acol$ and $\bcol$, \ie by contracting the bonds connecting them. Hence, the column for $AB$ unit cells has the same Schmidt rank for both $\parampump$ and $\paramtriv$. 
		
		The number of entangled modes from Eq.~\eqref{NumberModesESExact} corresponds to a maximal Schmidt rank
		\begin{equation}\label{MaxSchmidtRank}
		\max_{1 \leq L \leq N_y -1} \rank_L(\aoned) = 2^{N_y}
		\end{equation}
		for the state $\aoned$, and similarly for $\boned$. The bound of Eq.~\eqref{BondDimensionOBCES} therefore implies that the vertical bond dimensions $\dvert{A}$ and $\dvert{B}$ grow exponentially as a function of $N_y$. This is a generic feature of our construction and \emph{not} related to the topology of the state: It originates in the non-locality of the inverse FT which couples all states in the exponentially growing Hilbert space of one column. Indeed, a generic quantum state requires an exponentially growing bond dimension to be represented \emph{exactly} as a TNS. 
		
		Despite the faster decay of the single-particle entanglement energies for the topologically trivial state than for the Chern insulator in Fig.~\ref{fig:ESPumpingPEPS}(b), this does not allow us to make any statement about differences in the growth of the vertical bond dimension required for an \emph{approximative} PEPS for the two systems.

	\section{Fermionic PEPS for two-dimensional higher order TI\label{sec:2DHOTI}}
	
	In this section, we study a Gaussian fermionic PEPS for the topological quadrupole model from Ref.~\cite{benalcazar2017quantizedScience}, which is reviewed in Sec.~\ref{sec:Quadrupole}. The PEPS is defined in Sec.~\ref{sec:QuadrupolePEPS}, where we also provide its parent Hamiltonian. In Sec.~\ref{sec:3DPEPS} we discuss a three-dimensional PEPS with chiral hinge states obtained from a dipole pumping interpolation of the quadrupole model~\cite{schindler2018higher, benalcazar2017quantized}.
	
	\begin{figure*}[t]
	\includegraphics[width = \linewidth] {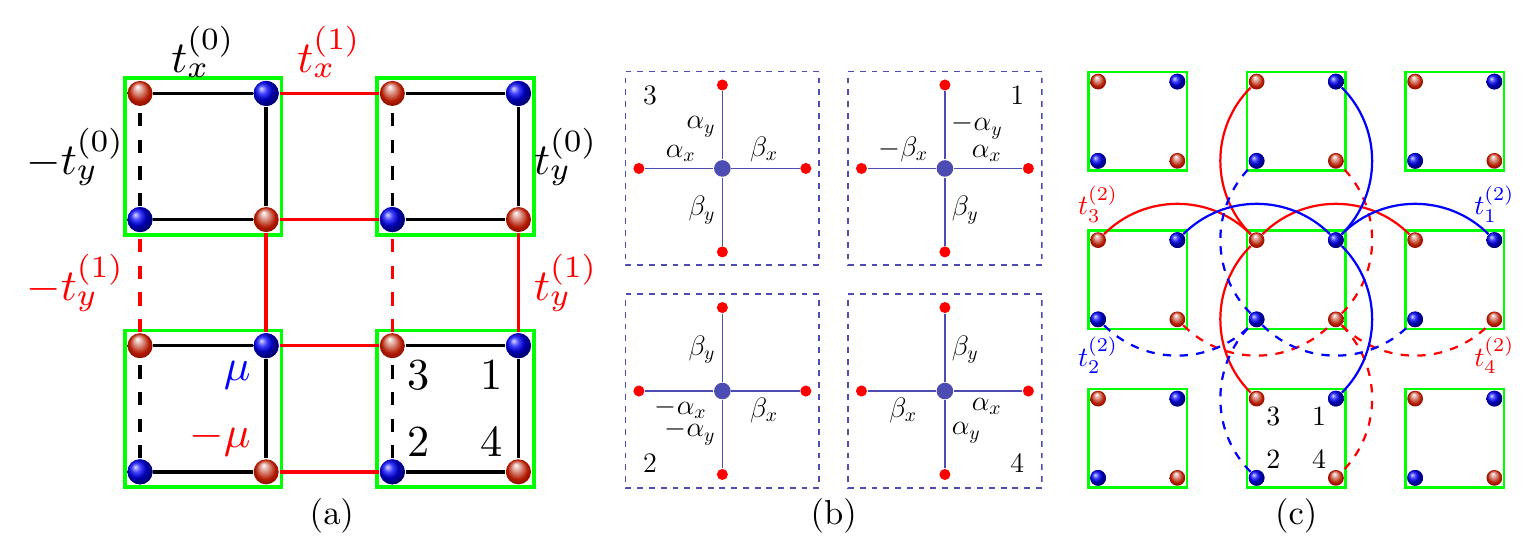}
	\caption{(a) Sketch of the quadrupole model from Ref.~\cite{benalcazar2017quantized}. A unit cell (marked with a green square) contains four sites. The horizontal and vertical nearest-neighbour hoppings $t_x^{(0)}$ and $t_y^{(0)}$ within unit cells are sketched in black, whereas the nearest-neighbour hoppings $t_x^{(1)}$ and $t_y^{(1)}$ between unit cells are sketched in red. The couplings corresponding to dashed bonds carry a negative sign to ensure a flux $\pi$ through every plaquette. Lattice sites marked in blue and red have a chemical potential $\mu$ and $-\mu$, respectively. (b)~Sketch of a unit cell of the quadrupole PEPS from Eq.~\eqref{QuadrupolePEPS}. For each site 1, 2, 3, or 4 marked by a dashed blue square, the blue/red circles denote its physical/virtual fermions. Each edge connecting a physical and virtual fermion is labeled with the amplitude of the local tensor when these two fermions are in state $\ket{1}$ and all other virtual fermions of the same site are in state $\ket{0}$. (c) Next-to-nearest neighbour hopping terms of Eq.~\eqref{NNNHoppingQuadrupolePEPS} in the parent Hamiltonian $H_{\mathrm{PEPS}}$.\label{fig:QuadrupolePEPS}}
	\end{figure*}
		
		\subsection{Second order quadrupole insulator\label{sec:Quadrupole}}
		A two-dimensional second-order topological phase with a quantized bulk quadrupole moment was recently proposed theoretically~\cite{benalcazar2017quantizedScience} and has subsequently been observed experimentally in mechanical~\cite{serra2018observation}, photonic~\cite{mittal2018photonic, hassan2018cornerphotonic, xie2018visualizationphotonic, yang2019gapped} and electrical~\cite{peterson2018quantized,imhof2018topolectrical} systems. This phase is described by a microscopic free-fermionic model with one spinless fermionic mode per site and a unit cell of $2 \times 2$ sites depicted in Fig.~\ref{fig:QuadrupolePEPS}(a). We consider the system at half-filling where only the lowest two bands are occupied. With open boundary conditions, the nearest-neighbour Hamiltonian is given by
		\begin{multline}\label{QuadrupoleHam}
		H_{\mathrm{Quad}} = \sum_{\mathbf{x}} \sum _{j = 0,1} \Big[t_x^{(j)} \left(\pmode^{\dagger}_{1, \mathbf{x}}\pmode_{3, \mathbf{x} + j\mathbf{\hat{x}}} +\pmode^{\dagger}_{4, \mathbf{x}}\pmode_{2, \mathbf{x}+ j\mathbf{\hat{x}}}\right) \\
		+t_y^{(j)} \left(\pmode^{\dagger}_{1, \mathbf{x}}\pmode_{4, \mathbf{x} + j\mathbf{\hat{y}}} -\pmode^{\dagger}_{3, \mathbf{x}}\pmode_{2, \mathbf{x}+ j\mathbf{\hat{y}}} \right) + h.c.\Big],
		\end{multline} 
		where $\mathbf{\hat{x}}$ and $\mathbf{\hat{y}}$ denote the unit vectors in the horizontal and vertical direction, respectively. The positions of the unit cells are $\mathbf{x} = x\mathbf{\hat{x}} +y\mathbf{\hat{y}}$ with $0 \leq x \leq N_x -1$ and $0 \leq y \leq N_y -1$ on a lattice with $N_x$ and $N_y$ unit cells in the horizontal and vertical direction. For $\tau = 1, \dotsc, 4$, $\pmode^{\dagger}_{\tau, \mathbf{x}}$ denotes the creation operator for a fermion in orbital $\tau$ and unit cell at position $\mathbf{x}$. The Hamiltonian of Eq.~\eqref{QuadrupoleHam} contains four nearest-neighbour hopping amplitudes $t_x^{(0)}$, $t_y^{(0)}$, $t_x^{(1)}$ and $t_y^{(1)}$, where the subscripts $x$ and $y$ refer to hoppings in the horizontal and vertical direction, and the superscripts $(0)$ and $(1)$ indicate hopping between sites on the same and adjacent unit cells, respectively. The signs of the hopping amplitudes ensure that there is a flux $\pi$ through every plaquette of the square lattice. 
		
		The quadrupole model of Eq.~\eqref{QuadrupoleHam} is in a second-order topological OAI phase protected by the horizontal and vertical mirror symmetries $M_x$ and $M_y$ if $ t_x^{(0)}/t_x^{(1)} \in \interval[open]{-1}{1} $ and $t_y^{(0)}/t_y^{(1)} \in  \interval[open]{-1}{1}$~\cite{benalcazar2017quantizedScience}. In this phase, corners host gapless protected states at the intersection of two gapped edges. Moreover, the system possesses a quantized bulk quadrupole moment, edge dipole moment and corner charge~\cite{benalcazar2017quantized}. When the hopping amplitudes $t_x^{(0)}$ and $t_y^{(0)}$ within unit cells vanish in the OAI phase, the model is in a dimerized OAI phase where every site in the bulk is contained in a decoupled plaquette shifted from the unit cell by one site in the horizontal and vertical directions. Hence, the corner modes are fully localized on the four corner sites of a rectangular patch. On the other hand, for $ |t_x^{(0)}/t_x^{(1)}| > 1 $ or $| t_y^{(0)}/t_y^{(1)}| > 1 $ the model of Eq.~\eqref{QuadrupoleHam} is in a trivial phase with gapped edges and corners. If the hoppings $t_x^{(1)}$ and $t_y^{(1)}$ between adjacent unit cells vanish, the system is in a trivial dimerized phase where each unit cell decouples from the rest of the system.

		Similarly to the charge pumping cycle discussed in Sec.~\ref{sec:SSHToChern}, a dipole pumping cycle interpolating between the trivial and OAI dimerized phases of the quadrupole model can be defined by adding the chemical potential $\mu$ to all sites on the sublattices $1$ and $2$ and $-\mu$ to all sites on the sublattices $3$ and $4$~\cite{benalcazar2017quantized}. Therefore, the staggering pattern breaks the symmetries protecting the topological phase but preserves the $C_2$ rotation symmetry (see Fig.~\ref{fig:QuadrupolePEPS}(a)). The dipole pumping cycle is obtained from $C_4$ symmetric hopping amplitudes $t^{(0)} \equiv t_x^{(0)} = t_y^{(0)}$ and $t^{(1)} \equiv t_x^{(1)} = t_y^{(1)}$ and a chemical potential $\mu$ evolving according to the same cyclic interpolation of Eq.~\eqref{DimerizedInterpolationHamiltonian} as for the SSH charge pump.

		The properties of the dipole pumping cycle for the quadrupole model are analogous to those of the charge pumping cycle for the SSH model. In particular, the system remains in a dimerized state throughout the interpolation since at all times $t \in [-\pi,\pi]$ only one of the two hopping amplitudes $t^{(0)}(t)$ and $t^{(1)}(t)$ is non-zero. At $t = \pm\pi$ and $t = 0$, the system is in an atomic state with only the orbitals 1 and 2 occupied at $t = \pm\pi$ and only the orbitals 3 and 4 occupied at $t = 0$. For $-\pi < t < 0$, the hopping within unit cells is non-zero, whereas the hopping between unit cells is non-zero for $0 < t < \pi$. In both cases, charge gets transferred by the changing chemical potential. At $t = -\pi/2$ and $t = \pi /2$, the chemical potential vanishes such that the mirror symmetries are restored and the Hamiltonian corresponds to the trivial and OAI dimerized phase of the quadrupole model, respectively.

		In the same manner as charge pumping relates the SSH model to a Chern insulator, dipole pumping induces a model with chiral hinge states from the quadrupole model~\cite{schindler2018higher, benalcazar2017quantized}. The chiral hinge model is a three-dimensional second order topological insulator whose one-dimensional protected boundary modes occur at the intersection of a pair of two-dimensional faces. The topology of the hinge model obeys a $\mathbb{Z}_2$ classification protected by the product $M_xM_y\mathcal{T}$ of the horizontal and vertical mirror symmetries and time reversal $\mathcal{T}$~\cite{schindler2018higher}.

		\subsection{PEPS for the quadrupole model\label{sec:QuadrupolePEPS}}
		
		In this subsection, we provide a Gaussian fermionic PEPS for the ground state of the quadrupole model. After giving the details of the construction in Sec.~\ref{sec:QuadrupolePEPSConstruction}, we compute its parent Hamiltonian in Sec.~\ref{sec:QuadrupolePEPSPH} and discuss its ES in Sec.~\ref{sec:QuadrupolePEPSES}.
		 
			\subsubsection{Construction\label{sec:QuadrupolePEPSConstruction}}
			
			In order to construct a Gaussian PEPS for the ground state of the quadrupole model, we use similar ideas as in the SSH model MPS derived in Sec.~\ref{sec:fMPS}. In analogy to the particle-hole transformation of Eq.~\eqref{PHTrafo} on the $B$ sublattice of the SSH chain, we define new physical modes $\pmodenew_{\tau, \mathbf{x}}$ by performing a particle-hole transformation on the sublattices 3 and 4 while leaving the sublattices 1 and 2 unaffected,
			\begin{subequations}\label{PHTrafoQuadrupole}
			\begin{gather}
			\pmodenew_{1, \mathbf{x}} = \pmode_{1, \mathbf{x}}, \, 
			\pmodenew_{2, \mathbf{x}} = \pmode_{2, \mathbf{x}}, \\
			\pmodenew_{3, \mathbf{x}} = \pmode^{\dagger}_{3, \mathbf{x}}, \, 
			\pmodenew_{4, \mathbf{x}} = \pmode^{\dagger}_{4, \mathbf{x}}.
			\end{gather}
			\end{subequations}
			The vacuum $\ket{\Omega}$ of the new modes, satisfying $\pmodenew_{\tau, \mathbf{x}}\ket{\Omega} = 0$ for $\tau = 1, \dotsc, 4$, contains exactly $2 N_xN_y$ physical particles as required for the quadrupole model at half-filling. Thus, the particle-hole transformation allows us to express the quadrupole PEPS in terms of a separate and parity-even local tensor for each lattice site. Parity-evenness is required in order to ensure that the state is independent of the order of contractions in the network~\cite{BultinckPEPS_2017}.
			
			The quadrupole PEPS has bond dimension $D_{\mathrm{Quad}} = 2$ and is constructed from four local tensors $\tensor*{A}{*^{[\tau]}_{}^i_{lurd}}$ for sites on the four sublattices $\tau = 1, \dotsc, 4$, respectively. Here, $i \in \{0,1\}$ corresponds to the physical index and $l, u, r, d \in \{0,1\}$ to the left, up, right and down virtual indices, respectively. The physical basis states $\ket{0} = \ket{\Omega}$ and $\ket{1} = \pmodenew_{\tau, \mathbf{x}}^{\dagger}\ket{\Omega}$ are obtained from the transformed mode operators of Eq.~\eqref{PHTrafoQuadrupole}. 
			
			Due to the close relation between the SSH and the quadrupole models, we are guided in our ansatz for the local tensors of the quadrupole PEPS by the MPS tensors of Eq.~\eqref{MPSMyInterpol}. This can be most easily seen from Fig.~\ref{fig:QuadrupolePEPS}(b). We obtain the quadrupole model by coupling neighboring sites both horizontally and vertically according to the pattern of an SSH chain. In the PEPS tensors, the couplings of the horizontal SSH chains are transmitted by the left and right virtual fermions associated to parameters $\alpha_x$, $\beta_x$ analogous to Eq.~\eqref{MPSMyInterpol}. Similarly, the vertical SSH chains are implemented using the top and bottom virtual fermions associated to the parameters $\alpha_y$, $\beta_y$. Using the analogy to Eq.~\eqref{MPSMyInterpol}, we therefore obtain local PEPS tensors whose non-vanishing elements are given in terms of five real parameters $\gamma$, $\alpha_x$, $\alpha_y$, $\beta_x$ and $\beta_y$ as
			\begin{subequations}\label{QuadrupolePEPS}
				\begin{gather}
				\tensor*{A}{*^{[1]}_{}^0_{0000}}  =\tensor*{A}{*^{[2]}_{}^0_{0000}} = \tensor*{A}{*^{[3]}_{}^0_{0000}} = \tensor*{A}{*^{[4]}_{}^0_{0000}} =  \gamma ,\\
				-\tensor*{A}{*^{[1]}_{}^1_{1000}} = \tensor*{A}{*^{[2]}_{}^1_{0010}} = \tensor*{A}{*^{[3]}_{}^1_{0010}} = \tensor*{A}{*^{[4]}_{}^1_{1000}} = \beta_x,\\
				\tensor*{A}{*^{[1]}_{}^1_{0001}} = \tensor*{A}{*^{[2]}_{}^1_{0100}} = \tensor*{A}{*^{[3]}_{}^1_{0001}} = \tensor*{A}{*^{[4]}_{}^1_{0100}} =\beta_y,\\			
				\tensor*{A}{*^{[1]}_{}^1_{0010}} = -\tensor*{A}{*^{[2]}_{}^1_{1000}} = \tensor*{A}{*^{[3]}_{}^1_{1000}} = \tensor*{A}{*^{[4]}_{}^1_{0010}} = \alpha_x,\\			
				-\tensor*{A}{*^{[1]}_{}^1_{0100}} = -\tensor*{A}{*^{[2]}_{}^1_{0001}} = \tensor*{A}{*^{[3]}_{}^1_{0100}} = \tensor*{A}{*^{[4]}_{}^1_{0000}} = \alpha_y.
				\end{gather}
			\end{subequations}
			All the other tensor elements are equal to zero. The phases in Eq.~\eqref{QuadrupolePEPS} were chosen such as to ensure that there is a flux $\pi$ through every plaquette. As sketched in Fig.~\ref{fig:QuadrupolePEPS}(b), the parameters $\beta_x$ and $\beta_y$ represent the coupling of the physical leg to the virtual legs corresponding to the horizontal and vertical bonds pointing into the unit cell, respectively. Similarly, $\alpha_x$ and $\alpha_y$ control the coupling of the physical leg to the virtual legs pointing out of the unit cell.
			
			The PEPS of Eq.~\eqref{QuadrupolePEPS} has a global $\mathrm{U}(1)$ symmetry analogous to Eq.~\eqref{LocalU1} for the SSH MPS. Indeed, the local tensors are invariant under a combination of $\mathrm{U}(1)$ rotations of the physical and virtual legs given by
			\begin{subequations}\label{U1SymmetryPEPS}
			\begin{multline}
			\tensor*{A}{*^{[\tau]}_{}^i_{lurd}} = \sum_j\sum_{l'u'r'd'} U(\varphi)_{ij} U(\varphi)^{\dagger}_{ll'}U(\varphi)^{\dagger}_{uu'} \times \\ \tensor*{A}{*^{[\tau]}_{}^j_{l'u'r'd'}}U(\varphi)^{\dagger}_{r'r} U(\varphi)^{\dagger}_{d'd}
			\end{multline}
			for sites on the sublattices $\tau = 1,2$, and 
			\begin{multline}
			\tensor*{A}{*^{[\tau]}_{}^i_{lurd}} = \sum_j\sum_{l'u'r'd'} U(\varphi)_{ij}^{\dagger} U(\varphi)_{ll'}U(\varphi)_{uu'} \times \\ \tensor*{A}{*^{[\tau]}_{}^j_{l'u'r'd'}}U(\varphi)_{r'r} U(\varphi)_{d'd}
			\end{multline}			
			\end{subequations}
			for sites on the sublattices $\tau = 3,4$. Here, $U (\varphi)= \big(\begin{smallmatrix}
			1 & 0 \\
			0 & e^{i\varphi}
			\end{smallmatrix}\big)$ is the $\mathrm{U}(1)$ rotation acting on a single fermion. Eq.~\eqref{U1SymmetryPEPS} implies that the virtual legs for the local tensors on the sublattices $\tau = 1,2$ and $\tau = 3,4$ transform as holes and particles. Hence, each pair of virtual legs associated with the same nearest-neighbour bond transforms oppositely under the $\mathrm{U}(1)$ rotation, such that the PEPS is invariant under the physical part of Eq.~\eqref{U1SymmetryPEPS}. The charge associated with this symmetry is 
			\begin{multline}
			\sum_{\mathbf{x}} \left[ \sum_{\tau = 1,2} \pmodenew^{\dagger}_{\tau, \mathbf{x}}\pmodenew_{\tau, \mathbf{x}} - \sum_{\tau = 3,4} \pmodenew^{\dagger}_{\tau, \mathbf{x}}\pmodenew_{\tau, \mathbf{x}}\right] \\
			 = \sum_{\mathbf{x}} \sum_{\tau = 1}^4 \pmode^{\dagger}_{\tau, \mathbf{x}}\pmode_{\tau, \mathbf{x}} - \frac{1}{2} N_x N_y
 			\end{multline} 
			such that the $\mathrm{U}(1)$ symmetry ensures that the state lies exactly at half-filling of the lattice just as in the one-dimensional case.

		\subsubsection{Parent Hamiltonian\label{sec:QuadrupolePEPSPH}}
			
		For all values of the parameters $\gamma$, $\alpha_x$, $\alpha_y$, $\beta_x$ and $\beta_y$, the PEPS from Eq.~\eqref{QuadrupolePEPS} can be expressed as a Gaussian TNS for free fermions. In Appendix~\ref{sec:AppendixQuadrupolePEPS}, we show that the PEPS with parameters $\alpha \equiv \alpha_x = \alpha_y$ and $\beta \equiv \beta_x = \beta_y$ is the unique ground state of an extended version of the quadrupole model $H_{\mathrm{PEPS}}$ with $C_4$ symmetric hoppings, with a staggered chemical potential that breaks $C_4$ symmetry and with an additional next-to-nearest neighbour hopping
		\begin{equation}\label{NNNHoppingQuadrupolePEPS}
		\sum_{\tau} \sum_{\mathbf{x}}t^{(2)}_{\tau} \left[ \pmode^{\dagger}_{\tau, \mathbf{x}}\pmode_{\tau, \mathbf{x}+ \mathbf{\hat{x}}} + \pmode^{\dagger}_{\tau, \mathbf{x}}\pmode_{\tau, \mathbf{x}+ \mathbf{\hat{y}}}+ h.c.\right]
		\end{equation}
		between sites on the same sublattice $\tau$ with amplitude $t^{(2)}_{\tau}$. The pattern of next-to-nearest neighbour hoppings is shown in Fig.~\ref{fig:QuadrupolePEPS}(c). The couplings of $H_{\mathrm{PEPS}}$ depend on the parameters of the PEPS as
		\begin{subequations}\label{PHQuadrupolePEPS}
			\begin{gather}
			\mu  = \frac{\gamma^4 - \alpha^4-\beta^4}{\alpha^4 + \beta^4 + \gamma^4},\\
			t^{(0)}_x = t^{(0)}_y =\frac{\sqrt{2} \beta^2 \gamma^2}{\alpha^4 + \beta^4 + \gamma^4},\\
			t^{(1)}_x = t^{(1)}_y =\frac{\sqrt{2}\alpha^2 \gamma^2}{\alpha^4 + \beta^4 + \gamma^4},\\
			t^{(2)}_{1} = t^{(2)}_{2} =  - t^{(2)}_{3} = - t^{(2)}_{4} = -\frac{1}{2}\frac{\alpha^2 \beta^2 }{\alpha^4 + \beta^4 + \gamma^4}.
			\end{gather}
		\end{subequations}
		
		Similarly to the parent Hamiltonian $H_{\mathrm{MPS}}$ of the SSH model MPS of Eq.~\eqref{MPSMyInterpol}, $H_{\mathrm{PEPS}}$ describes different phases depending on the values of the parameters $\alpha$, $\beta$ and $\gamma$. When two out of the three parameters vanish, the system is in an atomic insulator state: If $\alpha = \beta = 0$, the particles are localized on the orbitals $\tau = 3,4$, whereas they are localized on the orbitals $\tau = 1,2$ if $\alpha = \gamma = 0$ or $\beta = \gamma = 0$. In contrast, when $\alpha = 0$ and $\beta, \gamma \neq 0$, all hoppings of the parent Hamiltonian vanish except for the nearest-neighbour hopping within unit cells. Similarly, when $\beta = 0$ and $\alpha, \gamma \neq 0$, the only non-zero hopping is the nearest-neighbour hopping between unit cells. In both cases, the system is in a dimerized phase with a staggered chemical potential. Setting $\beta = \gamma $ and $\alpha = \gamma $, respectively, we recover the trivial and OAI dimerized phases of the quadrupole model with vanishing chemical potential. Finally, when all three parameters are non-zero, the system has a non-vanishing nearest and next-nearest neighbour hopping as well as a finite staggered chemical potential.
		
		\subsubsection{Corner states and entanglement spectrum\label{sec:QuadrupolePEPSES}}
		
		The characteristic $(d-1)$-dimensional gapless edge states of a conventional $d$-dimensional TI are reflected in the state's bulk ES~\cite{PhysRevLett.104.130502}. Similarly, the ES of the quadrupole model in its dimerized OAI phase hosts gapless corner states as long as the virtual cut is compatible with the protecting symmetries~\cite{wang2018entanglement}. In order to further characterize the different phases described by the PEPS from Eq.~\eqref{QuadrupolePEPS}, we therefore study the ES of the state defined on a torus \wrt a rectangular subsystem $\mathcal{A}_{L_x \times L_y}$ of $L_{x}$ and $L_{y}$ unit cells in the horizontal and vertical direction.
		
		\begin{figure}[t]
		\includegraphics[width = \linewidth] {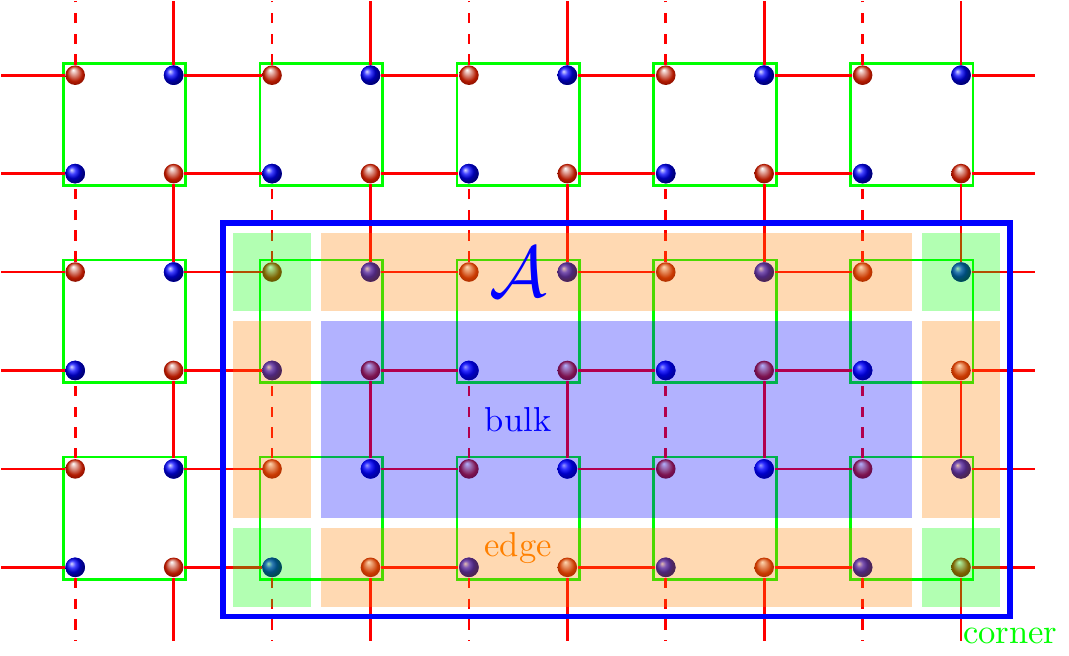}
		\caption{Quadrupole model in the dimerized OAI phase with a rectangular subsystem $\mathcal{A}$ marked by a blue rectangle. The four-site plaquettes coupled by the nearest-neighbour hopping $t^{(1)}$ (drawn in red) are shifted from the unit cells (denoted by green squares) by one site in both directions. The decoupled plaquettes in the bulk of $\mathcal{A}$ (marked in blue) do not contribute to the ES. In contrast, non-trivial levels in the ES come from both the plaquettes at the corners with a single site in $\mathcal{A}$ (marked in green), and from those at the edges with two sites in $\mathcal{A}$ (marked in orange).\label{fig:SketchBulkEdgeCornerESQuadrupolePEPS}}
		\end{figure}		
		
		For $\beta = 0$, the PEPS from Eq.~\eqref{QuadrupolePEPS} is in a dimerized phase where the system splits into four-site plaquettes shifted from the unit cell by one site in the horizontal and vertical direction. For $\alpha = \gamma$, we obtain the OAI dimerized phase of the quadrupole model, whereas for $\alpha \neq \gamma$ there is a non-zero staggered chemical potential that breaks the symmetries protecting the OAI phase. Due to the dimerization, with open boundaries the system has SSH chains with a staggered chemical potential at the edges, and corner sites which decouple from each other and the bulk. Correspondingly, the single-particle ES of the PEPS \wrt the subsystem $\mathcal{A}_{L_x \times L_y}$ has three distinct contributions from the bulk, the edges and the corners as sketched in Fig.~\ref{fig:SketchBulkEdgeCornerESQuadrupolePEPS}. The bulk consists of $(L_x -1) (L_y -1)$ plaquettes decoupled from the rest of the system, which contribute $4(L_x -1) (L_y -1)$ non-entangled modes with levels $|\lambda_{\mathrm{bulk}}| = 1$ (in the dimerized limit) to the single-particle ES. 
		
		On the other hand, the corner and edge sites belong to plaquettes crossed by the boundary of the subsystem $\mathcal{A}_{L_x \times L_y}$. In Appendix~\ref{sec:AppendixQuadrupolePEPS_ES} we show that the four boundaries provide $4 (L_x -1 + L_y -1)$ entangled modes with levels $\pm \lambda_{\mathrm{edge}}$, where
		\begin{subequations}\label{EntEnDimerizedQuadrupole}
		\begin{equation}\label{EntEnDimerizedQuadrupoleEdge}
		\lambda_{\mathrm{edge}} = \frac{\sqrt{\gamma ^8 + \alpha ^8}}{\gamma ^4 + \alpha ^4}.
		\end{equation}
		Since the number of these entangled modes grows linearly with the size of the edge, the boundary leads to an area law term in the entanglement entropy. 
		
		Furthermore, each corner site hosts exactly one mode corresponding to a level $\mp\lambda_{\mathrm{corner}}$ with	
		\begin{equation}\label{EntEnDimerizedQuadrupoleCorner}
		\lambda_{\mathrm{corner}} = \frac{\gamma ^4 - \alpha ^4}{\gamma ^4 + \alpha ^4}.
		\end{equation}
		\end{subequations}		
		Here, the negative sign holds for the top right and bottom left corner sites on the sublattices $\tau = 1$ and $\tau = 2$, respectively, whereas the positive signs applies to the top left and bottom right corner sites on the sublattices $\tau = 3$ and $\tau = 4$, respectively.
		
		For $\alpha = \gamma$, the PEPS of Eq.~\eqref{QuadrupolePEPS} describes the OAI phase of the quadrupole model. Indeed, in this case the edges have entanglement levels $\lambda_{\mathrm{edge}} = 1/\sqrt{2}$ and the four corners have degenerate levels $\lambda_{\mathrm{corner}} = 0$ corresponding to maximally entangled corner modes. On the other hand, for $\alpha = 0$, both $\lambda_{\mathrm{edge}} = 1$ and $\lambda_{\mathrm{corner}} = 1$ such that the TNS describes an atomic state.
		
		Finally, if the OAI dimerized phase is perturbed by a small non-zero value $\beta \neq 0$, one can check numerically that the corner modes acquire a finite splitting $\lambda_{\mathrm{corner}} \neq 0$ and are no longer perfectly localized at the corner sites. This confirms that the TNS with $\beta \neq 0$ is not in the OAI phase of the quadrupole model as expected from the breaking of the mirror symmetries and $C_4$ symmetry by the next-nearest neighbour hopping and the chemical potential.

	\subsection{3D chiral hinge PEPS from dipole pumping\label{sec:3DPEPS}}

	\begin{figure}[t]	
	\includegraphics[width = \linewidth] {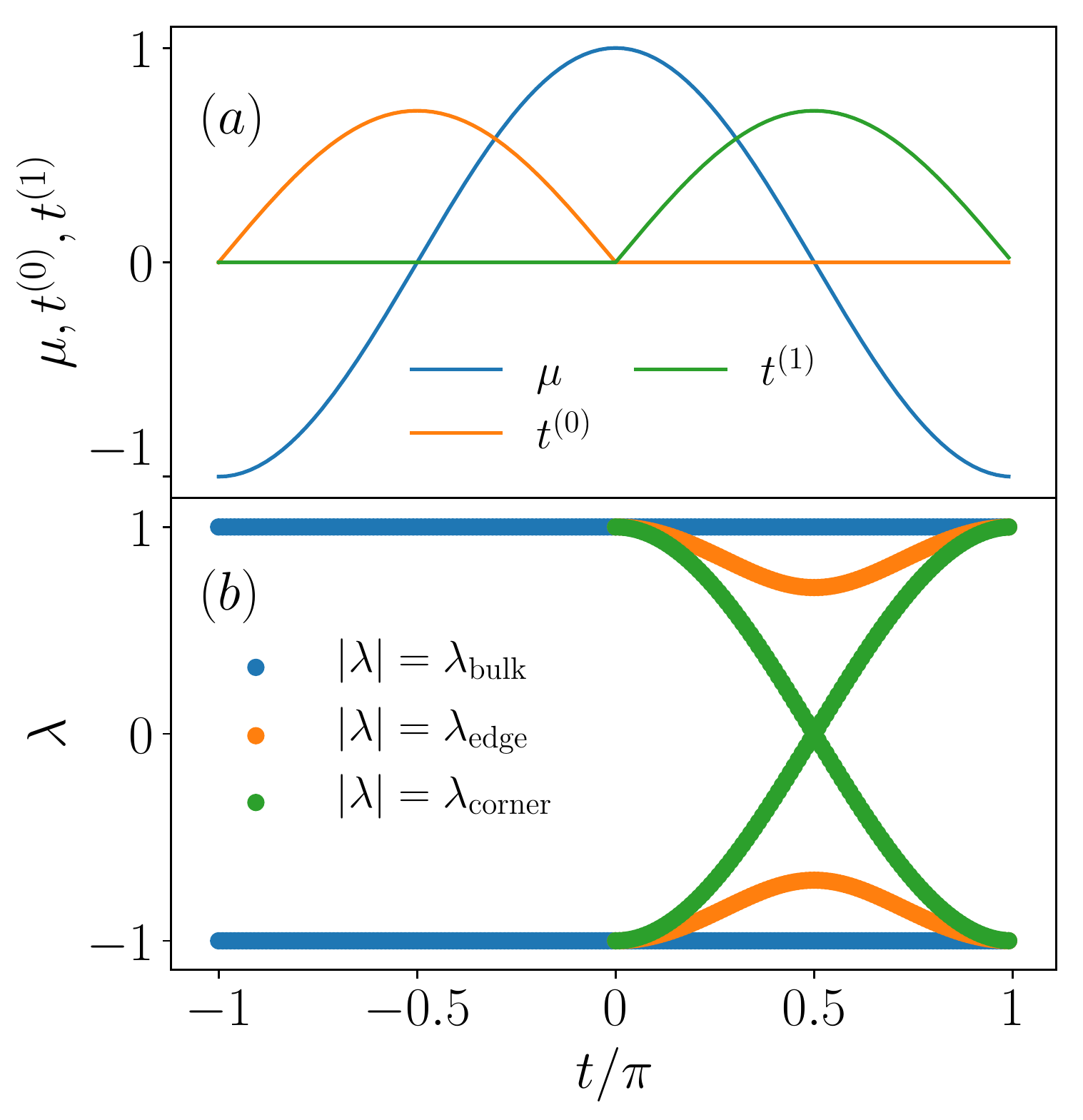}
	\caption{(a) Coupling constants of the parent Hamiltonian $H_{\mathrm{PEPS}}$ and (b) single-particle ES of the PEPS of Eq.~\eqref{QuadrupolePEPS} along the dipole pumping interpolation generated by the parametrisation $\parampump$. In (a), the vanishing of the next-nearest neighbour coupling $t^{(2)}_{\tau} =  0$ of $H_{\mathrm{PEPS}}$ is not shown. In~(b), we marked the bulk contribution in blue and for times $t \in [0, \pi]$ with $\beta (t) = 0$, the edge contribution in orange and the corner contribution in green. The ES was computed on a torus with $N_x = N_y = 10$ unit cells \wrt to the subsystem $\mathcal{A}_{L_x \times L_y}$ with $L_x = L_y = 5$.\label{fig:PHAndSPESQuadrupolePEPS}}
	\end{figure}
	
	\begin{figure*}[th]		
		\includegraphics[width = 0.99\linewidth]{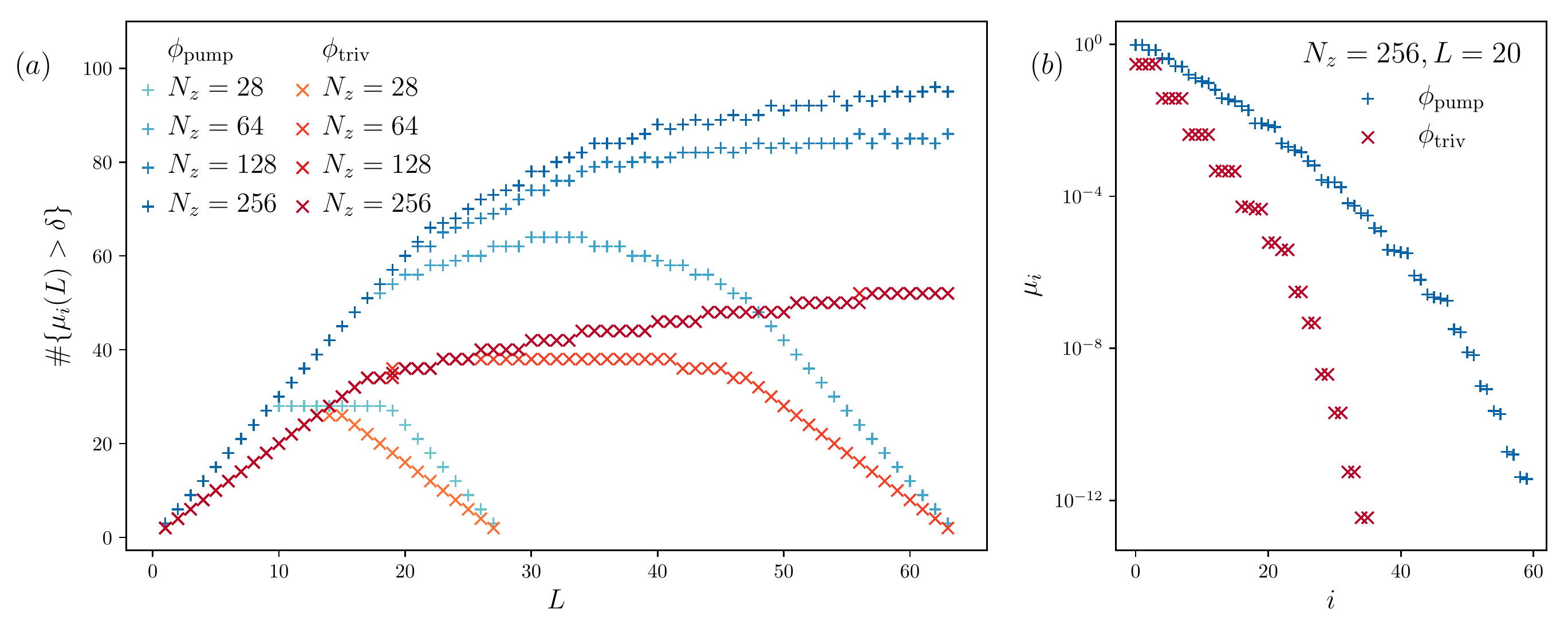}	
		\caption{Single-particle ES of a column of sites on the sublattice $\tau = 1$ of the three-dimensional real-space TNSs defined by two different parametrisations for the quadrupole PEPS of Eq.~\eqref{QuadrupolePEPS}: the charge pumping interpolation $\parampump$ of Eq.~\eqref{ParametrisationDimerized} leading to a chiral hinge insulator (blue) and the parametrisation $\paramtriv$ of Eq.~\eqref{MPSInterpolationTrivial} leading to a topologically trivial three-dimensional state (red). (a)~Number of modes in the single-particle ES with a finite entanglement corresponding to a value $\mu_i > \delta$ with $\delta = 10^{-13}$ the numerical accuracy for different system sizes $N_z$ as a function of the subsystem size $L$. Due to the mirror symmetry $M_{xy}$ of the quadrupole PEPS tensor on the sublattice $\tau = 1$, the number of entangled modes for both parametrisations is identical to the two-dimensional case shown in Fig.~\ref{fig:ESPumpingPEPS}(a). (b)~Single-particle ES for a column with $N_z = 256$ sites \wrt the subsystem $\mathcal{A}_{L}$ with $L = 20$. The double degeneracy in the ES corresponding to $\paramtriv$ is due to the decoupling of the left and down virtual legs since $\beta(t) = 0$ along $\paramtriv$ (see Appendix~\ref{sec:AppendixESUOne}). The spectra are similar, with the same number of entangled modes, but not strictly identical to those of the two-dimensional state shown in Fig.~\ref{fig:ESPumpingPEPS}(b).\label{fig:ColumnES_ChiralHingPEPS}}
	\end{figure*}

	The PEPS of Eq.~\eqref{QuadrupolePEPS} with parameters $\alpha = \alpha _x = \alpha _y$ and $\beta = \beta _x = \beta_y$ describes a dimerized dipole pumping cycle of the quadrupole model if $\alpha$, $\beta$ and $\gamma$ follow the parametrisation $\parampump$ which we found for the dimerized charge pumping cycle of the SSH model MPS. The coupling constants of the parent Hamiltonian $H_{\mathrm{PEPS}}$ derived from the parametrisation $\parampump$ are shown in Fig.~\ref{fig:PHAndSPESQuadrupolePEPS}(a). They differ from those of the dipole pumping Hamiltonian of Ref.~\cite{benalcazar2017quantized} only by a factor of $1/\sqrt{2}$ for the hopping amplitudes, which does not affect the topology of the interpolation. The single particle ES of the PEPS along $\parampump$ is shown in Fig.~\ref{fig:PHAndSPESQuadrupolePEPS}(b). In the first half of the cycle, the system is in a dimerized phase with each unit cell decoupled from the rest of the system. Hence, the single-particle ES contains only the bulk bands with  $\lambda = \pm 1$. However, in the second half of the cycle for $0 < t < \pi$, the contributions from the edges and corners can be clearly distinguished. The four corner modes connect the bands with $\lambda = \pm 1$ and are degenerate in the OAI dimerized phase of the quadrupole model obtained for $t = \pi / 2$. In the three-dimensional second-order TI of Ref.~\cite{schindler2018higher} obtained by the identification of the time $t$ along the dipole pumping cycle with the momentum $k_z$ in the third direction, these corner modes generate the chiral modes localized at the hinges.
	
	Following the steps described in Sec.~\ref{sec:ChernPEPS}, we may use the PEPS of Eq.~\eqref{QuadrupolePEPS} along the interpolation $\parampump$ to define a three-dimensional PEPS for the second order hinge TI. Moreover, we can construct a topologically trivial three-dimensional state from the interpolation $\paramtriv$ from Eq.~\eqref{MPSInterpolationTrivial}. These PEPSs have a finite bond dimension $D_{\mathrm{Quad}} = 2$ in the $x$ and $y$ directions. In the hybrid real-momentum space where the third dimension corresponds to the momentum $k_z$ or time $t$, the states have a finite bond dimension equal to one also in the third direction.
	
	On the other hand, due to the non-locality of the inverse FT, in real space their bond dimension $D_{z, \tau}$ in the third direction for sites on the sublattice $\tau$ for $\tau = 1, \dotsc, 4$ grows with the system size $N_z$. As in Sec.~\ref{sec:ChernPEPSBondDim}, we can estimate $D_{z, \tau}$ from the ES of a column of sites on the sublattice $\tau$ \wrt the subsystem $\mathcal{A}_L$ of the first $L$ sites.
	
	The number of entangled levels in the single-particle ES of a column of sites on the sublattice $\tau = 1$ of the three-dimensional PEPSs is shown in Fig.~\ref{fig:ColumnES_ChiralHingPEPS}(a). As we can see from the smallest system size $N_z = 28$, for both $\parampump$ and $\paramtriv$ the number of entangled modes is \emph{identical} to the two-dimensional case from Sec.~\ref{sec:ChernPEPS} when replacing $N_y$ with $N_z$. Moreover, the spectra displayed in Fig.~\ref{fig:ColumnES_ChiralHingPEPS}(b) are similar, although not identical, to the corresponding results for the two-dimensional PEPSs.   
	
	As we show in Appendix~\ref{sec:AppendixESUOne}, for $\parampump$ this number of entangled modes is the maximal number compatible with the symmetries of the quadrupole PEPS, here the $\mathrm{U}(1)$ symmetry from Eq.~\eqref{U1SymmetryPEPS}, and the mirror symmetry $M_{xy}$ of the local tensor on the sublattice $\tau = 1$. Indeed, the latter causes the decoupling of one superposition of the left and down virtual legs, and similarly for the up and right virtual legs. For $\paramtriv$, there is an additional reduction of the number of entangled modes since the left and down virtual legs decouple due to the vanishing parameter $\beta = 0$ along the trivial interpolation (see Appendix~\ref{sec:AppendixESUOne}).
	
	Analogously to Sec.~\ref{sec:ChernPEPSBondDim}, we therefore conclude that the bond dimension $D_{z, \tau}$ in the third direction grows exponentially with $N_z$ for $\tau = 1, 2, 3, 4$. Due to the mirror symmetries of the quadrupole model, $D_{z, \tau}$ has the same value as the vertical bond dimension of the two-dimensional PEPSs obtained from cyclic interpolations of the SSH model. The increase in spatial dimensionality therefore does not cause an increase of the bond dimension in the $(d+1)^{\text{st}}$ direction.
		
	\section{Conclusion\label{sec:Conclusion}}
	
	In this article, we showed how to use charge pumping to define TNSs for $(d+1)$-dimensional conventional or higher-order TIs starting from TNSs of TIs in $d$ space dimensions. To that end, we constructed a Gaussian fermionic MPS for the SSH model with bond dimension $D_{\mathrm{SSH}} = 2$ in $d = 1$ dimension, and a Gaussian fermionic PEPS for the topological quadrupole model with bond dimension $D_{\mathrm{Quad}} = 2$ in $d = 2$ dimensions. We proved that these TNSs have local gapped parent Hamiltonians with up to next-nearest neighbour hopping, and thereby showed that they describe the SSH model along a charge pumping cycle and the quadrupole model along a dipole pumping cycle, respectively. We employed these TNSs to construct a two-dimensional PEPS for a Chern insulator and a three-dimensional PEPS for a chiral hinge HOTI, respectively. The $(d+1)$-dimensional TNSs inherit the finite bond dimension $D_{\mathrm{SSH}}$ and $D_{\mathrm{Quad}}$ in the first $d$ dimensions, respectively. In a hybrid coordinate system where the $(d + 1)^{\mathrm{st}}$ dimension corresponds to momentum, the $(d+1)$-dimensional TNSs have a finite bond dimension in this direction. In contrast, we showed that the bond dimension in the $(d + 1)^{\mathrm{st}}$ direction grows exponentially in a real-space coordinate system. 
	
	Our results suggest several directions for future work. On one hand, it would be interesting to study if a real-space PEPS for the Chern insulator with a polynomially growing bond dimension can be found by truncating the Schmidt values of the real-space column in our construction. Such a result could potentially provide insight into the physical origin for the obstructions preventing the existence of chiral PEPSs with a finite bond dimension. On the other hand, we expect the TNSs constructed here to be useful for finite-size simulations despite their growing bond dimension, since the bond dimension is finite in all but one direction. By their local nature, they could be employed as the building block of interacting $(d+1)$-dimensional TIs obtained by Gutzwiller projection or parton constructions~\cite{PhysRevB.84.075128, PhysRevB.87.161113, PARAMESWARAN2013816, wu2019tensor}.  
	
	\section{Acknowledgement}
	
	We thank Ignacio Cirac, Claudius Hubig, Andras Molnar, Sarah Pinon, Antoine Sterdyniak and Erez Zohar for discussions. A.H. and N.S. acknowledge support by the European Research Council (ERC) under the European Union's Horizon 2020 research and innovation programme through the ERC Starting Grant WASCOSYS (No. 636201), and by the Deutsche Forschungsgemeinschaft (DFG) under Germany's Excellence Strategy (EXC-2111 – 390814868). A.H. and N.R. were supported by Grants No. ANR-17-CE30-0013-01 and No. ANR-16-CE30-0025. B.A.B. was supported by the Department of Energy Grant No. DE-SC0016239, the National Science Foundation EAGER Grant No. DMR 1643312, Simons Investigator Grant No. 404513, ONR Grant No.N00014-14-1-0330, the Packard Foundation, the Schmidt Fund for Innovative Research, ARO MURI W911NF-12-1-0461 and NSF-MRSEC DMR-1420541.

	\bibliography{HigherOrder}			

\begin{thebibliography}{68}%
\makeatletter
\providecommand \@ifxundefined [1]{%
 \@ifx{#1\undefined}
}%
\providecommand \@ifnum [1]{%
 \ifnum #1\expandafter \@firstoftwo
 \else \expandafter \@secondoftwo
 \fi
}%
\providecommand \@ifx [1]{%
 \ifx #1\expandafter \@firstoftwo
 \else \expandafter \@secondoftwo
 \fi
}%
\providecommand \natexlab [1]{#1}%
\providecommand \enquote  [1]{``#1''}%
\providecommand \bibnamefont  [1]{#1}%
\providecommand \bibfnamefont [1]{#1}%
\providecommand \citenamefont [1]{#1}%
\providecommand \href@noop [0]{\@secondoftwo}%
\providecommand \href [0]{\begingroup \@sanitize@url \@href}%
\providecommand \@href[1]{\@@startlink{#1}\@@href}%
\providecommand \@@href[1]{\endgroup#1\@@endlink}%
\providecommand \@sanitize@url [0]{\catcode `\\12\catcode `\$12\catcode
  `\&12\catcode `\#12\catcode `\^12\catcode `\_12\catcode `\%12\relax}%
\providecommand \@@startlink[1]{}%
\providecommand \@@endlink[0]{}%
\providecommand \url  [0]{\begingroup\@sanitize@url \@url }%
\providecommand \@url [1]{\endgroup\@href {#1}{\urlprefix }}%
\providecommand \urlprefix  [0]{URL }%
\providecommand \Eprint [0]{\href }%
\providecommand \doibase [0]{http://dx.doi.org/}%
\providecommand \selectlanguage [0]{\@gobble}%
\providecommand \bibinfo  [0]{\@secondoftwo}%
\providecommand \bibfield  [0]{\@secondoftwo}%
\providecommand \translation [1]{[#1]}%
\providecommand \BibitemOpen [0]{}%
\providecommand \bibitemStop [0]{}%
\providecommand \bibitemNoStop [0]{.\EOS\space}%
\providecommand \EOS [0]{\spacefactor3000\relax}%
\providecommand \BibitemShut  [1]{\csname bibitem#1\endcsname}%
\let\auto@bib@innerbib\@empty
%</preamble>
\bibitem [{\citenamefont {Benalcazar}\ \emph
  {et~al.}(2017{\natexlab{a}})\citenamefont {Benalcazar}, \citenamefont
  {Bernevig},\ and\ \citenamefont {Hughes}}]{benalcazar2017quantizedScience}%
  \BibitemOpen
  \bibfield  {author} {\bibinfo {author} {\bibfnamefont {W.~A.}\ \bibnamefont
  {Benalcazar}}, \bibinfo {author} {\bibfnamefont {B.~A.}\ \bibnamefont
  {Bernevig}}, \ and\ \bibinfo {author} {\bibfnamefont {T.~L.}\ \bibnamefont
  {Hughes}},\ }\href@noop {} {\bibfield  {journal} {\bibinfo  {journal}
  {Science}\ }\textbf {\bibinfo {volume} {357}},\ \bibinfo {pages} {61}
  (\bibinfo {year} {2017}{\natexlab{a}})}\BibitemShut {NoStop}%
\bibitem [{\citenamefont {Schindler}\ \emph
  {et~al.}(2018{\natexlab{a}})\citenamefont {Schindler}, \citenamefont {Cook},
  \citenamefont {Vergniory}, \citenamefont {Wang}, \citenamefont {Parkin},
  \citenamefont {Bernevig},\ and\ \citenamefont
  {Neupert}}]{schindler2018higher}%
  \BibitemOpen
  \bibfield  {author} {\bibinfo {author} {\bibfnamefont {F.}~\bibnamefont
  {Schindler}}, \bibinfo {author} {\bibfnamefont {A.~M.}\ \bibnamefont {Cook}},
  \bibinfo {author} {\bibfnamefont {M.~G.}\ \bibnamefont {Vergniory}}, \bibinfo
  {author} {\bibfnamefont {Z.}~\bibnamefont {Wang}}, \bibinfo {author}
  {\bibfnamefont {S.~S.}\ \bibnamefont {Parkin}}, \bibinfo {author}
  {\bibfnamefont {B.~A.}\ \bibnamefont {Bernevig}}, \ and\ \bibinfo {author}
  {\bibfnamefont {T.}~\bibnamefont {Neupert}},\ }\href@noop {} {\bibfield
  {journal} {\bibinfo  {journal} {Science advances}\ }\textbf {\bibinfo
  {volume} {4}},\ \bibinfo {pages} {eaat0346} (\bibinfo {year}
  {2018}{\natexlab{a}})}\BibitemShut {NoStop}%
\bibitem [{\citenamefont {Benalcazar}\ \emph
  {et~al.}(2017{\natexlab{b}})\citenamefont {Benalcazar}, \citenamefont
  {Bernevig},\ and\ \citenamefont {Hughes}}]{benalcazar2017quantized}%
  \BibitemOpen
  \bibfield  {author} {\bibinfo {author} {\bibfnamefont {W.~A.}\ \bibnamefont
  {Benalcazar}}, \bibinfo {author} {\bibfnamefont {B.~A.}\ \bibnamefont
  {Bernevig}}, \ and\ \bibinfo {author} {\bibfnamefont {T.~L.}\ \bibnamefont
  {Hughes}},\ }\href@noop {} {\bibfield  {journal} {\bibinfo  {journal}
  {Physical Review B}\ }\textbf {\bibinfo {volume} {96}},\ \bibinfo {pages}
  {245115} (\bibinfo {year} {2017}{\natexlab{b}})}\BibitemShut {NoStop}%
\bibitem [{\citenamefont {Langbehn}\ \emph {et~al.}(2017)\citenamefont
  {Langbehn}, \citenamefont {Peng}, \citenamefont {Trifunovic}, \citenamefont
  {von Oppen},\ and\ \citenamefont {Brouwer}}]{PhysRevLett.119.246401}%
  \BibitemOpen
  \bibfield  {author} {\bibinfo {author} {\bibfnamefont {J.}~\bibnamefont
  {Langbehn}}, \bibinfo {author} {\bibfnamefont {Y.}~\bibnamefont {Peng}},
  \bibinfo {author} {\bibfnamefont {L.}~\bibnamefont {Trifunovic}}, \bibinfo
  {author} {\bibfnamefont {F.}~\bibnamefont {von Oppen}}, \ and\ \bibinfo
  {author} {\bibfnamefont {P.~W.}\ \bibnamefont {Brouwer}},\ }\href {\doibase
  10.1103/PhysRevLett.119.246401} {\bibfield  {journal} {\bibinfo  {journal}
  {Phys. Rev. Lett.}\ }\textbf {\bibinfo {volume} {119}},\ \bibinfo {pages}
  {246401} (\bibinfo {year} {2017})}\BibitemShut {NoStop}%
\bibitem [{\citenamefont {Hasan}\ and\ \citenamefont
  {Kane}(2010)}]{RevModPhys.82.3045}%
  \BibitemOpen
  \bibfield  {author} {\bibinfo {author} {\bibfnamefont {M.~Z.}\ \bibnamefont
  {Hasan}}\ and\ \bibinfo {author} {\bibfnamefont {C.~L.}\ \bibnamefont
  {Kane}},\ }\href {\doibase 10.1103/RevModPhys.82.3045} {\bibfield  {journal}
  {\bibinfo  {journal} {Rev. Mod. Phys.}\ }\textbf {\bibinfo {volume} {82}},\
  \bibinfo {pages} {3045} (\bibinfo {year} {2010})}\BibitemShut {NoStop}%
\bibitem [{\citenamefont {Haldane}(1988)}]{PhysRevLett.61.2015}%
  \BibitemOpen
  \bibfield  {author} {\bibinfo {author} {\bibfnamefont {F.~D.~M.}\
  \bibnamefont {Haldane}},\ }\href {\doibase 10.1103/PhysRevLett.61.2015}
  {\bibfield  {journal} {\bibinfo  {journal} {Phys. Rev. Lett.}\ }\textbf
  {\bibinfo {volume} {61}},\ \bibinfo {pages} {2015} (\bibinfo {year}
  {1988})}\BibitemShut {NoStop}%
\bibitem [{\citenamefont {Bradlyn}\ \emph {et~al.}(2017)\citenamefont
  {Bradlyn}, \citenamefont {Elcoro}, \citenamefont {Cano}, \citenamefont
  {Vergniory}, \citenamefont {Wang}, \citenamefont {Felser}, \citenamefont
  {Aroyo},\ and\ \citenamefont {Bernevig}}]{bradlyn2017topological}%
  \BibitemOpen
  \bibfield  {author} {\bibinfo {author} {\bibfnamefont {B.}~\bibnamefont
  {Bradlyn}}, \bibinfo {author} {\bibfnamefont {L.}~\bibnamefont {Elcoro}},
  \bibinfo {author} {\bibfnamefont {J.}~\bibnamefont {Cano}}, \bibinfo {author}
  {\bibfnamefont {M.}~\bibnamefont {Vergniory}}, \bibinfo {author}
  {\bibfnamefont {Z.}~\bibnamefont {Wang}}, \bibinfo {author} {\bibfnamefont
  {C.}~\bibnamefont {Felser}}, \bibinfo {author} {\bibfnamefont
  {M.}~\bibnamefont {Aroyo}}, \ and\ \bibinfo {author} {\bibfnamefont {B.~A.}\
  \bibnamefont {Bernevig}},\ }\href@noop {} {\bibfield  {journal} {\bibinfo
  {journal} {Nature}\ }\textbf {\bibinfo {volume} {547}},\ \bibinfo {pages}
  {298} (\bibinfo {year} {2017})}\BibitemShut {NoStop}%
\bibitem [{\citenamefont {{Wieder}}\ and\ \citenamefont
  {{Bernevig}}(2018)}]{2018arXiv181002373W}%
  \BibitemOpen
  \bibfield  {author} {\bibinfo {author} {\bibfnamefont {B.~J.}\ \bibnamefont
  {{Wieder}}}\ and\ \bibinfo {author} {\bibfnamefont {B.~A.}\ \bibnamefont
  {{Bernevig}}},\ }\href@noop {} {\bibfield  {journal} {\bibinfo  {journal}
  {arXiv e-prints}\ ,\ \bibinfo {eid} {arXiv:1810.02373}} (\bibinfo {year}
  {2018})},\ \Eprint {http://arxiv.org/abs/1810.02373} {arXiv:1810.02373
  [cond-mat.mes-hall]} \BibitemShut {NoStop}%
\bibitem [{\citenamefont {Su}\ \emph {et~al.}(1980)\citenamefont {Su},
  \citenamefont {Schrieffer},\ and\ \citenamefont {Heeger}}]{PhysRevB.22.2099}%
  \BibitemOpen
  \bibfield  {author} {\bibinfo {author} {\bibfnamefont {W.~P.}\ \bibnamefont
  {Su}}, \bibinfo {author} {\bibfnamefont {J.~R.}\ \bibnamefont {Schrieffer}},
  \ and\ \bibinfo {author} {\bibfnamefont {A.~J.}\ \bibnamefont {Heeger}},\
  }\href {\doibase 10.1103/PhysRevB.22.2099} {\bibfield  {journal} {\bibinfo
  {journal} {Phys. Rev. B}\ }\textbf {\bibinfo {volume} {22}},\ \bibinfo
  {pages} {2099} (\bibinfo {year} {1980})}\BibitemShut {NoStop}%
\bibitem [{\citenamefont {Schindler}\ \emph
  {et~al.}(2018{\natexlab{b}})\citenamefont {Schindler}, \citenamefont {Wang},
  \citenamefont {Vergniory}, \citenamefont {Cook}, \citenamefont {Murani},
  \citenamefont {Sengupta}, \citenamefont {Kasumov}, \citenamefont {Deblock},
  \citenamefont {Jeon}, \citenamefont {Drozdov}, \citenamefont {Bouchiat},
  \citenamefont {Guéron}, \citenamefont {Yazdani}, \citenamefont {Bernevig},\
  and\ \citenamefont {Neupert}}]{SchindlerBismuth}%
  \BibitemOpen
  \bibfield  {author} {\bibinfo {author} {\bibfnamefont {F.}~\bibnamefont
  {Schindler}}, \bibinfo {author} {\bibfnamefont {Z.}~\bibnamefont {Wang}},
  \bibinfo {author} {\bibfnamefont {M.~G.}\ \bibnamefont {Vergniory}}, \bibinfo
  {author} {\bibfnamefont {A.~M.}\ \bibnamefont {Cook}}, \bibinfo {author}
  {\bibfnamefont {A.}~\bibnamefont {Murani}}, \bibinfo {author} {\bibfnamefont
  {S.}~\bibnamefont {Sengupta}}, \bibinfo {author} {\bibfnamefont {A.~Y.}\
  \bibnamefont {Kasumov}}, \bibinfo {author} {\bibfnamefont {R.}~\bibnamefont
  {Deblock}}, \bibinfo {author} {\bibfnamefont {S.}~\bibnamefont {Jeon}},
  \bibinfo {author} {\bibfnamefont {I.}~\bibnamefont {Drozdov}}, \bibinfo
  {author} {\bibfnamefont {H.}~\bibnamefont {Bouchiat}}, \bibinfo {author}
  {\bibfnamefont {S.}~\bibnamefont {Guéron}}, \bibinfo {author} {\bibfnamefont
  {A.}~\bibnamefont {Yazdani}}, \bibinfo {author} {\bibfnamefont {B.~A.}\
  \bibnamefont {Bernevig}}, \ and\ \bibinfo {author} {\bibfnamefont
  {T.}~\bibnamefont {Neupert}},\ }\href
  {https://doi.org/10.1038/s41567-018-0224-7} {\bibfield  {journal} {\bibinfo
  {journal} {Nature Physics}\ }\textbf {\bibinfo {volume} {14}},\ \bibinfo
  {pages} {918} (\bibinfo {year} {2018}{\natexlab{b}})}\BibitemShut {NoStop}%
\bibitem [{\citenamefont {Serra-Garcia}\ \emph {et~al.}(2018)\citenamefont
  {Serra-Garcia}, \citenamefont {Peri}, \citenamefont {S{\"u}sstrunk},
  \citenamefont {Bilal}, \citenamefont {Larsen}, \citenamefont {Villanueva},\
  and\ \citenamefont {Huber}}]{serra2018observation}%
  \BibitemOpen
  \bibfield  {author} {\bibinfo {author} {\bibfnamefont {M.}~\bibnamefont
  {Serra-Garcia}}, \bibinfo {author} {\bibfnamefont {V.}~\bibnamefont {Peri}},
  \bibinfo {author} {\bibfnamefont {R.}~\bibnamefont {S{\"u}sstrunk}}, \bibinfo
  {author} {\bibfnamefont {O.~R.}\ \bibnamefont {Bilal}}, \bibinfo {author}
  {\bibfnamefont {T.}~\bibnamefont {Larsen}}, \bibinfo {author} {\bibfnamefont
  {L.~G.}\ \bibnamefont {Villanueva}}, \ and\ \bibinfo {author} {\bibfnamefont
  {S.~D.}\ \bibnamefont {Huber}},\ }\href@noop {} {\bibfield  {journal}
  {\bibinfo  {journal} {Nature}\ }\textbf {\bibinfo {volume} {555}},\ \bibinfo
  {pages} {342} (\bibinfo {year} {2018})}\BibitemShut {NoStop}%
\bibitem [{\citenamefont {Xue}\ \emph {et~al.}(2019)\citenamefont {Xue},
  \citenamefont {Yang}, \citenamefont {Gao}, \citenamefont {Chong},\ and\
  \citenamefont {Zhang}}]{HoaranAcousticHOTI}%
  \BibitemOpen
  \bibfield  {author} {\bibinfo {author} {\bibfnamefont {H.}~\bibnamefont
  {Xue}}, \bibinfo {author} {\bibfnamefont {Y.}~\bibnamefont {Yang}}, \bibinfo
  {author} {\bibfnamefont {F.}~\bibnamefont {Gao}}, \bibinfo {author}
  {\bibfnamefont {Y.}~\bibnamefont {Chong}}, \ and\ \bibinfo {author}
  {\bibfnamefont {B.}~\bibnamefont {Zhang}},\ }\href
  {https://doi.org/10.1038/s41563-018-0251-x} {\bibfield  {journal} {\bibinfo
  {journal} {Nature Materials}\ }\textbf {\bibinfo {volume} {18}},\ \bibinfo
  {pages} {108} (\bibinfo {year} {2019})}\BibitemShut {NoStop}%
\bibitem [{\citenamefont {Ni}\ \emph {et~al.}(2019)\citenamefont {Ni},
  \citenamefont {Weiner}, \citenamefont {Alù},\ and\ \citenamefont
  {Khanikaev}}]{XiangAcoustic2Dchiral}%
  \BibitemOpen
  \bibfield  {author} {\bibinfo {author} {\bibfnamefont {X.}~\bibnamefont
  {Ni}}, \bibinfo {author} {\bibfnamefont {M.}~\bibnamefont {Weiner}}, \bibinfo
  {author} {\bibfnamefont {A.}~\bibnamefont {Alù}}, \ and\ \bibinfo {author}
  {\bibfnamefont {A.~B.}\ \bibnamefont {Khanikaev}},\ }\href
  {https://doi.org/10.1038/s41563-018-0252-9} {\bibfield  {journal} {\bibinfo
  {journal} {Nature Materials}\ }\textbf {\bibinfo {volume} {18}},\ \bibinfo
  {pages} {113} (\bibinfo {year} {2019})}\BibitemShut {NoStop}%
\bibitem [{\citenamefont {Mittal}\ \emph {et~al.}(2018)\citenamefont {Mittal},
  \citenamefont {Orre}, \citenamefont {Zhu}, \citenamefont {Gorlach},
  \citenamefont {Poddubny},\ and\ \citenamefont {Hafezi}}]{mittal2018photonic}%
  \BibitemOpen
  \bibfield  {author} {\bibinfo {author} {\bibfnamefont {S.}~\bibnamefont
  {Mittal}}, \bibinfo {author} {\bibfnamefont {V.~V.}\ \bibnamefont {Orre}},
  \bibinfo {author} {\bibfnamefont {G.}~\bibnamefont {Zhu}}, \bibinfo {author}
  {\bibfnamefont {M.~A.}\ \bibnamefont {Gorlach}}, \bibinfo {author}
  {\bibfnamefont {A.}~\bibnamefont {Poddubny}}, \ and\ \bibinfo {author}
  {\bibfnamefont {M.}~\bibnamefont {Hafezi}},\ }\href@noop {} {\bibfield
  {journal} {\bibinfo  {journal} {arXiv preprint arXiv:1812.09304}\ } (\bibinfo
  {year} {2018})}\BibitemShut {NoStop}%
\bibitem [{\citenamefont {Hassan}\ \emph {et~al.}(2018)\citenamefont {Hassan},
  \citenamefont {Kunst}, \citenamefont {Moritz}, \citenamefont {Andler},
  \citenamefont {Bergholtz},\ and\ \citenamefont
  {Bourennane}}]{hassan2018cornerphotonic}%
  \BibitemOpen
  \bibfield  {author} {\bibinfo {author} {\bibfnamefont {A.~E.}\ \bibnamefont
  {Hassan}}, \bibinfo {author} {\bibfnamefont {F.~K.}\ \bibnamefont {Kunst}},
  \bibinfo {author} {\bibfnamefont {A.}~\bibnamefont {Moritz}}, \bibinfo
  {author} {\bibfnamefont {G.}~\bibnamefont {Andler}}, \bibinfo {author}
  {\bibfnamefont {E.~J.}\ \bibnamefont {Bergholtz}}, \ and\ \bibinfo {author}
  {\bibfnamefont {M.}~\bibnamefont {Bourennane}},\ }\href@noop {} {\bibfield
  {journal} {\bibinfo  {journal} {arXiv preprint arXiv:1812.08185}\ } (\bibinfo
  {year} {2018})}\BibitemShut {NoStop}%
\bibitem [{\citenamefont {Xie}\ \emph {et~al.}(2018)\citenamefont {Xie},
  \citenamefont {Su}, \citenamefont {Wang}, \citenamefont {Su}, \citenamefont
  {Shen}, \citenamefont {Zhan}, \citenamefont {Lu}, \citenamefont {Wang},\ and\
  \citenamefont {Chen}}]{xie2018visualizationphotonic}%
  \BibitemOpen
  \bibfield  {author} {\bibinfo {author} {\bibfnamefont {B.-Y.}\ \bibnamefont
  {Xie}}, \bibinfo {author} {\bibfnamefont {G.-X.}\ \bibnamefont {Su}},
  \bibinfo {author} {\bibfnamefont {H.-F.}\ \bibnamefont {Wang}}, \bibinfo
  {author} {\bibfnamefont {H.}~\bibnamefont {Su}}, \bibinfo {author}
  {\bibfnamefont {X.-P.}\ \bibnamefont {Shen}}, \bibinfo {author}
  {\bibfnamefont {P.}~\bibnamefont {Zhan}}, \bibinfo {author} {\bibfnamefont
  {M.-H.}\ \bibnamefont {Lu}}, \bibinfo {author} {\bibfnamefont {Z.-L.}\
  \bibnamefont {Wang}}, \ and\ \bibinfo {author} {\bibfnamefont {Y.-F.}\
  \bibnamefont {Chen}},\ }\href@noop {} {\bibfield  {journal} {\bibinfo
  {journal} {arXiv preprint arXiv:1812.06263}\ } (\bibinfo {year}
  {2018})}\BibitemShut {NoStop}%
\bibitem [{\citenamefont {Yang}\ \emph {et~al.}(2019)\citenamefont {Yang},
  \citenamefont {Jia}, \citenamefont {Wu}, \citenamefont {Hang}, \citenamefont
  {Jiang},\ and\ \citenamefont {Xie}}]{yang2019gapped}%
  \BibitemOpen
  \bibfield  {author} {\bibinfo {author} {\bibfnamefont {Y.}~\bibnamefont
  {Yang}}, \bibinfo {author} {\bibfnamefont {Z.}~\bibnamefont {Jia}}, \bibinfo
  {author} {\bibfnamefont {Y.}~\bibnamefont {Wu}}, \bibinfo {author}
  {\bibfnamefont {Z.-H.}\ \bibnamefont {Hang}}, \bibinfo {author}
  {\bibfnamefont {H.}~\bibnamefont {Jiang}}, \ and\ \bibinfo {author}
  {\bibfnamefont {X.}~\bibnamefont {Xie}},\ }\href@noop {} {\bibfield
  {journal} {\bibinfo  {journal} {arXiv preprint arXiv:1903.01816}\ } (\bibinfo
  {year} {2019})}\BibitemShut {NoStop}%
\bibitem [{\citenamefont {Peterson}\ \emph {et~al.}(2018)\citenamefont
  {Peterson}, \citenamefont {Benalcazar}, \citenamefont {Hughes},\ and\
  \citenamefont {Bahl}}]{peterson2018quantized}%
  \BibitemOpen
  \bibfield  {author} {\bibinfo {author} {\bibfnamefont {C.~W.}\ \bibnamefont
  {Peterson}}, \bibinfo {author} {\bibfnamefont {W.~A.}\ \bibnamefont
  {Benalcazar}}, \bibinfo {author} {\bibfnamefont {T.~L.}\ \bibnamefont
  {Hughes}}, \ and\ \bibinfo {author} {\bibfnamefont {G.}~\bibnamefont
  {Bahl}},\ }\href@noop {} {\bibfield  {journal} {\bibinfo  {journal} {Nature}\
  }\textbf {\bibinfo {volume} {555}},\ \bibinfo {pages} {346} (\bibinfo {year}
  {2018})}\BibitemShut {NoStop}%
\bibitem [{\citenamefont {Imhof}\ \emph {et~al.}(2018)\citenamefont {Imhof},
  \citenamefont {Berger}, \citenamefont {Bayer}, \citenamefont {Brehm},
  \citenamefont {Molenkamp}, \citenamefont {Kiessling}, \citenamefont
  {Schindler}, \citenamefont {Lee}, \citenamefont {Greiter}, \citenamefont
  {Neupert},\ and\ \citenamefont {Thomale}}]{imhof2018topolectrical}%
  \BibitemOpen
  \bibfield  {author} {\bibinfo {author} {\bibfnamefont {S.}~\bibnamefont
  {Imhof}}, \bibinfo {author} {\bibfnamefont {C.}~\bibnamefont {Berger}},
  \bibinfo {author} {\bibfnamefont {F.}~\bibnamefont {Bayer}}, \bibinfo
  {author} {\bibfnamefont {J.}~\bibnamefont {Brehm}}, \bibinfo {author}
  {\bibfnamefont {L.~W.}\ \bibnamefont {Molenkamp}}, \bibinfo {author}
  {\bibfnamefont {T.}~\bibnamefont {Kiessling}}, \bibinfo {author}
  {\bibfnamefont {F.}~\bibnamefont {Schindler}}, \bibinfo {author}
  {\bibfnamefont {C.~H.}\ \bibnamefont {Lee}}, \bibinfo {author} {\bibfnamefont
  {M.}~\bibnamefont {Greiter}}, \bibinfo {author} {\bibfnamefont
  {T.}~\bibnamefont {Neupert}}, \ and\ \bibinfo {author} {\bibfnamefont
  {R.}~\bibnamefont {Thomale}},\ }\href@noop {} {\bibfield  {journal} {\bibinfo
   {journal} {Nature Physics}\ }\textbf {\bibinfo {volume} {14}},\ \bibinfo
  {pages} {925} (\bibinfo {year} {2018})}\BibitemShut {NoStop}%
\bibitem [{\citenamefont {Serra-Garcia}\ \emph {et~al.}(2019)\citenamefont
  {Serra-Garcia}, \citenamefont {S\"usstrunk},\ and\ \citenamefont
  {Huber}}]{PhysRevB.99.020304}%
  \BibitemOpen
  \bibfield  {author} {\bibinfo {author} {\bibfnamefont {M.}~\bibnamefont
  {Serra-Garcia}}, \bibinfo {author} {\bibfnamefont {R.}~\bibnamefont
  {S\"usstrunk}}, \ and\ \bibinfo {author} {\bibfnamefont {S.~D.}\ \bibnamefont
  {Huber}},\ }\href {\doibase 10.1103/PhysRevB.99.020304} {\bibfield  {journal}
  {\bibinfo  {journal} {Phys. Rev. B}\ }\textbf {\bibinfo {volume} {99}},\
  \bibinfo {pages} {020304} (\bibinfo {year} {2019})}\BibitemShut {NoStop}%
\bibitem [{\citenamefont {Bernevig}\ and\ \citenamefont
  {Hughes}(2013)}]{bernevig2013topological}%
  \BibitemOpen
  \bibfield  {author} {\bibinfo {author} {\bibfnamefont {B.~A.}\ \bibnamefont
  {Bernevig}}\ and\ \bibinfo {author} {\bibfnamefont {T.~L.}\ \bibnamefont
  {Hughes}},\ }\href@noop {} {\emph {\bibinfo {title} {Topological insulators
  and topological superconductors}}}\ (\bibinfo  {publisher} {Princeton
  university press},\ \bibinfo {year} {2013})\BibitemShut {NoStop}%
\bibitem [{\citenamefont {Fidkowski}(2010)}]{PhysRevLett.104.130502}%
  \BibitemOpen
  \bibfield  {author} {\bibinfo {author} {\bibfnamefont {L.}~\bibnamefont
  {Fidkowski}},\ }\href {\doibase 10.1103/PhysRevLett.104.130502} {\bibfield
  {journal} {\bibinfo  {journal} {Phys. Rev. Lett.}\ }\textbf {\bibinfo
  {volume} {104}},\ \bibinfo {pages} {130502} (\bibinfo {year}
  {2010})}\BibitemShut {NoStop}%
\bibitem [{\citenamefont {Li}\ and\ \citenamefont
  {Haldane}(2008)}]{PhysRevLett.101.010504}%
  \BibitemOpen
  \bibfield  {author} {\bibinfo {author} {\bibfnamefont {H.}~\bibnamefont
  {Li}}\ and\ \bibinfo {author} {\bibfnamefont {F.~D.~M.}\ \bibnamefont
  {Haldane}},\ }\href {\doibase 10.1103/PhysRevLett.101.010504} {\bibfield
  {journal} {\bibinfo  {journal} {Phys. Rev. Lett.}\ }\textbf {\bibinfo
  {volume} {101}},\ \bibinfo {pages} {010504} (\bibinfo {year}
  {2008})}\BibitemShut {NoStop}%
\bibitem [{\citenamefont {Wang}\ \emph {et~al.}(2018)\citenamefont {Wang},
  \citenamefont {Wang},\ and\ \citenamefont {Wang}}]{wang2018entanglement}%
  \BibitemOpen
  \bibfield  {author} {\bibinfo {author} {\bibfnamefont {Q.}~\bibnamefont
  {Wang}}, \bibinfo {author} {\bibfnamefont {D.}~\bibnamefont {Wang}}, \ and\
  \bibinfo {author} {\bibfnamefont {Q.-H.}\ \bibnamefont {Wang}},\ }\href@noop
  {} {\bibfield  {journal} {\bibinfo  {journal} {EPL (Europhysics Letters)}\
  }\textbf {\bibinfo {volume} {124}},\ \bibinfo {pages} {50005} (\bibinfo
  {year} {2018})}\BibitemShut {NoStop}%
\bibitem [{\citenamefont {Zhu}\ \emph {et~al.}(2019)\citenamefont {Zhu},
  \citenamefont {Loehr},\ and\ \citenamefont {Hughes}}]{zhu2019identifying}%
  \BibitemOpen
  \bibfield  {author} {\bibinfo {author} {\bibfnamefont {P.}~\bibnamefont
  {Zhu}}, \bibinfo {author} {\bibfnamefont {K.}~\bibnamefont {Loehr}}, \ and\
  \bibinfo {author} {\bibfnamefont {T.~L.}\ \bibnamefont {Hughes}},\
  }\href@noop {} {\enquote {\bibinfo {title} {Identifying higher order topology
  and fractional corner charge using entanglement spectra},}\ } (\bibinfo
  {year} {2019}),\ \Eprint {http://arxiv.org/abs/1910.10180} {arXiv:1910.10180
  [cond-mat.mes-hall]} \BibitemShut {NoStop}%
\bibitem [{\citenamefont {Takahashi}\ \emph {et~al.}(2019)\citenamefont
  {Takahashi}, \citenamefont {Tanaka},\ and\ \citenamefont
  {Murakami}}]{takahashi2019bulkedge}%
  \BibitemOpen
  \bibfield  {author} {\bibinfo {author} {\bibfnamefont {R.}~\bibnamefont
  {Takahashi}}, \bibinfo {author} {\bibfnamefont {Y.}~\bibnamefont {Tanaka}}, \
  and\ \bibinfo {author} {\bibfnamefont {S.}~\bibnamefont {Murakami}},\
  }\href@noop {} {\enquote {\bibinfo {title} {Bulk-edge and bulk-hinge
  correspondence in inversion-symmetric insulators},}\ } (\bibinfo {year}
  {2019}),\ \Eprint {http://arxiv.org/abs/1910.08290} {arXiv:1910.08290
  [cond-mat.mes-hall]} \BibitemShut {NoStop}%
\bibitem [{\citenamefont {Cirac}\ and\ \citenamefont
  {Verstraete}(2009)}]{Cirac_2009}%
  \BibitemOpen
  \bibfield  {author} {\bibinfo {author} {\bibfnamefont {J.~I.}\ \bibnamefont
  {Cirac}}\ and\ \bibinfo {author} {\bibfnamefont {F.}~\bibnamefont
  {Verstraete}},\ }\href {\doibase 10.1088/1751-8113/42/50/504004} {\bibfield
  {journal} {\bibinfo  {journal} {Journal of Physics A: Mathematical and
  Theoretical}\ }\textbf {\bibinfo {volume} {42}},\ \bibinfo {pages} {504004}
  (\bibinfo {year} {2009})}\BibitemShut {NoStop}%
\bibitem [{\citenamefont {Cirac}\ \emph {et~al.}(2011)\citenamefont {Cirac},
  \citenamefont {Poilblanc}, \citenamefont {Schuch},\ and\ \citenamefont
  {Verstraete}}]{PhysRevB.83.245134}%
  \BibitemOpen
  \bibfield  {author} {\bibinfo {author} {\bibfnamefont {J.~I.}\ \bibnamefont
  {Cirac}}, \bibinfo {author} {\bibfnamefont {D.}~\bibnamefont {Poilblanc}},
  \bibinfo {author} {\bibfnamefont {N.}~\bibnamefont {Schuch}}, \ and\ \bibinfo
  {author} {\bibfnamefont {F.}~\bibnamefont {Verstraete}},\ }\href {\doibase
  10.1103/PhysRevB.83.245134} {\bibfield  {journal} {\bibinfo  {journal} {Phys.
  Rev. B}\ }\textbf {\bibinfo {volume} {83}},\ \bibinfo {pages} {245134}
  (\bibinfo {year} {2011})}\BibitemShut {NoStop}%
\bibitem [{\citenamefont {Hastings}(2006)}]{PhysRevB.73.085115}%
  \BibitemOpen
  \bibfield  {author} {\bibinfo {author} {\bibfnamefont {M.~B.}\ \bibnamefont
  {Hastings}},\ }\href {\doibase 10.1103/PhysRevB.73.085115} {\bibfield
  {journal} {\bibinfo  {journal} {Phys. Rev. B}\ }\textbf {\bibinfo {volume}
  {73}},\ \bibinfo {pages} {085115} (\bibinfo {year} {2006})}\BibitemShut
  {NoStop}%
\bibitem [{\citenamefont {Wolf}\ \emph {et~al.}(2008)\citenamefont {Wolf},
  \citenamefont {Verstraete}, \citenamefont {Hastings},\ and\ \citenamefont
  {Cirac}}]{PhysRevLett.100.070502}%
  \BibitemOpen
  \bibfield  {author} {\bibinfo {author} {\bibfnamefont {M.~M.}\ \bibnamefont
  {Wolf}}, \bibinfo {author} {\bibfnamefont {F.}~\bibnamefont {Verstraete}},
  \bibinfo {author} {\bibfnamefont {M.~B.}\ \bibnamefont {Hastings}}, \ and\
  \bibinfo {author} {\bibfnamefont {J.~I.}\ \bibnamefont {Cirac}},\ }\href
  {\doibase 10.1103/PhysRevLett.100.070502} {\bibfield  {journal} {\bibinfo
  {journal} {Phys. Rev. Lett.}\ }\textbf {\bibinfo {volume} {100}},\ \bibinfo
  {pages} {070502} (\bibinfo {year} {2008})}\BibitemShut {NoStop}%
\bibitem [{\citenamefont {Schuch}\ \emph {et~al.}(2011)\citenamefont {Schuch},
  \citenamefont {P\'erez-Garc\'{\i}a},\ and\ \citenamefont
  {Cirac}}]{PhysRevB.84.165139}%
  \BibitemOpen
  \bibfield  {author} {\bibinfo {author} {\bibfnamefont {N.}~\bibnamefont
  {Schuch}}, \bibinfo {author} {\bibfnamefont {D.}~\bibnamefont
  {P\'erez-Garc\'{\i}a}}, \ and\ \bibinfo {author} {\bibfnamefont
  {I.}~\bibnamefont {Cirac}},\ }\href {\doibase 10.1103/PhysRevB.84.165139}
  {\bibfield  {journal} {\bibinfo  {journal} {Phys. Rev. B}\ }\textbf {\bibinfo
  {volume} {84}},\ \bibinfo {pages} {165139} (\bibinfo {year}
  {2011})}\BibitemShut {NoStop}%
\bibitem [{\citenamefont {Schuch}\ \emph {et~al.}(2010)\citenamefont {Schuch},
  \citenamefont {Cirac},\ and\ \citenamefont
  {P{\'e}rez-Garc{\'\i}a}}]{Schuch2010}%
  \BibitemOpen
  \bibfield  {author} {\bibinfo {author} {\bibfnamefont {N.}~\bibnamefont
  {Schuch}}, \bibinfo {author} {\bibfnamefont {I.}~\bibnamefont {Cirac}}, \
  and\ \bibinfo {author} {\bibfnamefont {D.}~\bibnamefont
  {P{\'e}rez-Garc{\'\i}a}},\ }\href@noop {} {\bibfield  {journal} {\bibinfo
  {journal} {Annals of Physics}\ }\textbf {\bibinfo {volume} {325}},\ \bibinfo
  {pages} {2153} (\bibinfo {year} {2010})}\BibitemShut {NoStop}%
\bibitem [{\citenamefont {Kraus}\ \emph {et~al.}(2010)\citenamefont {Kraus},
  \citenamefont {Schuch}, \citenamefont {Verstraete},\ and\ \citenamefont
  {Cirac}}]{Kraus}%
  \BibitemOpen
  \bibfield  {author} {\bibinfo {author} {\bibfnamefont {C.~V.}\ \bibnamefont
  {Kraus}}, \bibinfo {author} {\bibfnamefont {N.}~\bibnamefont {Schuch}},
  \bibinfo {author} {\bibfnamefont {F.}~\bibnamefont {Verstraete}}, \ and\
  \bibinfo {author} {\bibfnamefont {J.~I.}\ \bibnamefont {Cirac}},\ }\href@noop
  {} {\bibfield  {journal} {\bibinfo  {journal} {Phys. Rev. A}\ }\textbf
  {\bibinfo {volume} {81}},\ \bibinfo {pages} {052338} (\bibinfo {year}
  {2010})}\BibitemShut {NoStop}%
\bibitem [{\citenamefont {Yang}\ \emph {et~al.}(2015)\citenamefont {Yang},
  \citenamefont {Wahl}, \citenamefont {Tu}, \citenamefont {Schuch},\ and\
  \citenamefont {Cirac}}]{PhysRevLett.114.106803}%
  \BibitemOpen
  \bibfield  {author} {\bibinfo {author} {\bibfnamefont {S.}~\bibnamefont
  {Yang}}, \bibinfo {author} {\bibfnamefont {T.~B.}\ \bibnamefont {Wahl}},
  \bibinfo {author} {\bibfnamefont {H.-H.}\ \bibnamefont {Tu}}, \bibinfo
  {author} {\bibfnamefont {N.}~\bibnamefont {Schuch}}, \ and\ \bibinfo {author}
  {\bibfnamefont {J.~I.}\ \bibnamefont {Cirac}},\ }\href {\doibase
  10.1103/PhysRevLett.114.106803} {\bibfield  {journal} {\bibinfo  {journal}
  {Phys. Rev. Lett.}\ }\textbf {\bibinfo {volume} {114}},\ \bibinfo {pages}
  {106803} (\bibinfo {year} {2015})}\BibitemShut {NoStop}%
\bibitem [{\citenamefont {Kitaev}(2003)}]{Kitaev2003}%
  \BibitemOpen
  \bibfield  {author} {\bibinfo {author} {\bibfnamefont {A.}~\bibnamefont
  {Kitaev}},\ }\href {\doibase https://doi.org/10.1016/S0003-4916(02)00018-0}
  {\bibfield  {journal} {\bibinfo  {journal} {Annals of Physics}\ }\textbf
  {\bibinfo {volume} {303}},\ \bibinfo {pages} {2 } (\bibinfo {year}
  {2003})}\BibitemShut {NoStop}%
\bibitem [{\citenamefont {Levin}\ and\ \citenamefont
  {Wen}(2005)}]{PhysRevB.71.045110}%
  \BibitemOpen
  \bibfield  {author} {\bibinfo {author} {\bibfnamefont {M.~A.}\ \bibnamefont
  {Levin}}\ and\ \bibinfo {author} {\bibfnamefont {X.-G.}\ \bibnamefont
  {Wen}},\ }\href {\doibase 10.1103/PhysRevB.71.045110} {\bibfield  {journal}
  {\bibinfo  {journal} {Phys. Rev. B}\ }\textbf {\bibinfo {volume} {71}},\
  \bibinfo {pages} {045110} (\bibinfo {year} {2005})}\BibitemShut {NoStop}%
\bibitem [{\citenamefont {Verstraete}\ and\ \citenamefont
  {Cirac}(2004)}]{Verstraete:2004cf}%
  \BibitemOpen
  \bibfield  {author} {\bibinfo {author} {\bibfnamefont {F.}~\bibnamefont
  {Verstraete}}\ and\ \bibinfo {author} {\bibfnamefont {J.~I.}\ \bibnamefont
  {Cirac}},\ }\href@noop {} {\  (\bibinfo {year} {2004})},\ \Eprint
  {http://arxiv.org/abs/cond-mat/0407066} {arXiv:cond-mat/0407066 [cond-mat]}
  \BibitemShut {NoStop}%
%%CITATION = COND-MAT/0407066;%%
\bibitem [{\citenamefont {Verstraete}\ \emph {et~al.}(2006)\citenamefont
  {Verstraete}, \citenamefont {Wolf}, \citenamefont {Perez-Garcia},\ and\
  \citenamefont {Cirac}}]{VerstraetePRL2006}%
  \BibitemOpen
  \bibfield  {author} {\bibinfo {author} {\bibfnamefont {F.}~\bibnamefont
  {Verstraete}}, \bibinfo {author} {\bibfnamefont {M.~M.}\ \bibnamefont
  {Wolf}}, \bibinfo {author} {\bibfnamefont {D.}~\bibnamefont {Perez-Garcia}},
  \ and\ \bibinfo {author} {\bibfnamefont {J.~I.}\ \bibnamefont {Cirac}},\
  }\href {\doibase 10.1103/PhysRevLett.96.220601} {\bibfield  {journal}
  {\bibinfo  {journal} {Phys. Rev. Lett.}\ }\textbf {\bibinfo {volume} {96}},\
  \bibinfo {pages} {220601} (\bibinfo {year} {2006})}\BibitemShut {NoStop}%
\bibitem [{\citenamefont {Gu}\ \emph {et~al.}(2009)\citenamefont {Gu},
  \citenamefont {Levin}, \citenamefont {Swingle},\ and\ \citenamefont
  {Wen}}]{gu2009tensor}%
  \BibitemOpen
  \bibfield  {author} {\bibinfo {author} {\bibfnamefont {Z.-C.}\ \bibnamefont
  {Gu}}, \bibinfo {author} {\bibfnamefont {M.}~\bibnamefont {Levin}}, \bibinfo
  {author} {\bibfnamefont {B.}~\bibnamefont {Swingle}}, \ and\ \bibinfo
  {author} {\bibfnamefont {X.-G.}\ \bibnamefont {Wen}},\ }\href@noop {}
  {\bibfield  {journal} {\bibinfo  {journal} {Physical Review B}\ }\textbf
  {\bibinfo {volume} {79}},\ \bibinfo {pages} {085118} (\bibinfo {year}
  {2009})}\BibitemShut {NoStop}%
\bibitem [{\citenamefont {Wahl}\ \emph {et~al.}(2013)\citenamefont {Wahl},
  \citenamefont {Tu}, \citenamefont {Schuch},\ and\ \citenamefont
  {Cirac}}]{PhysRevLett.111.236805}%
  \BibitemOpen
  \bibfield  {author} {\bibinfo {author} {\bibfnamefont {T.~B.}\ \bibnamefont
  {Wahl}}, \bibinfo {author} {\bibfnamefont {H.-H.}\ \bibnamefont {Tu}},
  \bibinfo {author} {\bibfnamefont {N.}~\bibnamefont {Schuch}}, \ and\ \bibinfo
  {author} {\bibfnamefont {J.~I.}\ \bibnamefont {Cirac}},\ }\href {\doibase
  10.1103/PhysRevLett.111.236805} {\bibfield  {journal} {\bibinfo  {journal}
  {Phys. Rev. Lett.}\ }\textbf {\bibinfo {volume} {111}},\ \bibinfo {pages}
  {236805} (\bibinfo {year} {2013})}\BibitemShut {NoStop}%
\bibitem [{\citenamefont {Dubail}\ and\ \citenamefont
  {Read}(2015)}]{PhysRevB.92.205307}%
  \BibitemOpen
  \bibfield  {author} {\bibinfo {author} {\bibfnamefont {J.}~\bibnamefont
  {Dubail}}\ and\ \bibinfo {author} {\bibfnamefont {N.}~\bibnamefont {Read}},\
  }\href {\doibase 10.1103/PhysRevB.92.205307} {\bibfield  {journal} {\bibinfo
  {journal} {Phys. Rev. B}\ }\textbf {\bibinfo {volume} {92}},\ \bibinfo
  {pages} {205307} (\bibinfo {year} {2015})}\BibitemShut {NoStop}%
\bibitem [{\citenamefont {Read}(2017)}]{PhysRevB.95.115309}%
  \BibitemOpen
  \bibfield  {author} {\bibinfo {author} {\bibfnamefont {N.}~\bibnamefont
  {Read}},\ }\href {\doibase 10.1103/PhysRevB.95.115309} {\bibfield  {journal}
  {\bibinfo  {journal} {Phys. Rev. B}\ }\textbf {\bibinfo {volume} {95}},\
  \bibinfo {pages} {115309} (\bibinfo {year} {2017})}\BibitemShut {NoStop}%
\bibitem [{\citenamefont {Zaletel}\ and\ \citenamefont
  {Mong}(2012)}]{PhysRevB.86.245305}%
  \BibitemOpen
  \bibfield  {author} {\bibinfo {author} {\bibfnamefont {M.~P.}\ \bibnamefont
  {Zaletel}}\ and\ \bibinfo {author} {\bibfnamefont {R.~S.~K.}\ \bibnamefont
  {Mong}},\ }\href {\doibase 10.1103/PhysRevB.86.245305} {\bibfield  {journal}
  {\bibinfo  {journal} {Phys. Rev. B}\ }\textbf {\bibinfo {volume} {86}},\
  \bibinfo {pages} {245305} (\bibinfo {year} {2012})}\BibitemShut {NoStop}%
\bibitem [{\citenamefont {Thouless}(1983)}]{PhysRevB.27.6083}%
  \BibitemOpen
  \bibfield  {author} {\bibinfo {author} {\bibfnamefont {D.~J.}\ \bibnamefont
  {Thouless}},\ }\href {\doibase 10.1103/PhysRevB.27.6083} {\bibfield
  {journal} {\bibinfo  {journal} {Phys. Rev. B}\ }\textbf {\bibinfo {volume}
  {27}},\ \bibinfo {pages} {6083} (\bibinfo {year} {1983})}\BibitemShut
  {NoStop}%
\bibitem [{\citenamefont {K{\'a}roly}(2016)}]{karoly2016short}%
  \BibitemOpen
  \bibfield  {author} {\bibinfo {author} {\bibfnamefont {A.~J.}\ \bibnamefont
  {K{\'a}roly}},\ }\href@noop {} {\emph {\bibinfo {title} {A Short Course on
  Topological Insulators: Band-structure Topology and Edge States in One and
  Two Dimensions}}}\ (\bibinfo  {publisher} {Springer},\ \bibinfo {year}
  {2016})\BibitemShut {NoStop}%
\bibitem [{\citenamefont {Fannes}\ \emph {et~al.}(1992)\citenamefont {Fannes},
  \citenamefont {Nachtergaele},\ and\ \citenamefont
  {Werner}}]{fannes1992finitely}%
  \BibitemOpen
  \bibfield  {author} {\bibinfo {author} {\bibfnamefont {M.}~\bibnamefont
  {Fannes}}, \bibinfo {author} {\bibfnamefont {B.}~\bibnamefont
  {Nachtergaele}}, \ and\ \bibinfo {author} {\bibfnamefont {R.~F.}\
  \bibnamefont {Werner}},\ }\href@noop {} {\bibfield  {journal} {\bibinfo
  {journal} {Communications in mathematical physics}\ }\textbf {\bibinfo
  {volume} {144}},\ \bibinfo {pages} {443} (\bibinfo {year}
  {1992})}\BibitemShut {NoStop}%
\bibitem [{\citenamefont {Wahl}\ \emph {et~al.}(2014)\citenamefont {Wahl},
  \citenamefont {Ha\ss{}ler}, \citenamefont {Tu}, \citenamefont {Cirac},\ and\
  \citenamefont {Schuch}}]{PhysRevB.90.115133}%
  \BibitemOpen
  \bibfield  {author} {\bibinfo {author} {\bibfnamefont {T.~B.}\ \bibnamefont
  {Wahl}}, \bibinfo {author} {\bibfnamefont {S.~T.}\ \bibnamefont
  {Ha\ss{}ler}}, \bibinfo {author} {\bibfnamefont {H.-H.}\ \bibnamefont {Tu}},
  \bibinfo {author} {\bibfnamefont {J.~I.}\ \bibnamefont {Cirac}}, \ and\
  \bibinfo {author} {\bibfnamefont {N.}~\bibnamefont {Schuch}},\ }\href
  {\doibase 10.1103/PhysRevB.90.115133} {\bibfield  {journal} {\bibinfo
  {journal} {Phys. Rev. B}\ }\textbf {\bibinfo {volume} {90}},\ \bibinfo
  {pages} {115133} (\bibinfo {year} {2014})}\BibitemShut {NoStop}%
\bibitem [{\citenamefont {Deitmar}\ and\ \citenamefont
  {Echterhoff}(2014)}]{deitmar2014principles}%
  \BibitemOpen
  \bibfield  {author} {\bibinfo {author} {\bibfnamefont {A.}~\bibnamefont
  {Deitmar}}\ and\ \bibinfo {author} {\bibfnamefont {S.}~\bibnamefont
  {Echterhoff}},\ }\href@noop {} {\emph {\bibinfo {title} {Principles of
  harmonic analysis}}}\ (\bibinfo  {publisher} {Springer},\ \bibinfo {year}
  {2014})\BibitemShut {NoStop}%
\bibitem [{\citenamefont {Bultinck}\ \emph
  {et~al.}(2017{\natexlab{a}})\citenamefont {Bultinck}, \citenamefont
  {Williamson}, \citenamefont {Haegeman},\ and\ \citenamefont
  {Verstraete}}]{bultinck2017fermionic}%
  \BibitemOpen
  \bibfield  {author} {\bibinfo {author} {\bibfnamefont {N.}~\bibnamefont
  {Bultinck}}, \bibinfo {author} {\bibfnamefont {D.~J.}\ \bibnamefont
  {Williamson}}, \bibinfo {author} {\bibfnamefont {J.}~\bibnamefont
  {Haegeman}}, \ and\ \bibinfo {author} {\bibfnamefont {F.}~\bibnamefont
  {Verstraete}},\ }\href@noop {} {\bibfield  {journal} {\bibinfo  {journal}
  {Physical Review B}\ }\textbf {\bibinfo {volume} {95}},\ \bibinfo {pages}
  {075108} (\bibinfo {year} {2017}{\natexlab{a}})}\BibitemShut {NoStop}%
\bibitem [{\citenamefont {P\'erez-Garc\'{\i}a}\ \emph
  {et~al.}(2008)\citenamefont {P\'erez-Garc\'{\i}a}, \citenamefont {Wolf},
  \citenamefont {Sanz}, \citenamefont {Verstraete},\ and\ \citenamefont
  {Cirac}}]{PhysRevLett.100.167202}%
  \BibitemOpen
  \bibfield  {author} {\bibinfo {author} {\bibfnamefont {D.}~\bibnamefont
  {P\'erez-Garc\'{\i}a}}, \bibinfo {author} {\bibfnamefont {M.~M.}\
  \bibnamefont {Wolf}}, \bibinfo {author} {\bibfnamefont {M.}~\bibnamefont
  {Sanz}}, \bibinfo {author} {\bibfnamefont {F.}~\bibnamefont {Verstraete}}, \
  and\ \bibinfo {author} {\bibfnamefont {J.~I.}\ \bibnamefont {Cirac}},\ }\href
  {\doibase 10.1103/PhysRevLett.100.167202} {\bibfield  {journal} {\bibinfo
  {journal} {Phys. Rev. Lett.}\ }\textbf {\bibinfo {volume} {100}},\ \bibinfo
  {pages} {167202} (\bibinfo {year} {2008})}\BibitemShut {NoStop}%
\bibitem [{\citenamefont {Bravyi}(2005)}]{bravyi2005lagrangian}%
  \BibitemOpen
  \bibfield  {author} {\bibinfo {author} {\bibfnamefont {S.}~\bibnamefont
  {Bravyi}},\ }\href@noop {} {\bibfield  {journal} {\bibinfo  {journal}
  {Quantum Information \& Computation}\ }\textbf {\bibinfo {volume} {5}},\
  \bibinfo {pages} {216} (\bibinfo {year} {2005})}\BibitemShut {NoStop}%
\bibitem [{\citenamefont {Peschel}(2003)}]{peschel2003calculation}%
  \BibitemOpen
  \bibfield  {author} {\bibinfo {author} {\bibfnamefont {I.}~\bibnamefont
  {Peschel}},\ }\href@noop {} {\bibfield  {journal} {\bibinfo  {journal}
  {Journal of Physics A: Mathematical and General}\ }\textbf {\bibinfo {volume}
  {36}},\ \bibinfo {pages} {L205} (\bibinfo {year} {2003})}\BibitemShut
  {NoStop}%
\bibitem [{\citenamefont {{Schuch}}\ and\ \citenamefont
  {{Bauer}}(2019)}]{2019arXiv190610144S}%
  \BibitemOpen
  \bibfield  {author} {\bibinfo {author} {\bibfnamefont {N.}~\bibnamefont
  {{Schuch}}}\ and\ \bibinfo {author} {\bibfnamefont {B.}~\bibnamefont
  {{Bauer}}},\ }\href@noop {} {\bibfield  {journal} {\bibinfo  {journal} {arXiv
  e-prints}\ ,\ \bibinfo {eid} {arXiv:1906.10144}} (\bibinfo {year} {2019})},\
  \Eprint {http://arxiv.org/abs/1906.10144} {arXiv:1906.10144
  [cond-mat.str-el]} \BibitemShut {NoStop}%
\bibitem [{\citenamefont {Perez-Garcia}\ \emph {et~al.}(2006)\citenamefont
  {Perez-Garcia}, \citenamefont {Verstraete}, \citenamefont {Wolf},\ and\
  \citenamefont {Cirac}}]{perez2006matrix}%
  \BibitemOpen
  \bibfield  {author} {\bibinfo {author} {\bibfnamefont {D.}~\bibnamefont
  {Perez-Garcia}}, \bibinfo {author} {\bibfnamefont {F.}~\bibnamefont
  {Verstraete}}, \bibinfo {author} {\bibfnamefont {M.~M.}\ \bibnamefont
  {Wolf}}, \ and\ \bibinfo {author} {\bibfnamefont {J.~I.}\ \bibnamefont
  {Cirac}},\ }\href@noop {} {\bibfield  {journal} {\bibinfo  {journal} {arXiv
  preprint quant-ph/0608197}\ } (\bibinfo {year} {2006})}\BibitemShut {NoStop}%
\bibitem [{\citenamefont {Fishman}\ and\ \citenamefont
  {White}(2015)}]{PhysRevB.92.075132}%
  \BibitemOpen
  \bibfield  {author} {\bibinfo {author} {\bibfnamefont {M.~T.}\ \bibnamefont
  {Fishman}}\ and\ \bibinfo {author} {\bibfnamefont {S.~R.}\ \bibnamefont
  {White}},\ }\href {\doibase 10.1103/PhysRevB.92.075132} {\bibfield  {journal}
  {\bibinfo  {journal} {Phys. Rev. B}\ }\textbf {\bibinfo {volume} {92}},\
  \bibinfo {pages} {075132} (\bibinfo {year} {2015})}\BibitemShut {NoStop}%
\bibitem [{\citenamefont {Bultinck}\ \emph
  {et~al.}(2017{\natexlab{b}})\citenamefont {Bultinck}, \citenamefont
  {Williamson}, \citenamefont {Haegeman},\ and\ \citenamefont
  {Verstraete}}]{BultinckPEPS_2017}%
  \BibitemOpen
  \bibfield  {author} {\bibinfo {author} {\bibfnamefont {N.}~\bibnamefont
  {Bultinck}}, \bibinfo {author} {\bibfnamefont {D.~J.}\ \bibnamefont
  {Williamson}}, \bibinfo {author} {\bibfnamefont {J.}~\bibnamefont
  {Haegeman}}, \ and\ \bibinfo {author} {\bibfnamefont {F.}~\bibnamefont
  {Verstraete}},\ }\href {\doibase 10.1088/1751-8121/aa99cc} {\bibfield
  {journal} {\bibinfo  {journal} {Journal of Physics A: Mathematical and
  Theoretical}\ }\textbf {\bibinfo {volume} {51}},\ \bibinfo {pages} {025202}
  (\bibinfo {year} {2017}{\natexlab{b}})}\BibitemShut {NoStop}%
\bibitem [{\citenamefont {Zhang}\ \emph {et~al.}(2011)\citenamefont {Zhang},
  \citenamefont {Grover},\ and\ \citenamefont
  {Vishwanath}}]{PhysRevB.84.075128}%
  \BibitemOpen
  \bibfield  {author} {\bibinfo {author} {\bibfnamefont {Y.}~\bibnamefont
  {Zhang}}, \bibinfo {author} {\bibfnamefont {T.}~\bibnamefont {Grover}}, \
  and\ \bibinfo {author} {\bibfnamefont {A.}~\bibnamefont {Vishwanath}},\
  }\href {\doibase 10.1103/PhysRevB.84.075128} {\bibfield  {journal} {\bibinfo
  {journal} {Phys. Rev. B}\ }\textbf {\bibinfo {volume} {84}},\ \bibinfo
  {pages} {075128} (\bibinfo {year} {2011})}\BibitemShut {NoStop}%
\bibitem [{\citenamefont {Zhang}\ and\ \citenamefont
  {Vishwanath}(2013)}]{PhysRevB.87.161113}%
  \BibitemOpen
  \bibfield  {author} {\bibinfo {author} {\bibfnamefont {Y.}~\bibnamefont
  {Zhang}}\ and\ \bibinfo {author} {\bibfnamefont {A.}~\bibnamefont
  {Vishwanath}},\ }\href {\doibase 10.1103/PhysRevB.87.161113} {\bibfield
  {journal} {\bibinfo  {journal} {Phys. Rev. B}\ }\textbf {\bibinfo {volume}
  {87}},\ \bibinfo {pages} {161113} (\bibinfo {year} {2013})}\BibitemShut
  {NoStop}%
\bibitem [{\citenamefont {Parameswaran}\ \emph {et~al.}(2013)\citenamefont
  {Parameswaran}, \citenamefont {Roy},\ and\ \citenamefont
  {Sondhi}}]{PARAMESWARAN2013816}%
  \BibitemOpen
  \bibfield  {author} {\bibinfo {author} {\bibfnamefont {S.~A.}\ \bibnamefont
  {Parameswaran}}, \bibinfo {author} {\bibfnamefont {R.}~\bibnamefont {Roy}}, \
  and\ \bibinfo {author} {\bibfnamefont {S.~L.}\ \bibnamefont {Sondhi}},\
  }\href {\doibase https://doi.org/10.1016/j.crhy.2013.04.003} {\bibfield
  {journal} {\bibinfo  {journal} {Comptes Rendus Physique}\ }\textbf {\bibinfo
  {volume} {14}},\ \bibinfo {pages} {816 } (\bibinfo {year} {2013})},\ \bibinfo
  {note} {topological insulators / Isolants topologiques}\BibitemShut {NoStop}%
\bibitem [{\citenamefont {Wu}\ \emph {et~al.}(2019)\citenamefont {Wu},
  \citenamefont {Wang},\ and\ \citenamefont {Tu}}]{wu2019tensor}%
  \BibitemOpen
  \bibfield  {author} {\bibinfo {author} {\bibfnamefont {Y.-H.}\ \bibnamefont
  {Wu}}, \bibinfo {author} {\bibfnamefont {L.}~\bibnamefont {Wang}}, \ and\
  \bibinfo {author} {\bibfnamefont {H.-H.}\ \bibnamefont {Tu}},\ }\href@noop {}
  {\enquote {\bibinfo {title} {Tensor network representations of parton wave
  functions},}\ } (\bibinfo {year} {2019}),\ \Eprint
  {http://arxiv.org/abs/1910.11011} {arXiv:1910.11011 [cond-mat.str-el]}
  \BibitemShut {NoStop}%
\bibitem [{\citenamefont {Peschel}\ and\ \citenamefont
  {Eisler}(2009)}]{Peschel_2009}%
  \BibitemOpen
  \bibfield  {author} {\bibinfo {author} {\bibfnamefont {I.}~\bibnamefont
  {Peschel}}\ and\ \bibinfo {author} {\bibfnamefont {V.}~\bibnamefont
  {Eisler}},\ }\href {\doibase 10.1088/1751-8113/42/50/504003} {\bibfield
  {journal} {\bibinfo  {journal} {Journal of Physics A: Mathematical and
  Theoretical}\ }\textbf {\bibinfo {volume} {42}},\ \bibinfo {pages} {504003}
  (\bibinfo {year} {2009})}\BibitemShut {NoStop}%
\bibitem [{\citenamefont {Kitaev}(2006)}]{kitaev2006anyons}%
  \BibitemOpen
  \bibfield  {author} {\bibinfo {author} {\bibfnamefont {A.}~\bibnamefont
  {Kitaev}},\ }\href@noop {} {\bibfield  {journal} {\bibinfo  {journal} {Annals
  of Physics}\ }\textbf {\bibinfo {volume} {321}},\ \bibinfo {pages} {2}
  (\bibinfo {year} {2006})}\BibitemShut {NoStop}%
\bibitem [{\citenamefont {Botero}\ and\ \citenamefont
  {Reznik}(2003)}]{PhysRevA.67.052311}%
  \BibitemOpen
  \bibfield  {author} {\bibinfo {author} {\bibfnamefont {A.}~\bibnamefont
  {Botero}}\ and\ \bibinfo {author} {\bibfnamefont {B.}~\bibnamefont
  {Reznik}},\ }\href {\doibase 10.1103/PhysRevA.67.052311} {\bibfield
  {journal} {\bibinfo  {journal} {Phys. Rev. A}\ }\textbf {\bibinfo {volume}
  {67}},\ \bibinfo {pages} {052311} (\bibinfo {year} {2003})}\BibitemShut
  {NoStop}%
\bibitem [{\citenamefont {Giedke}\ and\ \citenamefont
  {Ignacio~Cirac}(2002)}]{PhysRevA.66.032316}%
  \BibitemOpen
  \bibfield  {author} {\bibinfo {author} {\bibfnamefont {G.}~\bibnamefont
  {Giedke}}\ and\ \bibinfo {author} {\bibfnamefont {J.}~\bibnamefont
  {Ignacio~Cirac}},\ }\href {\doibase 10.1103/PhysRevA.66.032316} {\bibfield
  {journal} {\bibinfo  {journal} {Phys. Rev. A}\ }\textbf {\bibinfo {volume}
  {66}},\ \bibinfo {pages} {032316} (\bibinfo {year} {2002})}\BibitemShut
  {NoStop}%
\bibitem [{\citenamefont {Eisert}\ \emph {et~al.}(2002)\citenamefont {Eisert},
  \citenamefont {Scheel},\ and\ \citenamefont
  {Plenio}}]{PhysRevLett.89.137903}%
  \BibitemOpen
  \bibfield  {author} {\bibinfo {author} {\bibfnamefont {J.}~\bibnamefont
  {Eisert}}, \bibinfo {author} {\bibfnamefont {S.}~\bibnamefont {Scheel}}, \
  and\ \bibinfo {author} {\bibfnamefont {M.~B.}\ \bibnamefont {Plenio}},\
  }\href {\doibase 10.1103/PhysRevLett.89.137903} {\bibfield  {journal}
  {\bibinfo  {journal} {Phys. Rev. Lett.}\ }\textbf {\bibinfo {volume} {89}},\
  \bibinfo {pages} {137903} (\bibinfo {year} {2002})}\BibitemShut {NoStop}%
\bibitem [{\citenamefont {Fiur\'a\ifmmode~\check{s}\else
  \v{s}\fi{}ek}(2002)}]{PhysRevLett.89.137904}%
  \BibitemOpen
  \bibfield  {author} {\bibinfo {author} {\bibfnamefont {J.}~\bibnamefont
  {Fiur\'a\ifmmode~\check{s}\else \v{s}\fi{}ek}},\ }\href {\doibase
  10.1103/PhysRevLett.89.137904} {\bibfield  {journal} {\bibinfo  {journal}
  {Phys. Rev. Lett.}\ }\textbf {\bibinfo {volume} {89}},\ \bibinfo {pages}
  {137904} (\bibinfo {year} {2002})}\BibitemShut {NoStop}%
\bibitem [{\citenamefont {Zohar}\ \emph {et~al.}(2015)\citenamefont {Zohar},
  \citenamefont {Burrello}, \citenamefont {Wahl},\ and\ \citenamefont
  {Cirac}}]{ZOHAR2015385}%
  \BibitemOpen
  \bibfield  {author} {\bibinfo {author} {\bibfnamefont {E.}~\bibnamefont
  {Zohar}}, \bibinfo {author} {\bibfnamefont {M.}~\bibnamefont {Burrello}},
  \bibinfo {author} {\bibfnamefont {T.~B.}\ \bibnamefont {Wahl}}, \ and\
  \bibinfo {author} {\bibfnamefont {J.~I.}\ \bibnamefont {Cirac}},\ }\href
  {\doibase https://doi.org/10.1016/j.aop.2015.10.009} {\bibfield  {journal}
  {\bibinfo  {journal} {Annals of Physics}\ }\textbf {\bibinfo {volume}
  {363}},\ \bibinfo {pages} {385 } (\bibinfo {year} {2015})}\BibitemShut
  {NoStop}%
\bibitem [{\citenamefont {Alexandradinata}\ \emph {et~al.}(2011)\citenamefont
  {Alexandradinata}, \citenamefont {Hughes},\ and\ \citenamefont
  {Bernevig}}]{PhysRevB.84.195103}%
  \BibitemOpen
  \bibfield  {author} {\bibinfo {author} {\bibfnamefont {A.}~\bibnamefont
  {Alexandradinata}}, \bibinfo {author} {\bibfnamefont {T.~L.}\ \bibnamefont
  {Hughes}}, \ and\ \bibinfo {author} {\bibfnamefont {B.~A.}\ \bibnamefont
  {Bernevig}},\ }\href {\doibase 10.1103/PhysRevB.84.195103} {\bibfield
  {journal} {\bibinfo  {journal} {Phys. Rev. B}\ }\textbf {\bibinfo {volume}
  {84}},\ \bibinfo {pages} {195103} (\bibinfo {year} {2011})}\BibitemShut
  {NoStop}%
\end{thebibliography}%
	\appendix
 	
 		\section{Covariance matrices of Gaussian fermionic states\label{sec:AppendixDefCM}}

	Eigenstates and thermal states of free-fermion systems are given by Gaussian states which satisfy Wick's theorem. They are hence fully characterised by their covariance matrix (CM) of two-point correlation functions~\cite{bravyi2005lagrangian}. In this appendix, we review the definitions of the complex and real CM for pure and mixed states in Appendix~\ref{sec:AppGaussianStateDef}. We summarise the relation with the entanglement spectrum in Appendix~\ref{sec:AppGaussianStateES} and provide the concrete expression for the CM of a Gaussian state parametrised as the exponential of a fermion bilinear in Appendix~\ref{sec:AppGaussianStateParam}.
	
	\subsection{Definitions\label{sec:AppGaussianStateDef}}	
	
	We consider a generic system of $N$ fermionic DOFs with annihilation operators $a_j$ and creation operators $a_j^{\dagger}$ for $ j = 1, \dotsc, N$. In a pure or mixed state of this system, its covariance matrix is defined as
	\begin{equation}\label{ComplexCM}
	G_{\mu\nu} = \frac{i}{2} \myexp{[\chi^{\dagger}_{\mu}, \chi_{\nu}]} = \begin{pmatrix}
	R^* & Q^*\\
	Q & R\\
	\end{pmatrix},
	\end{equation}
	where $\mu, \nu = 1, \dotsc, 2N$ and $\chi = ( a_1, \dotsc, a_N, a_1^{\dagger}, \dotsc, a_N^{\dagger})$ is the mode vector. The blocks $R$ and $Q$ of dimension $N\times N$ in Eq.~\eqref{ComplexCM} are anti-Hermitian and anti-symmetric, respectively, \ie $R^{\dagger} = - R$ and $Q^{T} = - Q$, such that $G$ is anti-Hermitian. For a generic mixed state, $G$ satisfies the inequality $GG^{\dagger} \leq \frac{1}{4} \mathbbm{1}$ and its eigenvalues come in complex conjugate pairs $\pm\frac{i}{2}|\lambda_j|$ with $0\leq |\lambda _j| \leq 1$ for $1 \leq j \leq N$. For pure states, $GG^{\dagger} = \frac{1}{4} \mathbbm{1}$ such that its eigenvalues are given by $|\lambda _j| = 1$.
	
	For a state with charge conservation, the levels $-1 \leq \lambda _j \leq 1$ can be computed directly including their sign~\cite{Peschel_2009}: Since $\myexp{a^{\dagger}_{j} a^{\dagger}_{k}} = 0$ for all $1 \leq j,k \leq N$, the off-diagonal block $Q^*$ vanishes in the complex CM $G$ of Eq.~\eqref{ComplexCM}, and the eigenvalues of the diagonal block $R^*$ are given by $\frac{i}{2}\lambda_j$. 
	
	For a generic system with pair creation and annihilation, it is convenient to define Majorana fermionic modes $c_{2j} = a_j^{\dagger} + a_j$, $c_{2j-1} = (-i) (a_j^{\dagger} - a_j)$ for $j = 1, \dotsc, N$ with $\{c_{\mu}, c_{\nu}\} = 2 \delta_{\mu\nu}$~\cite{kitaev2006anyons}. We denote the matrix corresponding to this basis change by 	
	\begin{equation}\label{BasisChangeMajoranaComplex}
	S_{\mu\nu} = \begin{cases}
	\delta_{\nu, \mu/2 } + \delta_{\nu, \mu/2 + N} &\quad \mu\text{ even}\\
	i(\delta_{\nu, (\mu+1)/2 } - \delta_{\nu, (\mu+1)/2 + N}) &\quad \mu\text{ odd}
	\end{cases}
	\end{equation}
	with $	c_{\mu} = \sum_{\nu} S_{\mu\nu}\chi_{\nu} $ and $S^{\dagger} = 2 S^{-1}$. In terms of the Majorana modes, the CM
	\begin{equation}\label{RealCM}
	\Gamma_{\mu\nu} = \frac{i}{2} \myexp{[c_{\mu}, c_{\nu}]} = \left[ S^{*} \,G \, S^T\right]_{\mu\nu} 
	\end{equation}	
	of size $2N \times 2N$ is real and anti-symmetric. Hence, each of its singular values $|\lambda_j|$ with $1 \leq j \leq N$ is doubly degenerate. Moreover, $\Gamma$ satisfies $\Gamma\Gamma^{\dagger} \leq \mathbbm{1}$ with equality for a pure state.

	\subsection{Relation to entanglement spectrum\label{sec:AppGaussianStateES}} 
	
	We frequently consider the restriction of a pure quantum state $\ket{\psi}$ to a subsystem $\mathcal{A}$ of the entire system, which is described by the reduced density matrix $\rho_{\mathcal{A}} = \tr_{\bar{\mathcal{A}}}[\ket{\psi}\bra{\psi}]$ obtained by tracing over the DOFs in the  complement $\bar{\mathcal{A}}$ of $\mathcal{A}$. The many body entanglement Hamiltonian $H_{\mathrm{Ent}}$ \wrt this partition is given by the logarithm of the reduced density matrix as~\cite{PhysRevLett.101.010504}
	\begin{equation}
	\rho_{\mathcal{A}} = \frac{1}{Z} e^{-H_{\mathrm{Ent}}}, 
	\end{equation}
	where $ Z = \tr \big(e^{-H_{\mathrm{Ent}}}\big)$. 
	
	If $\ket{\psi}$ is a Gaussian state with CM $G$, then $\rho_{\mathcal{A}}$ is Gaussian with its CM $G_{\mathcal{A}} = \left(G_{\mu\nu}\right)_{\mu, \nu \in \mathcal{A}}$ given by the restriction of $G$ to the modes of $\mathcal{A}$. In this case, the many-body entanglement Hamiltonian is a bilinear function of the fermionic mode operators defined by a square matrix referred to as the single-particle entanglement Hamiltonian. The eigenvalues $\beta_j$ of the single-particle entanglement Hamiltonian are related to the eigenvalues $\pm\frac{i}{2}|\lambda_j|$ of $G_{\mathcal{A}}$ as $|\lambda_j| = \tanh\frac{\beta_j}{2}$, where $1 \leq j \leq L$ and $L$ is the number of modes in the subsystem $\mathcal{A}$~\cite{peschel2003calculation}. We therefore refer to the collection
	\begin{equation}\label{DefSPES}
	\{|\lambda_j|\}_{1 \leq j \leq L}
	\end{equation}
	as the single particle ES of $\rho_{\mathcal{A}}$. 
	
	The single particle ES can also be computed from the Majorana CM $\Gamma$ of Eq.~\eqref{RealCM} for the Gaussian state $\ket{\psi}$. One obtains the levels $|\lambda_j|$ for $1 \leq j \leq L$ directly as the singular values of the block $(\Gamma_{\mu\nu})_{\mu, \nu \in \mathcal{A}}$ associated with the DOFs in $\mathcal{A}$. On the other hand, the levels can be computed indirectly from the off-diagonal block $(\Gamma_{\mu\nu})_{\mu \in \mathcal{A}, \nu \in \bar{\mathcal{A}}}$ describing the correlations between $\mathcal{A}$ and its complement $\bar{\mathcal{A}}$. Indeed, the two blocks are coupled by the constraint $\Gamma \Gamma ^T = \mathbbm{1}$ originating in the purity of the state $\ket{\psi}$. Concretely, the singular values of the off-diagonal block are given by
	\begin{equation}\label{OffDiagonalSPES}
	\mu_j = \sqrt{1 - \lambda_j^2}
	\end{equation}
	for $1 \leq j \leq L$~\cite{PhysRevA.67.052311}. This allows to infer the single-particle entanglement energies for weakly entangled modes $|\lambda_j| \approx 1$ with an improved numerical accuracy.
	
	\subsection{Gaussian projections and Schur complements\label{sec:AppGaussianStateSchur}}
	
	In the construction of GfTNSs to be discussed in Appendix~\ref{sec:AppendixGfPEPSConstruction}, we will frequently encounter the following situation: Let us take a total system of $N + M$ fermionic modes with mode operators $a_j, a_j^{\dagger}$ for $ j = 1, \dotsc, N + M$, and consider the subsystem defined by the $M$ last modes $j = N + 1, \dotsc, N + M$. Let $\ket{Q}$ be a Gaussian pure state of the total system and $\ket{\omega}$ a Gaussian pure state of the subsystem of the last $M$ modes. Then, the projection
	\begin{equation}\label{ProjectionGaussian}
	\ket{\psi} = \langle \omega | Q \rangle,
	\end{equation}
	which is a state of only the first $N$ modes, is again a Gaussian state. The Majorana CMs $\Gamma _{\psi}$, $\Gamma _{Q}$ and $\Gamma _{\omega}$ of the three Gaussian states are given by the expression in Eq.~\eqref{RealCM}, where the expectation values are taken in the states $\ket{\psi}$, $\ket{Q}$ and $\ket{\omega}$, respectively.
	
	$\Gamma _{\psi}$ can be computed from $\Gamma _{Q}$ and $\Gamma _{\omega}$ using a Schur complement coming from the Gaussian integration over the last $M$ modes~\cite{bravyi2005lagrangian, PhysRevA.66.032316, PhysRevLett.89.137903, PhysRevLett.89.137904}. Indeed, we can write the anti-symmetric Majorana CM $\Gamma _{Q}$ of the un-projected state of the total system as
	\begin{equation}
	\Gamma _{Q} = \begin{pmatrix}
	\mathbf{A} & \mathbf{B} \\
	- \mathbf{B}^T & \mathbf{D}
	\end{pmatrix},
	\end{equation}
	where the block $\mathbf{A}$ of size $2N \times 2N$ refers to the Majorana modes corresponding to $a_j, a_j^{\dagger}$ for $ j = 1, \dotsc, N$, and the block $\mathbf{D}$ of size $2M \times 2M$ refers to the Majorana modes corresponding to $a_j, a_j^{\dagger}$ for $ j = N + 1, \dotsc, N + M$. The block $\mathbf{B}$ of size $2N \times 2M$ encodes the correlations between the first $N$ and last $M$ modes. In terms of these blocks and the CM $\Gamma _{\omega}$ of size $2M \times 2M$, the CM of the projected state is 
	\begin{equation}\label{SchurComplementIntro}
	\Gamma _{\psi} = \mathbf{A} + \mathbf{B}\, \left[ \mathbf{D} + \Gamma _{\omega}\right]^{-1} \,\mathbf{B}^T .
	\end{equation}
	Mathematically, this is the Schur complement of the block corresponding to the last $M$ modes.

	\subsection{Parametrisation of Gaussian states\label{sec:AppGaussianStateParam}}
	
	We consider a normalised Gaussian state with even fermionic parity which, anticipating Appendix~\ref{sec:AppendixGfPEPSConstruction}, we will denote by $\ket{\psi} = \ket{Q}$. $\ket{Q}$ is parametrised in terms of the anti-symmetric complex matrix $M$ of dimension $N \times N$ as~\cite{bravyi2005lagrangian, ZOHAR2015385}
	\begin{equation}\label{GaussianMapApp}
	\ket{Q} = \mathcal{N} \times e^{\sum_{ij} M_{ij} a^{\dagger}_i a^{\dagger}_j}\ket{\Omega},
	\end{equation}
	where $\ket{\Omega}$ is the fermionic vacuum and $\mathcal{N}$ is a normalisation factor. 
	
	An relevant case is given by states whose parametrisation matrices have non-zero entries only in the first row and column (or more generally, in the $n^{\text{th}}$ row and column). In particular, we will see in Appendix~\ref{sec:AppendixSSHMPS} and Appendix~\ref{sec:AppendixQuadrupolePEPS} that the local tensors of the MPS of Eq.~\eqref{MPSMyInterpol} and the PEPS of Eq.~\eqref{QuadrupolePEPS} define Gaussian states of this form. In this case, $M_{ij}M_{kl} \, a^{\dagger}_i a^{\dagger}_j a^{\dagger}_k a^{\dagger}_l = 0$ for all $1 \leq i,j,k,l \leq N$ since all non-vanishing terms contain the factor $(a^{\dagger}_1)^2 = 0$ ($(a^{\dagger}_n)^2 = 0$ in the general case). Therefore, the series expression of the exponential in Eq.~\eqref{GaussianMapApp} terminates after first order such that
	\begin{equation}\label{GaussianStateFirstOrder}
	\ket{Q} = \mathcal{N} \left[1 + \sum_{i,j = 1}^{N}M_{ij} a^{\dagger}_i a^{\dagger}_j\right]\ket{\Omega},
	\end{equation}
	leading to the CM
	\begin{equation}\label{ComplexCMFirstOrder}
	G = \frac{i}{2}\begin{pmatrix}
	-\mathbbm{1}_N + 8 \mathcal{N}^2 \times M^{\dagger} M & -4 \mathcal{N}^2 \times M^{\dagger}\\
	-4 \mathcal{N}^2 \times M &  \mathbbm{1}_N - 8 \mathcal{N}^2 \times M M^{\dagger}
	\end{pmatrix},
	\end{equation}
	where the normalisation is $\mathcal{N}^{-2} = 1 + 2\tr \big(M M^{\dagger}\big)$.

		\section{Construction of GfTNS\label{sec:AppendixGfPEPSConstruction}}
		
	Gaussian fermionic tensor network states (GfTNS) describe free fermion systems. They have the advantage that the contraction of the network can be performed in terms of covariance matrices, allowing efficient computations even for large systems. In this appendix, we review the construction of Gaussian TNSs in one and two spatial dimensions. We begin in Appendix~\ref{sec:AppendixGfPEPSConstruction1} by illustrating the construction of a TNS via fiducial states using the simple example of a bosonic MPS, where we do not need to take care of fermionic signs in tensor products. In Appendix~\ref{sec:AppendixGfPEPSConstruction2}, we move to the case of fermionic physical and virtual particles. In Appendix~\ref{sec:AppendixSSHMPS}, the concepts introduced here are illustrated using the pedagogical example of the SSH model MPS from Sec.~\ref{sec:fMPS_Construction}.
	
	\subsection{Fiducial state approach\label{sec:AppendixGfPEPSConstruction1}}
	
	We recall that the construction of a TNS can be performed in several different but equivalent ways. Within the most well-known approach, the local information on the state is encoded in the local tensor with physical and virtual legs. For example, for a one-dimensional bosonic MPS $\ket{\psi_{\text{bMPS}}}$ the local tensor at position $x$ takes the form $A[x]^{i_x}_{l_xr_x}$, where $i_x$ is the physical index, $l_x$ is the left virtual index and $r_x$ is the right virtual index. The global state is obtained by contracting, for each nearest-neighbour bond, the two virtual legs associated with this bond which belong to neighboring tensors. This is done by identifying and summing over the corresponding virtual indices. For example, the contraction of two neighboring MPS tensors gives $\sum_{r_x, l_{x+1}} A[x]^{i_x}_{l_xr_x} \delta_{r_xl_{x+1}} A[x+1]^{i_{x+1}}_{l_{x+1}r_{x+1}}$.
	
	More formally, this can be expressed as follows: First, we create a total maximally entangled state in the virtual layer whose role it is to implement the contraction of bonds. The total maximally entangled state is the tensor product over all nearest-neighbour bonds of a maximally entangled state of the two virtual particles for this bond. In our MPS example, the maximally entangled state for the bond between sites $x$ and $x+1$ is
	\begin{equation}
	\ket{\omega_{x, x+1}} = \sum_{r_x, l_{x+1}} \delta_{r_xl_{x+1}} \ket{r_x,l_{x+1}}.
	\end{equation} 
	Secondly, at each site we translate the local tensor into a local projection map, which maps the virtual particles onto the physical particle at this site. The representation matrix of the local projection map is given by the local tensor. For instance, the MPS local projection map at site $x$ is
	\begin{equation}
	\hat{A}[x] = \sum_{i_x,l_x,r_x} A[x]^{i_x}_{l_xr_x} \ket{i_x} \bra{l_x, r_x}.
	\end{equation}
	The bond between sites $x$ and $x+1$ is contracted by applying the projection maps $\hat{A}[x]$ and $\hat{A}[x+1]$ to $\ket{\omega_{x, x+1}}$, giving 
	\begin{multline}
	\hat{A}[x]\hat{A}[x+1]\ket{\omega_{x, x+1}} = \sum_{\substack{i_x,i_{x+1}\\l_x,r_{x+1}}}  \ket{i_x, i_{x+1}} \bra{l_x, r_{x+1}} \times \\
	\sum_{r_x, l_{x+1}}A[x]^{i_x}_{l_xr_x} \delta_{r_xl_{x+1}} A[x+1]^{i_{x+1}}_{l_{x+1}r_{x+1}}.
	\end{multline} 
	The global state is then obtained by contracting all bonds, \ie by applying the product of all local projection maps to the total virtual maximally entangled state. For instance, the MPS on a chain with $N_x$ sites and periodic boundaries takes the form
	\begin{equation}\label{bMPS1}
	\ket{\psi_{\text{bMPS}}} = \prod _{x = 0} ^ {N_x -1} \hat{A}[x] \bigotimes _{x = 0} ^ {N_x -1} \ket{\omega_{x, x+1}}
	\end{equation} 
	where the lattice site indices are $x = 0, \dotsc, N_x -1$. This is a state of only the physical particles.
	
	For GfTNSs, we will deal with fermionic particles and therefore follow a slightly different, but equivalent, approach to construct the TNSs using so-called fiducial states (see Refs.~\cite{PhysRevB.90.115133, ZOHAR2015385} for pedagogical introductions). In this approach, we consider fiducial states instead of local projection maps, which contain only creation operators. The fiducial state on each lattice site lies in the joint Hilbert space of the physical and virtual particles on this site. Its basis coefficients are given by the entries of the local tensor. For example, the local fiducial states of the MPS are 
	\begin{equation}\label{FiducialStatebMPS}
	\ket{Q_x} = \sum_{i_x,l_x,r_x} A[x]^{i_x}_{l_xr_x} \ket{i_x} \ket{l_x, r_x}.
	\end{equation}
	Hence, the fiducial states are equivalent to the local projection maps or local tensors; in particular, they contain all the local information about the TNSs. The total state is obtained by projecting the tensor product of all local fiducial states on the total virtual maximally entangled state. For the MPS, we have
	\begin{equation}
	\ket{\psi_{\text{bMPS}}} = \bigotimes _{x = 0} ^ {N_x -1} \bra{\omega_{x, x+1}} \bigotimes _{x = 0} ^ {N_x -1} \ket{Q_x}
	\end{equation}
	which is equivalent to the expression in Eq.~\eqref{bMPS1}. We are now ready to generalize the fiducial state formalism to fermionic physical and virtual particles.

	\subsection{Fermionic particles\label{sec:AppendixGfPEPSConstruction2}}

	We consider a one- or two-dimensional lattice system of free fermions with $f$ fermionic modes per lattice site $\mathbf{x}$, which are associated to the physical creation operators $a^{\dagger}_{\tau, \mathbf{x}}$ with $\tau = 1, \dotsc, f$. A Gaussian fermionic TNS for this system with physical dimension $2^{f}$ and bond dimension $2^{\xi}$ is obtained by associating $\xi$ complex virtual fermionic modes with each physical lattice site $\mathbf{x}$ and nearest-neighbour direction $\alpha$, where $\alpha = L,R$ for one-dimensional MPSs with left and right nearest neighbour bonds and $\alpha = L, U, R, D$ for two-dimensional PEPSs on the square lattice with left, up, right and down nearest-neighbour bonds~\cite{Kraus, ZOHAR2015385}. We denote the creation operators for these virtual modes by $b^{\dagger}_{\alpha, j, \mathbf{x}}$ where $j = 1, \dotsc, \xi$ labels the different modes per bond and lattice site. For each lattice site, there are hence $n_{\mathrm{modes}} = f + 2\xi$ modes for a fermionic MPS and $n_{\mathrm{modes}} = f + 4\xi$ modes for a two-dimensional fermionic PEPS. The mode operators needed for the construction of the SSH MPS are discussed in the first paragraph of Appendix~\ref{sec:AppendixSSHMPS}.
	
	We collect the mode operators associated with one lattice site into a mode vector
	\begin{multline}\label{ModeVector}
	\chi_{\mathbf{x}} = \big(a_{1,\mathbf{x}}, a_{2,\mathbf{x}}, \dotsc, a_{f,\mathbf{x}}, b_{L,1,\mathbf{x}}, b_{L,2,\mathbf{x}}, \dotsc, b_{R,1,\mathbf{x}},\dotsc, \\
	a^{\dagger}_{1,\mathbf{x}} , \dotsc, b^{\dagger}_{L,1,\mathbf{x}}, \dotsc, b^{\dagger}_{R,1,\mathbf{x}}, \dotsc \big)^T
	\end{multline}
	of length $2n_{\mathrm{modes}}$. The physical and virtual mode operators obey canonical anti-commutation relations
	\begin{equation}
	\left\{ (\chi_{\mathbf{x}})_{\mu},  (\chi_{\mathbf{x'}})_{\nu} \right\} = \delta_{\mathbf{x}, \mathbf{x'} } \delta_{|\mu - \nu|,n_{\mathrm{modes}}}
	\end{equation}
	for $1 \leq \mu, \nu \leq 2n_{\mathrm{modes}}$. Their joint vacuum state $\ket{\Omega}$ satisfies
	\begin{equation}
	a_{\tau, \mathbf{x}}\ket{\Omega} = b_{\alpha, j, \mathbf{x}}\ket{\Omega} = 0
	\end{equation}
	for all $\mathbf{x}$, $\tau = 1, \dotsc, f$, $\alpha$ and $j = 1, \dotsc, \xi$.
	
	The local information about the TNS is contained in the local fiducial states $\ket{Q_{\mathbf{x}}}$, introduced in Eq.~\eqref{FiducialStatebMPS}, which are equivalent to the local tensors $A[\mathbf{x}]$ used in the main text. We can easily translate between the two approaches, since the basis elements of the local fiducial state are by definition equal to the local tensors (\cf Eq.~\eqref{FiducialStatebMPS}). For fermions, we write $\ket{Q_{\mathbf{x}}} = Q_{\mathbf{x}}\ket{\Omega}$ where $Q_{\mathbf{x}}$ is a polynomial of creation operators which acts on the vacuum to create the fiducial state. For fermionic MPSs and PEPSs in one and two dimensions we have~\cite{bultinck2017fermionic, BultinckPEPS_2017}
	\begin{gather}\label{RelationGaussianMapLocalTensor}
	A[\mathbf{x}]^i_{lr} = \left[\bra{r}\otimes \bra{i} \otimes \bra{l}\right] Q_{\mathbf{x}} \ket{\Omega},\\	
	A[\mathbf{x}]^i_{lurd} = \left[\bra{d} \otimes \bra{r} \otimes\bra{u}\otimes \bra{l} \otimes \bra{i}\right] Q_{\mathbf{x}} \ket{\Omega},
	\end{gather} 
	where $\ket{i}$ with $ i = 0, \dotsc, 2^{f}-1$ is a basis for the Fock space associated with the physical mode operators $\{a^{\dagger}_{\tau, \mathbf{x}}\}_{1 \leq \tau \leq f}$, and $\ket{l}$ with $l = 0, \dotsc, 2^{\xi}-1$ is a basis of the Fock space associated with the left virtual mode operators $\{b^{\dagger}_{L, j, \mathbf{x}}\}_{1 \leq j \leq \xi}$ on site $\mathbf{x}$ (similarly $\ket{u}$ for the up mode operators $\{b^{\dagger}_{U, j, \mathbf{x}}\}_{1 \leq j \leq \xi}$, $\ket{r}$ for the right mode operators $\{b^{\dagger}_{R, j, \mathbf{x}}\}_{1 \leq j \leq \xi}$ and $\ket{d}$ for the down mode operators $\{b^{\dagger}_{D, j, \mathbf{x}}\}_{1 \leq j \leq \xi}$). For the SSH model MPS, the fiducial states obtained thus are given in Eq.~\eqref{FiducialMapsMPS}.
	
	A Gaussian fermionic TNS has the property that all local fiducial states $\ket{Q_{\mathbf{x}}}$ satisfy Wick's theorem. In this case, the global physical state $\ket{\psi}$ is also Gaussian~\cite{Kraus, bravyi2005lagrangian}. We denote by $\Gamma _{Q}$ the Majorana CM of the product of all fiducial states, $\prod_{\mathbf{x}}Q_{\mathbf{x}}\ket{\Omega}$, also referred to as total fiducial state. From now onwards, we consider GfTNSs with parity-even local tensors as discussed in Sec.~\ref{sec:fMPS_fMPSAtHalfFilling}, whose local fiducial states therefore have an even number of physical and virtual fermions. The maps $Q_{\mathbf{x}}$ can thus be expressed as in Eq.~\eqref{GaussianMapApp} as the exponential of a quadratic form of the physical and virtual creation operators on the site $\mathbf{x}$, which are contained in the last $n_{\mathrm{modes}}$ entries $\big[(\chi_{\mathbf{x}})_{n_{\mathrm{modes}} + m}\big]_{1 \leq m \leq n_{\mathrm{modes}} }$ of the mode vector of Eq.~\eqref{ModeVector}~\cite{bravyi2005lagrangian, ZOHAR2015385}. Concretely, the local fiducial state is parametrised as  
	\begin{equation}\label{ParametrisationQLocal}
	Q_{\mathbf{x}} = \exp \left[\sum_{m,m' =  1}^{n_{\mathrm{modes}}} (M_{\mathbf{x}})_{mm'} (\chi_{\mathbf{x}})_{n_{\mathrm{modes}} + m} (\chi_{\mathbf{x}})_{n_{\mathrm{modes}} + m'} \right],
	\end{equation}
	with an anti-symmetric square matrix $M_{\mathbf{x}}$ of dimension $n_{\mathrm{modes}}$. For the SSH model MPS, this matrix is given in Eq.~\eqref{CoefficientMatrixSSH}.
	
	In order to illustrate the concepts introduced above, let us construct a Gaussian maximally entangled state of the virtual fermions for each nearest-neighbour bond $\langle \mathbf{x'} \mathbf{x''} \rangle$. Let us denote by $\alpha'$ and $\alpha''$ the type of virtual fermion involved in the bond $\langle \mathbf{x'} \mathbf{x''} \rangle$ on the site $\mathbf{x'}$ and $\mathbf{x''}$, respectively. For instance, if $\mathbf{x''} = \mathbf{x'} + \mathbf{\hat{x}}$, then $\alpha' = R$ and $\alpha'' = L$. A fermionic maximally entangled state for this bond is then given by~\cite{Kraus}
	\begin{equation}\label{MaxEntState}
	\ket{\omega_{\mathbf{x'} \mathbf{x''}}} = \prod_{j = 1}^{\xi} \frac{1}{\sqrt{2}} \left(1 + b^{\dagger}_{\alpha', j, \mathbf{x'}} b^{\dagger}_{\alpha'', j, \mathbf{x''}}\right) \ket{\Omega}. 
	\end{equation} 		
	Due to the fermionic anti-commutation relations, this expression is not symmetric under exchange of $(\alpha',\mathbf{x'})$ and $(\alpha'', \mathbf{x''})$; we say that the bond points from the initial site $\mathbf{x'}$ to the final site $\mathbf{x''}$.
	
	The state $\ket{\omega_{\mathbf{x'} \mathbf{x''}}}$ is a Gaussian state satisfying Wick's theorem. To see this, we first consider the simple case of a one-dimensional MPS with $\xi = 1$ virtual fermion per nearest-neighbour bond and lattice site. In this case, the virtual maximally entangled state from Eq.~\eqref{MaxEntState} for the bond $\langle \mathbf{x}, \mathbf{x} + \mathbf{\hat{x}} \rangle$ becomes
	\begin{multline}\label{MaxEntStateSingleFermion}
	\ket{\omega_{\mathbf{x}, \mathbf{x} + \mathbf{\hat{x}}}} = \frac{1}{\sqrt{2}}\left[1 + b^{\dagger}_{R,\mathbf{x}} b^{\dagger}_{L,\mathbf{x} + \mathbf{\hat{x}}}\right] \ket{\Omega}=\\
	 \frac{1}{\sqrt{2}}\left[1 + \left(b^{\dagger}_{R,\mathbf{x}} \, b^{\dagger}_{L,\mathbf{x} + \mathbf{\hat{x}}}\right) \frac{1}{2}\begin{pmatrix}
	 0 & 1 \\
	 -1 & 0
	 \end{pmatrix} \begin{pmatrix}
	b^{\dagger}_{R,\mathbf{x}} \\
	b^{\dagger}_{L,\mathbf{x} + \mathbf{\hat{x}}}
	 \end{pmatrix}\right] \ket{\Omega},
	\end{multline}
	where the bond points from site $\mathbf{x}$ to site $\mathbf{x} + \mathbf{\hat{x}}$. This is a Gaussian state of the form of Eq.~\eqref{GaussianStateFirstOrder} parametrised by the anti-symmetric matrix $M = (i / 2) \sigma_2$. Hence, its complex CM $G$ can be computed from Eq.~\eqref{ComplexCMFirstOrder} and is given by 
	\begin{equation}
	G = - \frac{1}{2} \begin{pmatrix}
	0 & \sigma_2 \\
	-\sigma_2 & 0
	\end{pmatrix},
	\end{equation}
	where $\sigma_2 = \big(\begin{smallmatrix}
	0 & -i\\
	i & 0
	\end{smallmatrix}\big)$ denotes the second Pauli matrix. According to Eq.~\eqref{RealCM}, the corresponding Majorana CM $\Gamma$ is obtained by conjugation with the matrix $S$, and is found to be
	\begin{equation}\label{RealCMBond}
	\Gamma = \begin{pmatrix}
	0 & \sigma_1 \\
	-\sigma_1 & 0
	\end{pmatrix},
	\end{equation}
	where $\sigma_1 = \big(\begin{smallmatrix}
	0 & 1\\
	1 & 0
	\end{smallmatrix}\big)$ denotes the first Pauli matrix. For a two-dimensional TNS with $\xi = 1$, the real CM of the virtual state for the vertical nearest-neighbour bond $\langle \mathbf{x}, \mathbf{x} - \mathbf{\hat{y}} \rangle$ is also given by Eq.~\eqref{RealCMBond}. In this case, we choose the bond to be oriented downwards from $\mathbf{x}$ to $\mathbf{x} - \mathbf{\hat{y}} $ to comply with our ordering of the virtual legs as $L, U, R, D$. For TNS with more virtual fermions, $\xi > 1$, the virtual maximally entangled state from Eq.~\eqref{MaxEntState} is a tensor product of multiple states of the form of Eq.~\eqref{MaxEntStateSingleFermion}. Hence, the corresponding real CM is a direct sum of multiple copies of the $\Gamma$ from Eq.~\eqref{RealCMBond}. We denote by $\Gamma _{\omega}$ the Majorana CM of the total virtual maximally entangled state $\otimes _{\langle \mathbf{x'} \mathbf{x''} \rangle} \ket{\omega_{\mathbf{x'} \mathbf{x''}}}$. Since the total virtual maximally entangled state is a tensor product over all bonds, $\Gamma _{\omega}$ is a direct sum of multiple copies of the $\Gamma$ from Eq.~\eqref{RealCMBond}. 
	
	As explained in Appendix~\ref{sec:AppendixGfPEPSConstruction1}, the global physical state $\ket{\psi}$ is obtained from the constituents introduced above by projecting the total fiducial state onto the virtual maximally entangled state,
	\begin{equation}\label{GaussianTNS}
	\ket{\psi} = \left[ \bigotimes _{\langle \mathbf{x'} \mathbf{x''} \rangle}  \bra {\omega_{\mathbf{x'} \mathbf{x''}}}\right] \prod_ {\mathbf{x}} Q_{\mathbf{x}} \ket{\Omega}.
	\end{equation}
	We recall from Appendix~\ref{sec:AppGaussianStateSchur} that for Gaussian states, this projection can be formulated in terms of CMs and gives rise to a Schur complement. This is the approach we take in the following.

	Let us see how we can apply the general Schur complement expression of Eq.~\eqref{SchurComplementIntro} in order to compute the Majorana CM $\Gamma _{\ket{\psi}}$ for the physical state from the CMs $\Gamma _{Q}$ and $\Gamma _{\omega}$ defined above for the fiducial and virtual maximally entangled states. We introduce the symbols $p$ and $v$ to collectively refer to all Majorana mode operators for the physical and virtual fermions. Therefore, $\left(\Gamma _{Q}\right)_{pp}$ and $\left(\Gamma _{Q}\right)_{vv}$ denote the blocks of the CM of the total fiducial state that describe the reduced state of only the physical and virtual \glspl*{dof}, respectively. On the other hand, the blocks $\left(\Gamma _{Q}\right)_{pv} = - \left(\Gamma _{Q}\right)_{vp}^T$ encode the correlations between physical and virtual fermions. In Eq.~\eqref{GaussianTNS}, we are projecting the fiducial state of physical and virtual modes onto a maximally entangled state of the virtual modes in order to obtain a state of only the physical modes. According to Eq.~\eqref{SchurComplementIntro}, this is represented by the Schur complement of the virtual block $(vv)$ given by
	\begin{equation}\label{SchurComplementGeneral}
	\Gamma _{\ket{\psi}} = \left(\Gamma _{Q}\right)_{pp} + \left(\Gamma _{Q}\right)_{pv} \left[\left(\Gamma _{Q}\right)_{vv} + \Gamma _{\omega} \right] ^{-1} \left(\Gamma _{Q}\right)_{pv}^T.
	\end{equation}
	Note that the CM on the LHS of this equation tracks every physical mode independently. Hence, its dimension is proportional to the system size and becomes large for our cases of interest. We will now specialize Eq.~\eqref{SchurComplementGeneral} to translation invariant states, where we can achieve a massive reduction of the size of the physical CM down to the number of Bloch bands independent of the system size. 
	
	\section{Translation invariant GfTNSs\label{sec:AppendixTIGfTNS}}
	
	In this appendix, we specialize the formalism reviewed in Appendix~\ref{sec:AppendixGfPEPSConstruction} to translation invariant GfTNSs, where the contraction of the network can often be performed analytically. In Appendix~\ref{sec:AppendixGfPEPSCM}, we compute the CM of a translation invariant GfTNS, which we then use in Appendix~\ref{sec:AppendixGfPEPSPH} to construct a parent Hamiltonian for the state.
	
	\subsection{Covariance matrix\label{sec:AppendixGfPEPSCM}}
	
	For translation-invariant Gaussian TNSs, we introduce the Fourier transform (FT) of the physical and virtual mode operators as
	\begin{subequations}\label{FTGfTNS}
	\begin{gather}
	a_{\tau, \mathbf{k}} = \sum_{\mathbf{x}} \mathcal{F}_{\mathbf{k}, \mathbf{x}} a_{\tau, \mathbf{x}},\\
	b_{\alpha, j, \mathbf{k}} = \sum_{\mathbf{x}} \mathcal{F}_{\mathbf{k}, \mathbf{x}} b_{\alpha, j, \mathbf{x}}
	\end{gather}
	for all $\tau = 1, \dotsc, f$, $j = 1, \dotsc, \xi$, $\alpha = L,R$ for MPSs and $\alpha = L, U, R, D$ for two-dimensional PEPSs. Here, the FT in two spatial dimensions with $N_x$ and $N_y$ sites in the horizontal and vertical direction, respectively, position vector $\mathbf{x} = (x,y)$ and momentum vector $\mathbf{k} = (k_x, k_y)$, is given by  
	\begin{equation}
	\mathcal{F}_{\mathbf{k}, \mathbf{x}} = \frac{1}{\sqrt{N_x N_y}} e^{- i \mathbf{k} \mathbf{x}}.
	\end{equation}
	\end{subequations}
	The momenta in the horizontal ($k_x$) and vertical ($k_y$) direction take values $k_x = \frac{2\pi j}{N_x}$ with $0 \leq j \leq N_x -1$ and $k_y = \frac{2\pi j}{N_y}$ with $0 \leq j \leq N_y -1$. The FT for a single spatial direction is analogous. 
	
	We define the FT of the mode vector $\chi_{\mathbf{x}}$ from Eq.~\eqref{ModeVector} to be 
	\begin{equation}
	\chi_{\mathbf{k}} = \sum_{\mathbf{x}} \mathcal{F}_{\mathbf{k}, \mathbf{x}} \chi_{\mathbf{x}}.
	\end{equation}
	Therefore,
	\begin{multline}\label{ModeVectorFT}
	\chi_{\mathbf{k}} = \big(a_{1,\mathbf{k}}, a_{2,\mathbf{k}}, \dotsc, a_{f,\mathbf{k}}, b_{L,1,\mathbf{k}}, b_{L,2,\mathbf{k}}, \dotsc, b_{R,1,\mathbf{k}},\dotsc, \\
	a^{\dagger}_{1,\mathbf{-k}} , \dotsc, b^{\dagger}_{L,1,\mathbf{-k}}, \dotsc, b^{\dagger}_{R,1,\mathbf{-k}}, \dotsc \big)^T
	\end{multline}
	mixes mode operators at momenta $\mathbf{k}$ and  $-\mathbf{k}$ similarly to a Nambu spinor.
	
	Due to the translation invariance of the GfTNS, the FT brings the CMs of the physical state, the total fiducial state and the total virtual maximally entangled state into a block diagonal form. We denote the Majorana CM of the total fiducial state \wrt the Fourier transform of the mode operators by
	\begin{equation}
	\left(\Gamma _{Q}\right) _{\mu,\nu ; \mathbf{k}\mathbf{q}} = \sum_{\mathbf{x}, \mathbf{x'}} \mathcal{F}_{\mathbf{k}, \mathbf{x}}^* \left(\Gamma _{Q}\right) _{\mu,\nu ; \mathbf{x}\mathbf{x'}} \mathcal{F}_{\mathbf{q}, \mathbf{x'}} = \delta_{\mathbf{k}\mathbf{q}} \big(\tilde{\Gamma} _{Q}(\mathbf{k})\big)_{\mu,\nu},
	\end{equation}
	where the last equality defines the Majorana CM $\tilde{\Gamma} _{Q}(\mathbf{k})$ restricted to the block of momentum $\mathbf{k}$. Note that the size of $\tilde{\Gamma} _{Q}(\mathbf{k})$ is given by twice the number of Bloch bands, $2 \times n_{\mathrm{modes}}$, and therefore no longer grows with the system size. Analogous statements hold for the physical CM $\Gamma_{\ket{\psi}}$ with the Fourier block matrix $\MCMF _{\ket{\psi}} (\mathbf{k})$ of size $2f$, and for the CM of the total virtual maximally entangled state $\Gamma _{\omega}$ with Fourier block $\MCMF _{\omega}(\mathbf{k})$ of size $2(n_{\mathrm{modes}} - f)$.

	Since the TNS is translation invariant, the local fiducial state $\ket{Q_{\mathbf{x}}}$ is the same on every unit cell. Hence, the Fourier CM of the total fiducial state is localised at momentum $\mathbf{k} = 0$,
	\begin{equation}\label{CMMajoranaFourier}
	\tilde{\Gamma} _{Q} (\mathbf{k}) = \Gamma _{Q_\mathbf{x}} \times \delta_{\mathbf{k}, 0}. 
	\end{equation}
	Here, $\Gamma _{Q_\mathbf{x}}$ is the CM of the local fiducial state $\ket{Q_{\mathbf{x}}}$ on a single site of dimension $2n_{\mathrm{modes}}$. It has the block structure
	\begin{equation}\label{CMMajoranaBlock}
	\Gamma _{Q_\mathbf{x}} = \begin{pmatrix}
	\mathbf{A} & \mathbf{B} \\
	-\mathbf{B}^T & \mathbf{D}\\
	\end{pmatrix},
	\end{equation}
	where the real anti-symmetric blocks $\mathbf{A}$ and $\mathbf{D}$ of dimension $2f$ and $2(n_{\mathrm{modes}} -f)$ describe the physical and virtual subspaces, respectively, whereas the block $\mathbf{B}$ encodes the coupling between physical and virtual modes. 
	
	On the other hand, the CM of the total virtual maximally entangled state $\MCMF _{\omega}(\mathbf{k})$ has a non-trivial momentum dependence since the maximally entangled states from Eq.~\eqref{MaxEntState} connect different unit cells. It is a direct sum of the contributions from the different spatial directions. For a single virtual fermion with $\xi = 1$, the Majorana Fourier CM of the horizontal bonds oriented from left to right is the Fourier transform of the matrix $\Gamma$ from Eq.~\eqref{RealCMBond}. It reads~\cite{PhysRevLett.111.236805} 
	\begin{equation}\label{CMMajoranaFourierVirtual}
	\MCMF_{\omega} (k_x) = \begin{pmatrix}
	0 & -\sigma_1 e^{-i k_x}\\
	\sigma_1 e^{i k_x} & 0
	\end{pmatrix}.
	\end{equation} 
	This matrix is written in the basis of the FT of the Majorana operators constructed from the complex modes $(b_{L,\mathbf{x}}, b_{R,\mathbf{x}}, b_{L,\mathbf{x}}^{\dagger}, b_{R,\mathbf{x}}^{\dagger})$ where we have omitted the index $j$ since $\xi = 1$. The contribution from the vertical bonds is given by Eq.~\eqref{CMMajoranaFourierVirtual} with $k_x \mapsto - k_y$, since the vertical bonds are oriented downwards and hence in the direction of negative $k_y$. Therefore, the Fourier CM of the virtual bonds is anti-hermitian and satisfies the identities $\MCMF _{\omega}(\mathbf{k}) ^* =  \MCMF _{\omega}(-\mathbf{k})$ and $\MCMF _{\omega}(\mathbf{k}) ^T =  -\MCMF _{\omega}(-\mathbf{k})$.
	
	We are now in a position to compute the physical Majorana CM in momentum space by taking the FT of Eq.~\eqref{SchurComplementGeneral}. In momentum space, the RHS simplifies significantly due to the expression for the fiducial state CM given in Eqs.~\eqref{CMMajoranaFourier} and~\eqref{CMMajoranaBlock}. We thus find~\cite{Kraus}
	\begin{equation}\label{SchurComplementFourier}
	\MCMF _{\ket{\psi}} (\mathbf{k})= \mathbf{A} + \mathbf{B} \left[\mathbf{D} + \MCMF _{\omega}(\mathbf{k}) \right] ^{-1} \mathbf{B}^T.
	\end{equation}
	The physical state is well-defined if the matrix inversion can be carried out, \ie if the determinant
	\begin{equation}\label{DeterminantSchurComplement}
	q(\mathbf{k}) = \det \left[\mathbf{D} + \MCMF _{\omega}(\mathbf{k}) \right]
	\end{equation} 
	is not equal to zero. We emphasize that the matrices in Eq.~\eqref{SchurComplementFourier} have a constant size given by the number of Bloch bands, such that Eq.~\eqref{SchurComplementFourier} can typically be evaluated analytically. In Appendix~\ref{sec:AppendixSSHMPS}, we use Eq.~\eqref{SchurComplementFourier} to evaluate the Bloch CM of the SSH model MPS from Sec.~\ref{sec:SSHToChern}, and in Appendix~\ref{sec:AppendixQuadrupolePEPS}, we use Eq.~\eqref{SchurComplementFourier} to evaluate the Bloch CM of the quadrupole model PEPS of Sec.~\ref{sec:QuadrupolePEPS}.

	\subsection{Parent Hamiltonian\label{sec:AppendixGfPEPSPH}}
	
	A translation invariant Gaussian fermionic TNS $\ket{\psi}$ has an infinite number of parent Hamiltonians for which it is an exact ground state. For any non-negative scalar function $\epsilon (\mathbf{k}) \geq 0$ on the Brillouin zone,  
	\begin{equation}\label{GaussianPH}
	H_{\epsilon} = \frac{i}{4} \sum _{\mathbf{k}} \epsilon(\mathbf{k})\,  \left(\chi^{(p)}_{\mathbf{k}}\right)^{\dagger} \, \left[S^T \MCMF _{\ket{\psi}} (\mathbf{k}) S^* \right]\,  \chi^{(p)}_{\mathbf{k}}
	\end{equation}
	is a parent Hamiltonian for the Gaussian fermionic TNS $\ket{\psi}$~\cite{Kraus, PhysRevLett.111.236805}. Here, 
	\begin{equation}
	\chi^{(p)}_{\mathbf{k}} = \left( a_{1,\mathbf{k}}, \dotsc, a_{f,\mathbf{k}}, a^{\dagger}_{1,-\mathbf{k}}, \dotsc,  a^{\dagger}_{f,-\mathbf{k}} \right)^T
	\end{equation}
	is the physical part of the mode vector of Eq.~\eqref{ModeVectorFT}, and $S$ is the transformation matrix defined in Eq.~\eqref{BasisChangeMajoranaComplex} from the Majorana fermions to the complex fermions.
	
	The properties of the parent Hamiltonian $H_{\epsilon}$ depend on the dispersion function $\epsilon (\mathbf{k})$. If $\epsilon (\mathbf{k}) > 0$ is strictly positive throughout the Brillouin zone, $H_{\epsilon}$ is gapped. Moreover, if all matrix entries of the product $\epsilon (\mathbf{k}) \MCMF _{\ket{\psi}} (\mathbf{k})$ are polynomials in $e^{\pm i k_x}$ and $e^{\pm i k_y}$, the parent Hamiltonian is strictly local.
	
	A natural, \emph{but not unique}, choice for the dispersion function is given by $\epsilon (\mathbf{k}) = q(\mathbf{k})$ from Eq.~\eqref{DeterminantSchurComplement}. Indeed, since $D$ and $\MCMF _{\omega}(\mathbf{k})$ are anti-hermitian and of even dimension (see Eqs.~\eqref{CMMajoranaBlock} and~\eqref{CMMajoranaFourierVirtual}), it follows that $q(\mathbf{k})$ is real. If moreover $q(\mathbf{k})$ is strictly positive throughout the Brillouin zone, implying that the PEPS has exponentially decaying real-space correlations, the parent Hamiltonian $H_q$ is gapped and strictly local with all terms acting on at most $2\xi$ successive unit cells~\cite{PhysRevLett.111.236805, PhysRevB.90.115133}.
	
	\section{GfTNSs with conserved particle number\label{sec:AppendixGfTNSParticleNumber}}
	
	In the main text, we consider Gaussian fermionic TNSs with a conserved particle number: the ground states of both the SSH model and the quadrupole model lie at half filling. The TNSs are written in a basis related to the physical basis by a staggered particle-hole conjugation (\cf Eq.~\eqref{PHTrafo} for the SSH MPS and Eq.~\eqref{PHTrafoQuadrupole} for the quadrupole PEPS). Hence, the $\mathrm{U}(1)$ symmetry of the local tensors, which imposes the conservation of the physical particle number, also takes a staggered form (\cf Eq.~\eqref{LocalU1} for the SSH MPS and Eq.~\eqref{U1SymmetryPEPS} for the quadrupole PEPS). In this appendix, we rephrase this $\mathrm{U}(1)$ symmetry in the language of fiducial states, and show that it enforces many vanishing elements for the CM of the local fiducial state. These are equivalent to the vanishing of the off-diagonal block $Q = 0$ of the complex CM of a state with conserved particle number (see Appendix~\ref{sec:AppGaussianStateDef}), but expressed in the basis after the staggered particle-hole transformation. We focus on the one-dimensional case since the computation for two-dimensional GfTNSs is analogous. 
	
	For one-dimensional Gaussian fermionic MPS, we consider a $\mathrm{U}(1)$ symmetry of the local tensor of the general form
	\begin{multline}\label{U1LocalTensorGeneral}
	A[\mathbf{x}]^i_{lr} = \sum_{i' l' r'} \left(\bigotimes_{\tau = 0}^f
	U(\eta_{\tau, \mathbf{x}} \varphi)\right) _{ii'} \left(\bigotimes_{j = 0}^{\xi} 
	U(\eta_{L,j,\mathbf{x}} \varphi)\right) _{ll'} \\
	\left(\bigotimes_{j = 0}^{\xi} 
	U(\eta_{R,j, \mathbf{x}} \varphi)\right) _{rr'}	A[\mathbf{x}]^{i'}_{l'r'}
	\end{multline}
	where $\eta_{\tau, \mathbf{x}}, \eta_{L,j, \mathbf{x}}, \eta_{R,j, \mathbf{x}} \in \{\pm 1\}$ for $\tau = 1, \dotsc, f$ and $j = 1, \dotsc, \xi$. Here,
	\begin{equation}\label{U1MatrixRep}
	U(\varphi) = 	
	\begin{pmatrix}
	1 & 0\\
	0 & e^{i\varphi}
	\end{pmatrix}
	\end{equation}
	is the $\mathrm{U}(1)$ rotation acting on a single spinless fermion. Positive and negative values for $\eta$ indicate that the corresponding physical or virtual modes transform as particles and holes, respectively. The symmetries of Eq.~\eqref{LocalU1} for the local tensors on the $A$ and $B$ sublattice of the SSH charge pumping MPS are examples with $f = \xi = 1$.
	
	Since the elements of the local tensor $A[\mathbf{x}]$ are the basis coefficients of the local fiducial state $\ket{Q_{\mathbf{x}}}$ (\cf Eq.\eqref{RelationGaussianMapLocalTensor}), Eq.~\eqref{U1LocalTensorGeneral} is equivalent to the invariance of the local fiducial state under the $\mathrm{U}(1)$ symmetry
	\begin{multline}\label{U1LocalFidStateGeneral}
	\hat{U}_{\mathbf{x}} (\varphi) = \prod_{\tau = 0}^f
	\hat{U}_{\tau, \mathbf{x}} (\eta_{\tau, \mathbf{x}} \varphi) \times \\
	\prod_{j = 0}^{\xi} 
	\hat{U}_{L,j,\mathbf{x}}(\eta_{L,j,\mathbf{x}} \varphi)
	\hat{U}_{R,j,\mathbf{x}}(\eta_{R,j, \mathbf{x}} \varphi)
	\end{multline}
	whose many-body basis representation is given in Eq.~\eqref{U1LocalTensorGeneral}. Here, each individual operator $\hat{U}$ in the product acts on exactly one fermion. For instance, the operator acting on the physical fermion $\tau$ is given by
	\begin{equation}\label{U1SingleMode2ndQuantisation}
	\hat{U}_{\tau, \mathbf{x}} (\eta_{\tau, \mathbf{x}}\varphi) = e^{i\eta_{\tau, \mathbf{x}}\varphi}a^{\dagger}_{\tau,\mathbf{x}}a_{\tau,\mathbf{x}} + a_{\tau,\mathbf{x}}a^{\dagger}_{\tau,\mathbf{x}}
	\end{equation}
	and similarly for the virtual fermions. This agrees with the matrix representation of the $\mathrm{U}(1)$ rotation acting on a single mode given in Eq.~\eqref{U1MatrixRep}.
	
	We observe that the $\mathrm{U}(1)$ operator of the physical fermion $\tau$ from Eq.~\eqref{U1SingleMode2ndQuantisation} satisfies $a_{\tau,\mathbf{x}} \hat{U}_{\tau, \mathbf{x}} (\eta_{\tau, \mathbf{x}}\varphi) = e^{i\eta_{\tau, \mathbf{x}}\varphi} \hat{U}_{\tau, \mathbf{x}} (\eta_{\tau, \mathbf{x}}\varphi) a_{\tau,\mathbf{x}}$ and $a_{\tau,\mathbf{x}}^{\dagger} \hat{U}_{\tau, \mathbf{x}} (\eta_{\tau, \mathbf{x}}\varphi) = e^{-i\eta_{\tau, \mathbf{x}}\varphi} \hat{U}_{\tau, \mathbf{x}} (\eta_{\tau, \mathbf{x}}\varphi) a_{\tau,\mathbf{x}}^{\dagger}$. Extending this to all of the modes in the mode operator of Eq.~\eqref{ModeVector}, we find that
	\begin{equation}\label{CommutationU1Modes}
	\left(\chi _{\mathbf{x}}\right)_{\mu} \hat{U}_{\mathbf{x}} (\varphi)  = e^{i \varphi \left(\eta_{\mathbf{x}}\right)_{\mu}}  \hat{U}_{\mathbf{x}} (\varphi) \left(\chi _{\mathbf{x}}\right)_{\mu},
	\end{equation} 
	where $1 \leq \mu, \nu \leq 2n_{\mathrm{modes}}$. Here, we collected all the parameters $\eta$ into a vector of length $2n_{\mathrm{modes}}$ using the same ordering as in the mode vector of Eq.~\eqref{ModeVector},
	\begin{multline}\label{ModeListPH}
	\eta_{\mathbf{x}} = \big( \eta_{1, \mathbf{x}}, \eta_{2, \mathbf{x}},\dotsc, \eta_{f, \mathbf{x}}, \eta_{L,1, \mathbf{x}}, \eta_{L,2, \mathbf{x}},\dotsc,  \eta_{R,1, \mathbf{x}}, \dotsc\\
	-\eta_{1, \mathbf{x}}, \dotsc, - \eta_{L,1, \mathbf{x}}, \dotsc, - \eta_{R,1, \mathbf{x}}, \dotsc, \big),
	\end{multline}
	where the last $n_{\mathrm{modes}}$ entries describe the creation operators which transform with the opposite sign under the $\mathrm{U}(1)$ symmetry. 
	
	We are now ready to infer the consequences of the local fiducial state's invariance under the $\mathrm{U}(1)$ symmetry from Eq.~\eqref{U1LocalFidStateGeneral} for its complex CM. Indeed, due to the relation of Eq.~\eqref{CommutationU1Modes}, correlation functions transform under the symmetry as
	\begin{multline}
	\bra{Q_{\mathbf{x}}} \left( \chi_{\mathbf{x}}^{\dagger}\right)_{\nu} \left( \chi_{\mathbf{x}}\right)_{\mu} \ket{Q_{\mathbf{x}}} = \\
	\bra{Q_{\mathbf{x}}} \hat{U}^{\dagger}_{\mathbf{x}} (\varphi) \left( \chi_{\mathbf{x}}^{\dagger}\right)_{\nu} \left( \chi_{\mathbf{x}}\right)_{\mu} \hat{U}_{\mathbf{x}} (\varphi) \ket{Q_{\mathbf{x}}} = \\
	e^{i \varphi \left[ \left(\eta_{\mathbf{x}}\right)_{\mu} - \left(\eta_{\mathbf{x}}\right)_{\nu}\right]} \times \bra{Q_{\mathbf{x}}} \left( \chi_{\mathbf{x}}^{\dagger}\right)_{\nu} \left( \chi_{\mathbf{x}}\right)_{\mu} \ket{Q_{\mathbf{x}}}.
	\end{multline}
	Therefore, they vanish unless $\left(\eta_{\mathbf{x}}\right)_{\mu} = \left(\eta_{\mathbf{x}}\right)_{\nu}$. This shows that the symmetry of Eq.~\eqref{U1LocalTensorGeneral} forces the vanishing of half the elements of the complex CM of the local fiducial state,
	\begin{gather}\label{ComplexCMLocalU1}
	\left(G_{\mathbf{x}}\right)_{\mu \nu} = 0 \quad \text{if }\left(\eta_{\mathbf{x}}\right)_{\mu} \neq \left(\eta_{\mathbf{x}}\right)_{\nu}
	\end{gather}
 	where $1 \leq \mu, \nu \leq 2n_{\mathrm{modes}}$.  	
	\section{Covariance matrix for SSH MPS\label{sec:AppendixSSHMPS}}
In this appendix, we illustrate the formalism of GfTNS introduced in Appendix~\ref{sec:AppendixGfPEPSConstruction} and~\ref{sec:AppendixTIGfTNS} using the example of the MPS from Eq.~\eqref{MPSMyInterpol} describing charge pumping in the SSH model. After expressing the state as a GfTNS in Appendix~\ref{sec:AppendixSSHMPSGfTNS}, we demonstrate the computation of its Bloch CM and parent Hamiltonian in Appendix~\ref{sec:AppendixSSHMPSPH}. 

\subsection{Expression as GfTNS\label{sec:AppendixSSHMPSGfTNS}}

Our goal is to express the SSH charge pumping MPS, which is already fully defined by Eq.~\eqref{MPSMyInterpol} in the language of local tensors, as a GfTNS using the formalism from Appendix~\ref{sec:AppendixGfPEPSConstruction}. Since the MPS has physical dimension $2$ and bond dimension $2$, it is described by $f = 1$ physical fermion and $\xi = 1$ virtual fermion per nearest-neighbour bond and lattice site (\cf the first paragraph of Appendix~\ref{sec:AppendixGfPEPSConstruction2}). On the $A$ sublattice, the annihilation operators for the physical, left virtual and right virtual modes are $a_{A,x}$, $b_{L,A,x}$ and $b_{R,A,x}$, respectively, and similarly for the $B$ sublattice. Following the recipe given above, we need to find the local fiducial states $\ket{Q_{A, x}} = Q_{A,x} \ket{\Omega}$ and $\ket{Q_{B,x}} = Q_{B,x} \ket{\Omega}$ that match the local tensors $A$ and $B$ from Eq.~\eqref{MPSMyInterpol} at each unit cell $x$. The link between the local tensors and fiducial states is then given by Eq.~\eqref{RelationGaussianMapLocalTensor} stating that the local tensor entries are the basis coefficients of the fiducial state. For example, the tensor element $A^1_{01} = \beta$ tells us that the local fiducial map $Q_{A,x}$ contains a term $ \beta a^{\dagger}_{A,x} b^{\dagger}_{R,A,x}$. Performing this matching for every non-zero tensor entry, we find that the fiducial maps are given by
\begin{subequations}\label{FiducialMapsMPS}
	\begin{gather}
	Q_{A,x} = \gamma + \alpha a^{\dagger}_{A,x} b^{\dagger}_{L,A,x} + \beta a^{\dagger}_{A,x} b^{\dagger}_{R,A,x},\\
	Q_{B,x} = \gamma + \beta a^{\dagger}_{B,x} b^{\dagger}_{L,B,x} - \alpha a^{\dagger}_{B,x} b^{\dagger}_{R,B,x}	.
	\end{gather}
\end{subequations} 

In a second step, we want to write the local fiducial states as Gaussian states satisfying Wick's theorem in order to express the MPS as a GfTNS. Since the local tensors are parity even, the local fiducial states $\ket{Q_{A,x}}$ and $\ket{Q_{B,x}}$ can be parametrised as in Eq.~\eqref{ParametrisationQLocal} using the exponential of anti-symmetric coefficient matrices $M_{A,x}$ and $M_{B,x}$. In the case of the SSH pumping MPS, this is very simple, since the fiducial states we derived in Eq.~\eqref{FiducialMapsMPS} have the form of Eq.~\eqref{GaussianStateFirstOrder}, allowing us to directly read off $M_{A,x}$ and $M_{B,x}$ (imposing antisymmetry). We find that the coefficient matrices are given by
\begin{subequations}\label{CoefficientMatrixSSH}
\begin{gather}
M_{A,x} = \frac{1}{2}\begin{pmatrix}
0 & a & b \\
-a & 0 & 0\\
-b & 0 & 0
\end{pmatrix},\\
M_{B,x} = \frac{1}{2} \begin{pmatrix}
0 & b & -a \\
-b & 0 & 0\\
a & 0 & 0
\end{pmatrix}.
\end{gather}
\end{subequations}
Here, we defined the quotients $a = \alpha / \gamma $ and $b = \beta / \gamma $ of the parameters used in Eq.~\eqref{MPSMyInterpol}, and we absorbed the remaining factor $\gamma$ into the normalisation constant $\mathcal{N}$ in Eq.~\eqref{GaussianStateFirstOrder}. The complex CMs of these local fiducial states are computed from $M_{A,x}$ and $M_{B,x}$ using Eq.~\eqref{ComplexCMFirstOrder}, and are then transformed to the Majorana representation using Eq.~\eqref{RealCM}. One finds that the Majorana CM $\Gamma_{Q_{A,x}}$ is
\begin{equation}\label{MajoranaCMSSHSite}
\Gamma_{Q_{A,x}} = \frac{1}{c}
\begin{pmatrix}
\left( c-2\right) i \sigma_2 & 4a\sigma_1 & 4b\sigma_1\\
-4a\sigma_1 & \left( 8a^2 -c \right) i \sigma_2 & 8ab i \sigma_2\\
-4b\sigma_1 & 8ab i \sigma_2 & \left( 8b^2-c\right) i \sigma_2
\end{pmatrix},
\end{equation}
where we introduced the shorthand notation $c \equiv 1 + 4a^2 + 4b^2$, and the basis is the Majorana basis derived from $(a_{A,x}, b_{L,A,x}, b_{R,A,x}, a^{\dagger}_{A,x}, b^{\dagger}_{L,A,x}, b^{\dagger}_{R,A,x})$. The Majorana CM $\Gamma_{Q_{B,x}}$ in the Majorana basis derived from $(a_{B,x}, b_{L,B,x}, b_{R,B,x}, a^{\dagger}_{B,x}, b^{\dagger}_{L,B,x}, b^{\dagger}_{R,B,x})$ is given by Eq.~\eqref{MajoranaCMSSHSite} with the replacements $a \mapsto b$ and $b \mapsto -a$. We have thus successfully expressed the MPS from Eq.~\eqref{MPSMyInterpol} as a GfTNS.

The virtual maximally entangled states for the SSH MPS are of the form discussed in Appendix~\ref{sec:AppendixGfPEPSConstruction2}. In particular, the Majorana CM $\Gamma_{\ket{\omega_{A,x, B,x}}}$ of the state $\ket{\omega_{A,x, B,x}}$ within a unit cell is given by Eq.~\eqref{RealCMBond} in the Majorana basis obtained from $(b_{R,A, x}, b_{L, B,x}, b_{R, A,x}^{\dagger}, b_{L, B, x}^{\dagger})$. Similarly, the Majorana CM of the state $\ket{\omega_{B,x, A, x+1}}$ between unit cells is given by Eq.~\eqref{RealCMBond} in the Majorana basis obtained from $(b_{R,B, x}, b_{L, A,x+1}, b_{R, B,x}^{\dagger}, b_{L, A, x+1}^{\dagger})$.

\subsection{Bloch CM and parent Hamiltonian\label{sec:AppendixSSHMPSPH}}

Now that we have expressed the MPS from Eq.~\eqref{MPSMyInterpol} as a GfTNS, we want to use the power of the formalism introduced in Appendix~\ref{sec:AppendixTIGfTNS} to derive its CM and parent Hamiltonian on a chain with periodic boundary conditions. We will compute the Bloch CM by evaluating Eq.~\eqref{SchurComplementFourier}, and from there obtain the parent Hamiltonian via Eq.~\eqref{GaussianPH}.

\subsubsection{CM for a unit cell\label{sec:AppendixSSHMPSPHCMUnitCell}}

The expression for the Bloch CM from Eq.~\eqref{SchurComplementFourier} is valid for a translation-invariant GfTNS. In particular, the CM $\Gamma _{Q_x}$ with its blocks $\mathbf{A}$, $\mathbf{B}$ and $\mathbf{D}$ from Eq.~\eqref{CMMajoranaFourier} is the Majorana CM of the fiducial state of a \emph{unit cell}, not a single site. In order to proceed, we therefore need to derive the fiducial state of an entire unit cell by contracting the virtual bond within a unit cell. In the language of GfTNSs, this is described by projecting the fiducial states $Q_{A, x} Q_{B, x}  \ket{\Omega}$ of the two sites in one unit cell onto the virtual maximally entangled state connecting them,
\begin{equation}
\bra{\omega_{AB}} \left[Q_{A,x} Q_{B,x}  \ket{\Omega} \right].
\end{equation}
This projection is a special case of Eq.~\eqref{ProjectionGaussian}. Hence, the CM $\Gamma _{Q_x}$ of the resulting state is given, as in Eq.~\eqref{SchurComplementIntro}, by the Schur complement of the of the block of size $4 \times 4$ corresponding to the Majorana mode operators constructed from $(b_{R,A}, b_{L,B}, b^{\dagger}_{R,A}, b^{\dagger}_{L,B})$,
\begin{equation}\label{SchurComplementSSH}
\Gamma _{Q_x} = \mathbf{A'} + \mathbf{B'}\, \left[ \mathbf{D'} +\Gamma_{\ket{\omega_{A,x, B,x}}}\right]^{-1} \,\mathbf{B'}^T. 
\end{equation}
Here, $\mathbf{A'}$ and $\mathbf{D'}$ are the blocks on the diagonal of the direct sum $\Gamma_{Q_{A,x}} \oplus \Gamma_{Q_{B,x}}$ corresponding to the Majorana modes derived from $(a_{A,x}, a_{B,x}, b_{L,A,x}, b_{R,B,x}, a^{\dagger}_{A,x}, a^{\dagger}_{B,x}, b^{\dagger}_{L,A,x}, b^{\dagger}_{R,B,x})$ and $(b_{R,A,x}, b_{L,B,x}, b^{\dagger}_{R,A,x}, b^{\dagger}_{L,B,x})$, respectively. Correspondingly, $\mathbf{B'}$ is the off-diagonal block.

$\Gamma _{Q_x}$ has the block form	
\begin{equation}
\Gamma _{Q_x} = \begin{pmatrix}
\mathbf{A} & \mathbf{B} \\
-\mathbf{B}^T & \mathbf{D}\\
\end{pmatrix},
\end{equation} 
where $\mathbf{A}$ and $\mathbf{D}$ are real anti-symmetric blocks of size $4 \times 4$, and $\mathbf{B}$ is a real block of size $4 \times 4$. We find that the diagonal blocks $\mathbf{A}$ and $\mathbf{D}$ are of the form
\begin{equation}\label{DiagonalBlockD1}
Z^{(1)} (r,s) =  
\begin{pmatrix}
0 & -r & 0 & -s \\
r & 0 & -s & 0 \\
0 & s^* & 0 & -r \\
s^* & 0 & r & 0 \\
\end{pmatrix}
\end{equation}
where $r$ and $s$ are parameters. Specifically, the physical and virtual blocks are
\begin{subequations}\label{MajoranaCMUnitCellSSH}
\begin{gather}
\mathbf{A} = Z^{(1)} (r_{p}, s_{p}),\\
\mathbf{D} = Z^{(1)} (r_v, s_v)
\end{gather}
with
\begin{gather}
r_{p} = \frac{1 - a^4 - b^4}{1 + 2 a^2 + a^4 + b^4},\\
s_{p} = \frac{2b^2}{1 + 2 a^2 + a^4 + b^4},\\
r_{v} = \frac{1 - a^4 + b^4}{1 + 2 a^2 + a^4 + b^4},\\
s_{v} = \frac{2 a^2 b^2}{1 + 2 a^2 + a^4 + b^4}.
\end{gather}
\end{subequations}
Here, the denominator is a consequence of the matrix inverse in Eq.~\eqref{SchurComplementSSH}. In addition, the block containing the coupling between physical and virtual fermions is 
\begin{equation}
\mathbf{B} = a
\begin{pmatrix}
0 & r_{p}+1 & 0 & -s_{p} \\
r_{p}+1 & 0 & s_{p} & 0 \\
0 & - s_{p} & 0 & -r_{p}-1 \\
 s_{p} & 0 & -r_{p}-1 & 0 \\
\end{pmatrix}.
\end{equation}

\subsubsection{Bloch CM}
We can now directly compute the Fourier Majorana CM $\MCMF_{\ket{\psi}}(k_x)$ of the physical state defined by the MPS on a chain with $N_x$ sites and periodic boundary conditions. $\MCMF_{\ket{\psi}}(k_x)$ is given by the Schur complement in Eq.~\eqref{SchurComplementFourier}, where the Fourier Majorana CM $\MCMF_{\omega} (k_x)$ for the virtual bonds is given by Eq.~\eqref{CMMajoranaFourierVirtual}, and $\mathbf{A}$, $\mathbf{B}$ and $\mathbf{D}$ are given in the previous section. 

For the SSH pumping MPS, Eq.~\eqref{SchurComplementFourier} is an matrix equation of size $4\times 4$ since there are two physical Bloch bands. The matrix inverse can be evaluated analytically using the special parametrisation $Z^{(1)} (r,s)$ from Eq.~\eqref{DiagonalBlockD1}. Indeed, the Fourier Majorana CM for the virtual bonds from Eq.~\eqref{CMMajoranaFourierVirtual} can be written as $\MCMF_{\omega} (k_x) = Z^{(1)} (0, e^{ik_x})$. One easily checks that
\begin{subequations}
\begin{gather}
Z^{(1)} (r,s) + Z^{(1)} (r',s') = Z^{(1)} (r + r', s + s'), \\
\label{DetZ1} \det \left[Z^{(1)}(r,s)\right] = \left(r^2 + s s^*\right)^2, \\
\left(Z^{(1)}(r,s)\right)^{-1} = - \frac{Z^{(1)}(r,s)}{\sqrt{\det \left[Z^{(1)}(r,s)\right]}}  .
\end{gather}
\end{subequations}
Using these identities, the matrix inverse in Eq.~\eqref{SchurComplementFourier} can be performed by hand, 
\begin{multline}
\left( \mathbf{D} + \MCMF_{\omega} (k_x) \right)^{-1} = \left(Z^{(1)} (r_v, s_v + e^{ik_x})\right)^{-1} \\
=- \frac{Z^{(1)}(r_v, s_v + e^{ik_x})}{\sqrt{r_v^2 + \left(s_v + e^{ik_x}\right)^2}},
\end{multline}
 and we evaluate the determinant $q(k_x)$ from Eq.~\eqref{DeterminantSchurComplement} as
\begin{multline}\label{DeterminantSchurSSHMPS}
q(k_x) = \left[r_v^2 + \left(s_v + e^{ik_x}\right)^2\right]^2 \\
= \frac{4 \left(1 + a^4+b^4 +2 a^2 b^2 \cos k_x\right)^2}{\left(\left(a^2+1\right)^2+b^4\right)^2}.
\end{multline}
Unless $\gamma = 0$ and $|\alpha| = |\beta|$, $q(k_x)$ is strictly positive such that the MPS is well-defined everywhere except for these parameter values.

After the matrix inversion in Eq.~\eqref{SchurComplementFourier}, the remaining matrix multiplications in Eq.~\eqref{SchurComplementFourier} can be performed using a computer algebra system. We thus find that the Fourier Majorana CM $\MCMF_{\ket{\psi}}(k_x)$ of the physical state is again of the form of Eq.~\eqref{DiagonalBlockD1},
\begin{subequations}\label{FourierMaforanaCMSSH}
\begin{equation}
\MCMF_{\ket{\psi}}(k_x) = Z^{(1)} \left(r(k_x),s(k_x)\right)
\end{equation}
with parameters
\begin{gather}\label{CMSSHMPS}
r(k_x) = \frac{1 - a^4- b^4 - 2 a^2 b^2 \cos k_x}{1 + a^4+b^4+2 a^2 b^2 \cos k_x},\\
s(k_x) = \frac{2 \left(b^2+a^2 e^{-i k_x}\right)}{1 + a^4+b^4+2 a^2 b^2 \cos k_x}.
\end{gather}
\end{subequations}

\subsubsection{Parent Hamiltonian}

Using Eq.~\eqref{GaussianPH}, we can now find a Bloch parent Hamiltonian $H_{\epsilon}$ for the MPS from its Majorana Fourier CM which we computed in Eq.~\eqref{FourierMaforanaCMSSH}. In order to gain a physical understanding of the parent Hamiltonian, we express it in terms of the original complex physical modes before the particle-hole transformation of Eq.~\eqref{PHTrafo} given by
\begin{subequations}
\begin{gather}
\pmode_{A, \mathbf{k}} = \pmodenew_{A, \mathbf{k}}, \\
\pmode_{B, \mathbf{k}} = \pmodenew^{\dagger}_{B, - \mathbf{k}}.
\end{gather}
\end{subequations}
after the FT of Eq.~\eqref{FTGfTNS}. We find
\begin{equation}
H_{\epsilon} = \sum_{k_x} \epsilon(k_x) \begin{pmatrix}
\pmode_{A, k_x} \\
\pmode_{B, k_x}
\end{pmatrix}^{\dagger}
\begin{pmatrix}
r(k_x) & s(k_x) \\
s(k_x)^* & -r(k_x)
\end{pmatrix}
\begin{pmatrix}
\pmode_{A, k_x} \\
\pmode_{B, k_x}
\end{pmatrix},
\end{equation}
where the functions $s(k_x)$ and $r(k_x)$ are defined in Eq.~\eqref{FourierMaforanaCMSSH}.

In order for $H_{\epsilon}$ to be gapped and strictly local, we need to find a dispersion function $\epsilon (k_x)$ which is strictly positive such that $\epsilon (k_x)s(k_x)$ and $\epsilon (k_x)r(k_x)$ are polynomials in $e^{\pm i k_x}$. As explained in Appendix~\ref{sec:AppendixGfPEPSPH}, a natural choice is $\epsilon (k_x) = q (k_x)$. Indeed, $q (k_x)$ computed in Eq.~\eqref{DeterminantSchurSSHMPS} cancels the denominator of Eq.~\eqref{FourierMaforanaCMSSH}. Since $q (k_x)s(k_x)$ and $q (k_x)r(k_x)$ contain the factor $e^{\pm i k_x}$ up to second order, the parent Hamiltonian $H_{q}$ contains hopping terms between up to second-nearest neighbour unit cells.  

Due to the special structure of Eq.~\eqref{MajoranaCMUnitCellSSH}, we can in fact obtain a more short-ranged parent Hamiltonian from the choice 
\begin{equation}
\epsilon (k_x) = \frac{1 + a^4+b^4 + 2 a^2 b^2 \cos k_x}{a^4 + b^4 + 1}
\end{equation}
proportional to $\sqrt{q(k_x)}$, which is in turn proportional to the denominator of $r$ and $s$ from Eq.~\eqref{CMSSHMPS}. $\epsilon (k_x)$ is strictly positive for all parameter values that lead to a well-defined state. The factor $1/(a^4 + b^4 + 1)$ is a normalisation ensuring that the parent Hamiltonian matches Eq.~\eqref{DimerizedInterpolationHamiltonian} if the MPS is given by the parametrisation $\parampump$ of Eq.~\eqref{ParametrisationDimerized}. We see that the parent Hamiltonian $H_{\epsilon}$ with Bloch  representation 
\begin{multline}\label{CMSHHMPS}
H_{\epsilon}(k_x) = \frac{1}{a^4 + b^4 + 1}
\Bigg[ \left(1 - a^4-b^4 - 2 a^2 b^2 \cos k_x\right) \sigma_3 \\
+ 2 \left(b^2 + a^2 \cos k_x\right) \sigma_1 + 2 a^2 \sin k_x \sigma_2 \Bigg]
\end{multline}	 
\wrt the original complex physical modes has hopping only up to nearest-neighbour unit cells. 	
	\section{Column covariance matrix of real-space $(d+1)$-dimensional TNS from charge pumping of $d$-dimensional TNS with conserved particle number\label{sec:AppendixColumnCM}}

In Sec.~\ref{sec:PEPSFromMPS}, we introduced the tensors $\acol$ and $\bcol$ describing the real-space Chern PEPS restricted to a column of sites on the $A$ and $B$ sublattice at positions $\{(x,y)\}_{0 \leq y \leq N_y -1}$. $\acol$ and $\bcol$ are defined by the application of the inverse FT of Eq.~\eqref{InverseFTPHComplete} to the physical and horizontal virtual legs of a column of the hybrid Chern PEPS at positions $\{(x,k_y^{(j)})\}_{0 \leq j \leq N_y -1}$. In this appendix, using the representation of the SSH pumping MPS as a GfTNS from Appendix~\ref{sec:AppendixSSHMPS}, we compute $\acol$ and $\bcol$ explicitly in terms of their CMs. In order to demonstrate the generality of the result, we will consider a general $(d+1)$-dimensional TNS constructed from charge pumping of a $d$-dimensional TNS, which is assumed to have a conserved number of physical particles, and hence possess a $\mathrm{U}(1)$ symmetry of the form discussed in Appendix~\ref{sec:AppendixGfTNSParticleNumber}.

Thus, let $\ket{\psi_d(t)}$ be the Gaussian fermionic TNS in $d$ spatial dimensions along a cyclic interpolation parametrised by the time $t \in (- \pi, \pi]$. With the same basis as in Eq.~\eqref{ModeVector}, we collect the physical and virtual mode operators for one unit cell $\mathbf{x} \in \mathbb{Z}^d$ of $\ket{\psi_d(t)}$ into the mode vector
\begin{multline}\label{ModeVectorTime}
\chi_{\mathbf{x}} (t) = 
\Big(a_{1,\mathbf{x}} (t), a_{2,\mathbf{x}} (t),\dotsc, a_{f,\mathbf{x}} (t), \\
b_{L,1,\mathbf{x}} (t), b_{L,2,\mathbf{x}} (t), \dotsc, b_{R,1,\mathbf{x}} (t), \dotsc \\ 
a^{\dagger}_{1,\mathbf{x}} (t) , \dotsc, b^{\dagger}_{L,1,\mathbf{x}} (t), \dotsc,b^{\dagger}_{R,1,\mathbf{x}} (t), \dotsc \Big)^T
\end{multline}
of length $2n_{\mathrm{modes}}$. The physical and virtual mode operators now depend on the time $t$ along the interpolation. As explained in Appendix~\ref{sec:AppendixGfPEPSConstruction}, the $d$-dimensional TNS is defined by its Gaussian local fiducial state $Q_{\mathbf{x}} (t)\ket{\Omega}$, which is characterised by its complex CM $G_{\mathbf{x}} (t)$ of dimension $2n_{\mathrm{modes}}$.

We assume that the TNS has a conserved number of particles, such that the local fiducial state $Q_{\mathbf{x}} (t)\ket{\Omega}$ of $\ket{\psi_d(t)}$ has a $\mathrm{U}(1)$ symmetry of the form discussed in Appendix~\ref{sec:AppendixGfTNSParticleNumber}. This symmetry determines which physical and virtual modes correspond to holes and particles: for each $1 \leq \mu \leq n_{\mathrm{modes}}$, an entry $(\eta_{\mathbf{x}})_{\mu} = 1$ or $(\eta_{\mathbf{x}})_{\mu} = -1$ in the vector $\eta_{\mathbf{x}}$ from Eq.~\eqref{ModeListPH} indicates that the mode $\mu$ has a particle- or hole-like character, respectively. Note that $\eta_{\mathbf{x}}$ does not depend on the time $t$, such that the hole- or particle-like character of the modes remains unchanged along the interpolation.

We can now move to the hybrid $(d+1)$-dimensional TNS, which is defined by Eq.~\eqref{GSFromPumpingGeneral} of the main text. From Sec.~\ref{sec:HybridTNS} we recall that the local fiducial state (equivalent to the local tensor) of the hybrid state at the position $(\mathbf{x}, k_{d+1}^{(j)})$ is given by $Q_{\mathbf{x}} (t^{(j)})\ket{\Omega}$ containing the modes $\chi_{\mathbf{x}} (t^{(j)})$ from Eq.~\eqref{ModeVectorTime}. In particular, due to the tensor product in the $(d+ 1)^{\mathrm{st}}$ direction, virtual fermions in this direction are not needed.

We can now easily write the CM of one column of the hybrid $(d+1)$-dimensional TNS, given by the sites at positions $\{(\mathbf{x}, k_{d+1}^{(j)})\}_{0 \leq j \leq N_{d+1}-1}$. Indeed, due to the absence of virtual legs in the direction $d+1$, the contraction of the bonds in this direction of the hybrid column amounts to a trivial tensor product in the language of local tensors. In terms of fiducial states, this corresponds to a direct sum of CMs. Hence, the complex CM $G_{\mathbf{x}}^{\mathrm{hybrid}}$ of the column $\{(\mathbf{x}, k_{d+1}^{(j)})\}_{0 \leq j \leq N_{d+1}-1}$ of the hybrid state is block diagonal, 
\begin{equation}\label{HybridCMDirectSum}
\left(G_{\mathbf{x}}^{\mathrm{hybrid}}\right)_{\mu,\mu'; t^{(j)}, t^{(j')}} = \delta_{j , j'} \left(G_{\mathbf{x}} (t^{(j)})\right)_{\mu, \mu'}.
\end{equation}
Here, $G_{\mathbf{x}}^{\mathrm{hybrid}}$ is written in the basis $\{(\chi_{\mathbf{x}})_{\mu} (t^{(j)})\}_{1 \leq \mu \leq 2n_{\mathrm{modes}}, 0 \leq j \leq N_{d+1}-1}$ of all operators for the physical and virtual modes in the first $d$ directions in the column.

We now consider the $(d+1)$-dimensional real-space state restricted to a column $\{(\mathbf{x}, x_{d+1})\}_{0 \leq x_{d+1} \leq N_{d+1}-1}$, which is obtained by applying the inverse FT $\tilde{\mathcal{F}}$ in direction $d+1$ to the physical and virtual legs of the hybrid column. The complex CM $G_{\mathbf{x}}^{\mathrm{col}}$ of this at position $\mathbf{x}$ is therefore given by
\begin{equation}\label{IFTGHybrid}
G_{\mathbf{x}}^{\mathrm{col}} = \tilde{\mathcal{F}}^* G_{\mathbf{x}}^{\mathrm{hybrid}} \tilde{\mathcal{F}}^T.
\end{equation}
The inverse FT is the $d$-dimensional generalization of Eq.~\eqref{InverseFTPHComplete},
\begin{equation}\label{IFTGeneral}
\tilde{\mathcal{F}} _{\mu, \mu'; x_{d+1},t^{(j)}} = \frac{\delta _{\mu, \mu'}}{\sqrt{N_{d+1}}} e^{i \left(\eta_{\mathbf{x}}\right)_{\mu} x_{d+1} t^{(j)}}
\end{equation}
with $0 \leq j, x_{d+1} \leq N_{d+1} -1$.

From Eqs.~\eqref{HybridCMDirectSum}, \eqref{IFTGHybrid}, \eqref{IFTGeneral}, $G_{\mathbf{x}}^{\mathrm{col}}$ becomes 
\begin{multline}\label{GColGeneral}
\left(G_{\mathbf{x}}^{\mathrm{col}}\right)_{\mu, \mu'; x_{d+1}, x'_{d+1}} =  
\frac{1}{N_{d+1}} \\
\sum_{j = 0}^{N_{d+1} -1} 
e^{-i \left[\left(\eta_{\mathbf{x}}\right)_{\mu} x_{d+1} - \left(\eta_{\mathbf{x}}\right)_{\mu'} x'_{d+1}\right] t^{(j)}}
\left(G_{\mathbf{x}} (t^{(j)})\right)_{\mu, \mu'}.
\end{multline}
This expression can be further simplified due to the constraint of Eq.~\eqref{ComplexCMLocalU1} imposed on $G_{\mathbf{x}} (t^{(j)})$ by its $\mathrm{U}(1)$ symmetry, implying $G_{\mathbf{x}} (t^{(j)})_{\mu, \mu'} = 0$ unless $(\eta_{\mathbf{x}})_{\mu} = (\eta_{\mathbf{x}})_{\mu'}$. Indeed, we may thus define the matrix 
\begin{equation}
\tilde{G}_{\mathbf{x}} \left(t^{(j)}\right)_{\mu,\mu'} \equiv  G_{\mathbf{x}} \left((\eta_{\mathbf{x}})_{\mu} t^{(j)}\right)_{\mu,\mu'}
\end{equation}
that mixes elements of the complex CM of the $d$-dimensional state at times $ t^{(j)}$ and $-t^{(j)}$ according to whether the modes $\mu$ and $\mu'$ transform as particles or holes, respectively. The $2n_{\mathrm{modes}}$-dimensional blocks of the column CM $G_{\mathbf{x}}^{\mathrm{col}}$ are then given by the FT of this matrix,
\begin{equation}
\left(G_{\mathbf{x}}^{\mathrm{col}}\right)_{x_{d+1}, x'_{d+1}} =  \sum_{j = 0}^{N_{d+1} -1}
\frac{e^{-i \left[x_{d+1} - x'_{d+1}\right] t^{(j)}}}{N_{d+1}} 
\tilde{G}_{\mathbf{x}} (t^{(j)}).
\end{equation}
This expression is explicitly invariant under real-space translations $x_{d+1} \mapsto x_{d+1} +1$ acting on both the physical modes and the virtual modes in the first $d$ directions. Hence, the inverse FT of Eq.~\eqref{IFTGeneral} guarantees the translation invariance of the column CM $G_{\mathbf{x}}^{\mathrm{col}}$ in the direction $d+1$.

Finally, we want to investigate which form the global $\mathrm{U}(1)$ symmetry related to particle number conservation takes for $G_{\mathbf{x}}^{\mathrm{col}}$. This result will be used in Appendix~\ref{sec:AppendixESUOne}. The $\mathrm{U}(1)$ is inherited from the invariance of each fiducial state $Q_{\mathbf{x}} (t)\ket{\Omega}$ under the $\mathrm{U}(1)$ symmetry of Eq.~\eqref{U1LocalFidStateGeneral}, which is independent of $t$. From the expression of the latter in second quantization (\cf Eq.~\eqref{U1SingleMode2ndQuantisation}), we see that the generator of the global $\mathrm{U}(1)$ symmetry of the column state is
\begin{multline}\label{ExpValueGlobalU1Column}
\Bigg\langle\sum_{x_{d+1} = 0}^{N_{d+1}-1} \Bigg[ \sum_{\tau = 0} ^f \eta_{\tau, \mathbf{x}} a^{\dagger}_{\tau,(\mathbf{x}, x_{d+1})}a_{\tau,(\mathbf{x}, x_{d+1})} + \\
\sum_{\alpha} \sum_{j = 0} ^{\xi} \eta_{\alpha, j, \mathbf{x}} b^{\dagger}_{\alpha, j,(\mathbf{x}, x_{d+1})}b_{\alpha, j,(\mathbf{x}, x_{d+1})}	\Bigg] \Bigg\rangle = 0,
\end{multline} 
whose expectation value vanishes. 	
	\section{Disentangled modes in single-particle ES at fixed particle number\label{sec:AppendixESUOne}}

In this appendix, we show that as stated in Eq.~\eqref{NumberModesESExact}, the maximal number of entangled modes compatible with the $\mathrm{U}(1)$ symmetry of the SSH model MPS in the ES of the Chern PEPS column state is given by $\max\{3L, N_y\}$, where L is the number of sites in the subsystem. We proceed by deriving a lower bound on the number of disentangled modes in the ES of a state with a conserved particle number (and therefore an upper bound on the number of entangled modes). We first consider a generic state in Sec.~\ref{sec:AppendixESInsulator}, before specialising to the fiducial state of a real-space column of the $(d+1)$-dimensional TNS in Sec.~\ref{sec:AppendixESAcol}.

\subsection{Insulator with filling fraction $q$\label{sec:AppendixESInsulator}}

We consider a non-interacting system of $N$ fermionic DOFs with creation and annihilation operators $a^{\dagger}_j$, $a_j$ for $j = 1, \dotsc, N$. We define a bi-partition of the system into the subsystem $\mathcal{A}$ and its complement $\bar{\mathcal{A}}$, where the DOFs of $\mathcal{A}$ are described by the first $N_{\mathcal{A}}$ modes $j = 1, \dotsc, N_{\mathcal{A}}$. Let $\ket{\psi}$ be a pure state of this system with a conserved particle number and filling fraction $q$, such that the total number of occupied modes in $\ket{\psi}$ is $qN$. We denote by $n^{\lambda = 1}_{\mathcal{A}}$ the number of entanglement levels with the value $\lambda = 1$ in the single-particle ES of $\ket{\psi}$ restricted to $\mathcal{A}$. We now want to show that this number is bounded below by the filling fraction as
\begin{equation}\label{BoundDecoupledLevelsGeneral}
n^{\lambda = 1}_{\mathcal{A}} \geq \max \{ qN - \left(N - N_{\mathcal{A}}\right), \,0\}.
\end{equation}

\textit{Proof:} Let $H = \sum_{i,j = 1}^N h_{ij} a^{\dagger}_i  a_j$ be a non-interacting flat-band Hamiltonian whose ground state is $\ket{\psi}$, where $h$ is a Hermitian matrix of dimension $N \times N$. The occupied modes in $\ket{\psi}$ are given by the $qN$ orthogonal eigenstates $u^{(k)}$ of $h$ with energy -1, \ie
\begin{equation}
\sum_{j = 1}^N h_{ij} u^{(k)}_j = - u^{(k)}_i,
\end{equation}
where $k = 1, \dotsc, qN$. 

\textit{W.r.t.} the bi-partition into $\mathcal{A}, \bar{\mathcal{A}}$, each basis state $u^{(k)}$ falls into exactly one of the following three categories (assuming that the eigenstates are ordered accordingly): 

\begin{enumerate}[label = {(\arabic*)}]
	\item For $k = 1, \dots, m_1$, the states satisfy $u^{(k)}_i = 0$ for $i = N_{\mathcal{A}} + 1, \dotsc, N$ such that the corresponding occupied modes are composed of DOFs of the subsystem $\mathcal{A}$. According to Theorem 1 of Ref~\cite{PhysRevB.84.195103}, each such state leads to an entanglement level $\lambda = 1$ in the single-particle ES of $\ket{\psi}$ restricted to $\mathcal{A}$. Hence, $m_1 \leq n^{\lambda = 1}_{\mathcal{A}}$.
	
	\item For $k = m_1 + 1, \dots, m_2$, $u^{(k)}_i = 0$ for $i = 1, \dotsc, N_{\mathcal{A}}$ such that the corresponding occupied modes are localised in $\bar{\mathcal{A}}$.
	
	\item For $k = m_2 + 1, \dots, qN$, $u^{(k)}_i \neq 0$ both for some $i \in \{1, \dotsc, N_{\mathcal{A}} \}$ and some $i \in \{N_{\mathcal{A}} + 1, \dotsc, N\}$ such that the corresponding occupied modes are localised neither in $\mathcal{A}$ nor in $\bar{\mathcal{A}}$.
\end{enumerate}
The numbers $m_1, m_2$ are assumed to be maximal in the sense that no linear combination of eigenstates of the category (3) lies either purely in $\mathcal{A}$ or purely in $\bar{\mathcal{A}}$.

In the next paragraph, we will show that the number $qN - m_1$ of states from the categories (2) and (3) is no larger than the number $N - N_{\mathcal{A}}$ of DOFs in $\bar{\mathcal{A}}$. This proves the claim, since the number of states from the different categories can then be estimated as
\begin{equation}
qN = m_1 + (qN - m_1) \leq n^{\lambda = 1}_{\mathcal{A}} + (N - N_{\mathcal{A}}),
\end{equation}
leading to Eq.~\eqref{BoundDecoupledLevelsGeneral}.

As a final step, we need to show that $qN - m_1\leq N - N_{\mathcal{A}}$. This follows from the linear independence of the $qN - m_1$ vectors $\{\tilde{u}^{(k)} \}_{k > m_1}$ of dimension $N - N_{\mathcal{A}}$, where $\tilde{u}^{(k)} _i = u^{(k)}_{N_{\mathcal{A}} + i}$ for $i = N_{\mathcal{A}} + 1, \dotsc, N$ is the restriction of the eigenvector to the DOFs of $\bar{\mathcal{A}}$. Indeed, let us assume that we have a vanishing linear combination
\begin{equation}\label{LinInd}
0 = \sum_{k = m_1 +1}^{qN} \mu_k \tilde{u}^{(k)}
\end{equation}
with coefficients $\mu_k$. The scalar product with $\tilde{u}^{(l)}$ shows that $0 = \mu_l$ for $l = m_1 +1, \dotsc, m_2$: Indeed, the orthogonality of the $u^{(k)}$ implies $\sum_{i > N_{\mathcal{A}}} \tilde{u}^{(l)}_i\tilde{u}^{(k)}_i = \sum_{i \geq  1} u^{(l)}_i u^{(k)}_i =\delta_{kl}$. Eq.~\eqref{LinInd} therefore implies $0 = \sum_{k > m_2 }^{qN} \mu_k \tilde{u}^{(k)}$ where the sum runs only over category (3). Due to the maximality of $m_1, m_2$ as defined above just below point~(3), we then have $0 = \sum_{k > m_2 }^{qN} \mu_k u^{(k)}$. Otherwise, this would be a linear combination purely in $\mathcal{A}$ since the part in $\bar{\mathcal{A}}$ vanishes by assumption. We thus also get $0 = \mu_l$ for $l = m_2 +1, \dotsc, qN$, proving that $qN - m_1\leq N - N_{\mathcal{A}}$.

\subsection{Real-space column of $(d+1)$-dimensional \gls*{tns}\label{sec:AppendixESAcol}}

We now apply Eq.~\eqref{BoundDecoupledLevelsGeneral} to the case where $\ket{\psi}$ is the fiducial state of one real-space column of the $(d+1)$-dimensional pumping TNS, whose CM is computed in Eq.~\eqref{GColGeneral}. The column state has $N = n_{\text{modes}}N_{d+1}$ degrees of freedom, where $N_{d+1}$ is the number of sites in the direction $d+1$ and $n_{\text{modes}}$ the number of physical and virtual particles per lattice site of the $d$-dimensional TNS. Below we will show that in a suitable basis, the column state has a conserved particle number $qN = N_{d+1}n_{\eta_-}$, where $n_{\eta_-}$ is the number of values $-1$ in the first $n_{\text{modes}}$ entries of the vector $\eta_{\mathbf{x}}$ from Eq.~\eqref{ModeListPH}. We consider the single-particle ES of the column state \wrt the subsystem $\mathcal{A}_L$ of the first $L$ sites $0 \leq x_{d+1} \leq L-1$ of the column, which has $ N_{\mathcal{A}_L} = Ln_{\text{modes}}$ DOFs. By Eq.~\eqref{BoundDecoupledLevelsGeneral}, the number $n_{\mathcal{A}_L}^{\lambda = 1}$ of entanglement levels with the value $\lambda = 1$ in this spectrum is lower bounded as
\begin{equation}\label{LowerBoundESColumnGeneral}
n_{\mathcal{A}_L}^{\lambda = 1} \geq \max \{N_{d+1}n_{\eta_-} - (N_{d+1} - L)n_{\text{modes}}, \,0\}.
\end{equation}  

We now show that with a suitable particle-hole transformation we can find a single-particle basis \wrt which the column state has a conserved particle number $N_{d+1}n_{\eta_-}$. In Eq.~\eqref{GColGeneral} the state is expressed in a basis where the CM $G_{\mathbf{x}}^{\mathrm{col}}$ has a non-vanishing off-diagonal block $Q$ corresponding to superconducting correlations, such that only the parity of the particle number is conserved (\cf Eq.~\eqref{ComplexCM}). Below Eq.~\eqref{GColGeneral} we note that half the entries of $G_{\mathbf{x}}^{\mathrm{col}}$ vanish, namely those with $(\eta_{\mathbf{x}})_{\mu} \neq (\eta_{\mathbf{x}})_{\mu'}$. A new basis, in which the complex CM has only a diagonal block $R$ and hence a fixed particle number, is created as follows: For those modes $1 \leq m \leq n_{\text{modes}}$ with $(\eta_{\mathbf{x}})_{m} = -1$, we exchange annihilation and creation operators, since the latter have $(\eta_{\mathbf{x}})_{m + n_{\text{modes}}} = 1$. This is a particle-hole transformation corresponding to the mapping
\begin{equation}\label{PHColumnState}
a_{\tau,(\mathbf{x}, x_{d+1})} \mapsto a^{\dagger}_{\tau,(\mathbf{x}, x_{d+1})}
\end{equation}
for all physical modes with $\eta_{\tau, \mathbf{x}} = -1$ while leaving modes with $\eta_{\tau, \mathbf{x}} = 1$ unchanged, and similarly for the virtual modes. 

The number of particles in the column state is given by the expectation value from Eq.~\eqref{ExpValueGlobalU1Column} of the generator of its global $\mathrm{U}(1)$ symmetry. Note that this is not the physical particle number, but rather the number of particles in the system composed of the physical legs and the virtual legs in the first $d$ spatial dimensions. Under the transformation of Eq.~\eqref{PHColumnState}, the particle number expectation value from Eq.~\eqref{ExpValueGlobalU1Column} transforms as explained in Eq.~\eqref{GeneratorU1MPS}. We thus find that in the new basis of modes the particle number is $N_{d+1}n_{\eta_-}$ as claimed above.

\subsubsection{Chern PEPS}

Let us apply these results to the Chern \gls*{peps} derived from the \gls*{ssh} pumping \gls*{mps}. For simplicity, we restrict ourselves to a column $\acol$ of $A$ sites with $n_{\text{modes}} = 3$ DOFs per site and $\eta_{A} = 1$ and $\eta_{L, A} = \eta_{R,A} = -1$ (\cf the discussion below Eq.~\eqref{GeneratorU1MPS}). Therefore, $n_{\eta_-} = 2$ and Eq.~\eqref{LowerBoundESColumnGeneral} becomes
\begin{equation}\label{BoundChernPEPS}
n_{\mathcal{A}_L}^{\lambda = 1} \geq \max \{3L - N_y, \,0\}.
\end{equation}
This shows that Eq.~\eqref{NumberModesESExact} gives the maximal number of entangled modes compatible with the $\mathrm{U}(1)$ symmetry of the SSH model MPS.

For the PEPS defined by the trivial cycle $\paramtriv$ from Eq.~\eqref{MPSInterpolationTrivial}, the discussion above can be refined: since $\beta = 0$ throughout the interpolation, all right virtual modes decouple from the column tensor such that there trivially are $N_y$ entanglement levels $|\lambda| = 1$. Let us investigate if the $\mathrm{U}(1)$ symmetry causes additional decoupled levels in the system of the coupled physical and left virtual particles. This system has only two DOFs per site, namely the physical leg with $\eta_{A} = 1$ and the left virtual leg with $\eta_{L, A} = -1$. In this case, Eq.~\eqref{LowerBoundESColumnGeneral} gives a trivial lower bound for the number of decoupled modes with levels $\lambda = 1$, 
\begin{equation}\label{BoundTirivalPEPS}
n_{\mathcal{A}_L}^{\lambda = 1} \geq \max \{2L - N_y, \,0\} = 0 \quad \text{for } L \leq N_y/2.
\end{equation}
Hence, the $\mathrm{U}(1)$ symmetry does not cause any additional decoupled modes, and the number of entangled modes is given by $2L$ as discussed in the main text.

\subsubsection{Chiral hinge PEPS\label{sec:AppendixESChiralHingeState}}

The discussion for the three-dimensional chiral hinge PEPS from Sec.~\ref{sec:3DPEPS} is analogous, where a column of sites on the sublattice 1 has $n_{\text{modes}} = 5$ DOFs per site with $\eta_{1} = 1$ and $\eta_{L, 1} = \eta_{U, 1} = \eta_{R,1} = \eta_{D, 1} = -1$. Therefore, $n_{\eta_-} = 4$ and Eq.~\eqref{LowerBoundESColumnGeneral} implies $n_{\mathcal{A}_L}^{\lambda = 1} \geq \max \{5L - N_z, \,0\}$.

For the mirror-symmetric case with couplings $\alpha = \alpha _x = \alpha _y$ and $\beta = \beta_x = \beta _y$, this bound can be refined. Indeed, by defining the linear combinations $b_{\pm, 1}^{LD} = (b_{L, 1} \pm b_{D, 1}) / \sqrt{2}$ and $b_{\pm, 1}^{UR} = (b_{R, 1} \pm b_{R, 1}) / \sqrt{2}$, the local fiducial state $\ket{Q^{[1]}}$ from Eq.~\eqref{FiducialStatesQuadrupolePEPS} below can be written as
\begin{equation}
\ket{Q^{[1]}} = \Big[\gamma - \sqrt{2} \beta  a_{1} ^{\dagger} \left(b_{-, 1}^{LD}\right)^{\dagger} - \sqrt{2} \alpha a_{1}^{\dagger}\left(b_{-, 1}^{UR}\right)^{\dagger} \Big] \ket{\Omega}.
\end{equation}
Therefore, when considering only one column of sites on the sublattice 1, two virtual fermionic modes decouple from the local tensor on each site. Effectively, the remaining coupled system therefore has only $n_{\text{modes}} = 3$ DOFs per site with $\eta_{1} = 1$ and $\eta_{-, 1}^{LD} = \eta_{-, 1}^{UR} = -1$, such that the number of additional disentangled modes due to the $\mathrm{U}(1)$ symmetry can be estimated as above for the Chern PEPS. Hence, with the identification $N_y \mapsto N_z$, the bound on the number of disentangled modes is given by Eqs.~\eqref{BoundChernPEPS} and~\eqref{BoundTirivalPEPS} for the \glspl*{peps} derived from $\parampump$ and $\paramtriv$, respectively.
 	
	\section{Quadrupole PEPS as GfTNS\label{sec:AppendixQuadrupolePEPS}}

In this appendix we apply the formalism of GfTNSs to the quadrupole model pumping PEPS of Eq.~\eqref{QuadrupolePEPS}. In Appendix~\ref{sec:AppendixQuadrupolePEPS_FidStates}, we show how the state can be expressed as a GfTNS. This allows us to compute its Bloch CM and a Bloch parent Hamiltonian on the torus in Appendix~\ref{sec:AppendixQuadrupolePEPS_CMPH}. These sections are completely analogous to Appendix~\ref{sec:AppendixSSHMPS} for the SSH pumping MPS. We therefore refer the reader to this appendix for a detailed explanation of each step. Finally, in Appendix~\ref{sec:AppendixQuadrupolePEPS_ES} we discuss the ES of the PEPS when the parameters are chosen such that the state represents the OAI dimerized phase of the quadrupole model.

\subsection{Expression as GfTNS\label{sec:AppendixQuadrupolePEPS_FidStates}}

The quadrupole model pumping PEPS is defined in Eq.~\eqref{QuadrupolePEPS} of the main text in the language of local tensors. Here, we want to re-express this state in the formalism of GfTNSs. We recall that each unit cell consists of $2 \times 2$ lattice sites, and that the PEPS has physical dimension $2$ and bond dimension $2$. This corresponds to $f = 1$ physical fermion per lattice site and $\xi = 1$ virtual fermion per nearest-neighbour bond and lattice site, represented in Fig.~\ref{fig:QuadrupolePEPS}(b) by blue and red circles, respectively.

As explained in Appendix~\ref{sec:AppendixGfPEPSConstruction}, the local PEPS tensors $A^{[\tau]}$ on the four sublattices $\tau = 1,2,3,4$ from the main text correspond to local fiducial states $\ket{Q^{[\tau]}}$, whose basis coefficients are given by the local tensors (see Eq.~\eqref{RelationGaussianMapLocalTensor}). We write $a_{\tau}$ for the annihilation operator of the physical fermion and $b_{L, \tau}$, $b_{U, \tau}$, $b_{R, \tau}$, $b_{D, \tau}$ for the annihilation operators of the left, up, right, down virtual fermions on the sublattice $\tau$ (we dropped the unit cell index due to the translation invariance). Let $\ket{\Omega}$ denote the vacuum annihilated by all these operators. Applying Eq.~\eqref{RelationGaussianMapLocalTensor}, we see that the local fiducial states on the four sublattices derived from the local tensors of Eq.~\eqref{QuadrupolePEPS} are
\begin{subequations}\label{FiducialStatesQuadrupolePEPS}
\begin{gather}
\notag \ket{Q^{[1]}} = \Big[\gamma - \beta_x  a_{1} ^{\dagger}b_{L, 1}^{\dagger} - \alpha_y a_{1}^{\dagger}b_{U, 1}^{\dagger} + \alpha_x a_{1}^{\dagger}b_{R, 1}^{\dagger} \\
+\beta_y a_{1}^{\dagger}b_{D, 1}^{\dagger} \Big] \ket{\Omega},\\
\notag \ket{Q^{[2]}} = \Big[\gamma - \alpha_x  a_{2} ^{\dagger}b_{L, 2}^{\dagger} + \beta_y a_{2}^{\dagger}b_{U, 2}^{\dagger} + \beta_x a_{2}^{\dagger}b_{R, 2}^{\dagger} \\
-\alpha_y a_{2}^{\dagger}b_{D, 2}^{\dagger} \Big] \ket{\Omega},\\
\notag \ket{Q^{[3]}} = \Big[\gamma + \alpha_x  a_{3} ^{\dagger}b_{L, 3}^{\dagger} + \alpha_y a_{3}^{\dagger}b_{U, 3}^{\dagger} + \beta_x a_{3}^{\dagger}b_{R, 3}^{\dagger} \\
+\beta_y a_{3}^{\dagger}b_{D, 3}^{\dagger} \Big] \ket{\Omega},\\
\notag \ket{Q^{[4]}} = \Big[\gamma + \beta_x  a_{4} ^{\dagger}b_{L, 4}^{\dagger} + \beta_y a_{4}^{\dagger}b_{U, 4}^{\dagger} + \alpha_x a_{4}^{\dagger}b_{R, 4}^{\dagger} \\
+\alpha_y a_{4}^{\dagger}b_{D, 4}^{\dagger} \Big] \ket{\Omega}.
\end{gather}
\end{subequations}

These fiducial states are of the form of Eq.~\eqref{GaussianStateFirstOrder} with only zero- and second-order terms in the creation operators. Hence, the fiducial states are Gaussian and can be parametrised as in Eqs.~\eqref{GaussianMapApp} and~\eqref{ParametrisationQLocal} with antisymmetric coefficient matrices
\begin{subequations}
	\label{QuadrupolePEPS_Parametrisation}
	\begin{gather}
	M_1 = \frac{1}{2 \times 2^{1/4}}\begin{pmatrix}
	0 & -b_x & -a_y & a_x & b_y \\
	b_x & 0 & 0 & 0 & 0\\
	a_y & 0 & 0 & 0 & 0\\
	- a_x & 0 & 0 & 0 & 0\\
	- b_y & 0 & 0 & 0 & 0
	\end{pmatrix},\\
	M_2 = \frac{1}{2 \times 2^{1/4}}\begin{pmatrix}
	0 & -a_x & b_y & b_x & - a_y \\
	a_x & 0 & 0 & 0 & 0\\
	- b_y & 0 & 0 & 0 & 0\\
	- b_x & 0 & 0 & 0 & 0\\
	a_y & 0 & 0 & 0 & 0
	\end{pmatrix},\\
	M_3 = \frac{1}{2 \times 2^{1/4}}\begin{pmatrix}
	0 & a_x & a_y & b_x & b_y \\
	-a_x & 0 & 0 & 0 & 0\\
	-a_y & 0 & 0 & 0 & 0\\
	- b_x & 0 & 0 & 0 & 0\\
	- b_y & 0 & 0 & 0 & 0
	\end{pmatrix},\\
	M_4 = \frac{1}{2 \times 2^{1/4}}\begin{pmatrix}
	0 & b_x & b_y & a_x & a_y \\
	-b_x & 0 & 0 & 0 & 0\\
	-b_y & 0 & 0 & 0 & 0\\
	- a_x & 0 & 0 & 0 & 0\\
	- a_y & 0 & 0 & 0 & 0
	\end{pmatrix}.
	\end{gather}
\end{subequations}
Here, we defined the quotients $a_x = \alpha_x /\gamma$, $a_y = \alpha_y /\gamma$, $b_x = \beta_x / \gamma$ and $b_y = \beta_y / \gamma$ of the parameters from Eq.~\eqref{QuadrupolePEPS}. We absorbed the remaining factor $\gamma$ into the normalisation constant $\mathcal{N}$ in Eq.~\eqref{GaussianStateFirstOrder}.

We have thus successfully written the PEPS from Eq.~\eqref{QuadrupolePEPS} as a GfTNS. Before proceeding, we choose the orientation of the virtual bonds as follows: All horizontal bonds are oriented from left to right and all vertical bonds are oriented from top to bottom.

\subsection{Bloch CM and parent Hamiltonian\label{sec:AppendixQuadrupolePEPS_CMPH}}

Having expressed the quadrupole model pumping PEPS as a GfTNS, we now want to use the formalism from Appendix~\ref{sec:AppendixTIGfTNS} to compute the Bloch CM and a Bloch parent Hamiltonian for the state on a torus. We will compute the Bloch CM by evaluating Eq.~\eqref{SchurComplementFourier}, and from there obtain the parent Hamiltonian via Eq.~\eqref{GaussianPH}.

\subsubsection{CM of unit cell\label{sec:AppendixQuadrupolePEPS_CMUnitCell}}

The Majorana CM $\Gamma_{Q_\mathbf{x}}$ in Eq.~\eqref{SchurComplementFourier} refers to the fiducial state of a unit cell, not that of a single site. We therefore need to compute $\Gamma_{Q_\mathbf{x}}$ from the fiducial states for each individual sublattice given in Eq.~\eqref{FiducialStatesQuadrupolePEPS}. This is done by contracting the four virtual bonds within one unit cell. In the language of fiducial states, we project the tensor product $\ket{Q^{[1]}} \otimes \ket{Q^{[2]}} \otimes \ket{Q^{[3]}} \otimes\ket{Q^{[4]}}$ of the fiducial states for a unit cell on the product of the maximally entangled states of the four virtual bonds connecting the lattice sites within this unit cell. This is analogous to the computation leading to Eq.~\eqref{FourierMaforanaCMSSH}, and we refer the reader to Appendix~\ref{sec:AppendixSSHMPSPHCMUnitCell} for details. Here we will only state the result for the Majorana CM $\Gamma_{Q_\mathbf{x}}$.

\begin{figure}
	\includegraphics[width=0.8\linewidth]{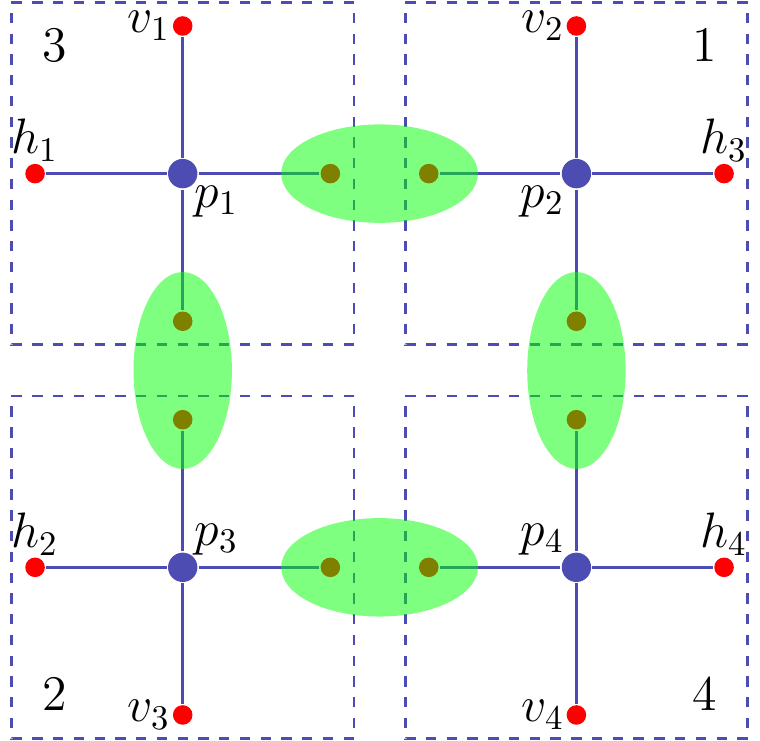}
	\caption{Unit cell of the quadrupole model pumping PEPS after projection of the virtual fermions corresponding to the bonds within the unit cell (marked by green ellipses) onto maximally entangled states. This results in a fiducial state with four physical fermions (blue circles) collected into the block $p$, four horizontal virtual fermions in the block $v_h$ and four virtual fermions in the block $v_v$ (red circles for both). The order of the individual fermions within these blocks is indicated by the labels next to each circle.\label{fig:SketchFiducialStateUnitCell}}
\end{figure}

The fiducial state of one unit cell, after contraction of the virtual bonds within the plaquette, describes four physical fermions and two virtual fermions per direction left, up right, down (see Fig.~\ref{fig:SketchFiducialStateUnitCell}). We collect the four physical, horizontal virtual and vertical virtual fermions into blocks labeled $p$, $v_h$ and $v_v$. The order of the individual fermions within these blocks is indicated in Fig.~\ref{fig:SketchFiducialStateUnitCell}. Since the Majorana CM $\Gamma_{Q_\mathbf{x}}$ of the fiducial state of a unit cell is anti-symmetric (see Appendix~\ref{sec:AppGaussianStateDef}), we can then separate it into $8 \times 8$ blocks as
\begin{equation}\label{CMPlaquetteQuadrupole}
\Gamma_{Q_\mathbf{x}} = \begin{pmatrix}
\mathbf{A}_p & \mathbf{B}_{pv_h} & \mathbf{B}_{pv_v} \\
- \mathbf{B}_{pv_h}^T & \mathbf{D}_{v_h} & \mathbf{B}_{v_hv_v} \\
- \mathbf{B}_{pv_v}^T & - \mathbf{B}_{v_hv_v}^T & \mathbf{D}_{v_v}
\end{pmatrix}.
\end{equation}
The blocks $\mathbf{A}_p$, $\mathbf{D}_{v_h}$ and $\mathbf{D}_{v_v}$ describe the reduced fiducial state of the physical, horizontal virtual and vertical virtual subsystem, respectively. The off-diagonal blocks describe the coupling between these three subsystems. Eq.~\eqref{CMPlaquetteQuadrupole} is a generalization of Eq.~\eqref{CMMajoranaBlock} with
\begin{subequations}
\begin{gather}
\mathbf{A} = \mathbf{A}_p,\\
\mathbf{B} = \left(  \mathbf{B}_{pv_h} \, \mathbf{B}_{pv_v}\right),\\
\mathbf{D} = \begin{pmatrix}
\mathbf{D}_{v_h} & \mathbf{B}_{v_hv_v} \\
- \mathbf{B}_{v_hv_v}^T & \mathbf{D}_{v_v}
\end{pmatrix}
\end{gather}
\end{subequations}
which we find useful since we want to treat horizontal and vertical virtual fermions separately. 

We are now ready to give the expression for the Majorana CM $\Gamma_{Q_\mathbf{x}}$ of the fiducial state of a unit cell. For simplicity, we set $\alpha_x = \alpha_y \equiv \alpha$ and $\beta_x = \beta_y \equiv \beta$ from now on, corresponding to parameters $a_x = a_y \equiv a$ and $b_x = b_y \equiv b$ in Eq.~\eqref{QuadrupolePEPS_Parametrisation}. The diagonal blocks $\mathbf{A}_p$, $\mathbf{D}_{v_h}$ and $\mathbf{D}_{v_v}$ of $\Gamma_{Q_\mathbf{x}}$ take the form
\begin{equation}\label{ZMatrix_2}
Z^{(2)} (r, s, u) = 
\begin{pmatrix}
0 & -r & 0 & -u & 0 & -s & 0 & 0 \\
r & 0 & -u & 0 & -s & 0 & 0 & 0 \\
0 & u^* & 0 & -r & 0 & 0 & 0 & -s \\
u^* & 0 & r & 0 & 0 & 0 & -s & 0 \\
0 & s^* & 0 & 0 & 0 & -r & 0 & u \\
s^* & 0 & 0 & 0 & r & 0 & u & 0 \\
0 & 0 & 0 & s^* & 0 & -u^* & 0 & -r \\
0 & 0 & s^* & 0 & -u^* & 0 & r & 0 \\
\end{pmatrix}
\end{equation}
with parameters $r \in \mathbb{R}$ and $s,u \in \mathbb{C}$. Specifically,
\begin{subequations}
	\begin{gather}
	\mathbf{A}_p = Z^{(2)}(r_p, - s_p, s_p),\\
	\mathbf{D}_{v_h} = Z^{(2)}(r_v, s_v,  s_v),\\
	\mathbf{D}_{v_v} = Z^{(2)}(r_v, -s_v, s_v)
	\end{gather}
\end{subequations}
with parameters
\begin{subequations}
\begin{align}
r_p &= -\frac{\sqrt{2}\left(2a^4 + b^4 - 1\right)}{4a^2 + 2\sqrt{2}a^4 + \sqrt{2}\left(1 + b^4\right)}, \\
 s_p & =  \frac{2b^2}{4a^2 + 2\sqrt{2}a^4 + \sqrt{2}\left(1 + b^4\right)}, \\ 
r_v &= \frac{2a^2 +\sqrt{2} \left( b^4 + 1\right)}{4a^2 + 2\sqrt{2}a^4 + \sqrt{2}\left(1 + b^4\right)},\\
s_v &= \frac{\sqrt{2} a^2 b^2}{4a^2 + 2\sqrt{2}a^4 + \sqrt{2}\left(1 + b^4\right)},
\end{align}
\end{subequations}
where the denominator stems from the matrix inversion in the Schur complements used to evaluate the projection on the virtual bonds within one unit cell. The off-diagonal blocks of $\Gamma_{Q_\mathbf{x}}$ are
\begin{widetext}
\begin{subequations}
	\begin{gather}
\label{Bvirtual}	\mathbf{B}_{v_hv_v} = 
\begin{pmatrix}
	0 & 1-r_v & 0 & s_v & 0 & -s_v & 0 & 0 \\
	r_v-1 & 0 & s_v & 0 & -s_v & 0 & 0 & 0 \\
	0 & s_v & 0 & 0 & 0 & 1-r_v & 0 & -s_v \\
	s_v & 0 & 0 & 0 & r_v-1 & 0 & -s_v & 0 \\
	0 & s_v & 0 & r_v-1 & 0 & 0 & 0 & s_v \\
	s_v & 0 & 1-r_v & 0 & 0 & 0 & s_v & 0 \\
	0 & 0 & 0 & s & 0 & s_v & 0 & 1-r_v \\
	0 & 0 & s_v & 0 & s_v & 0 & r_v-1 & 0 \\
\end{pmatrix},\\
\mathbf{B}_{pv_h} = \frac{a}{2^{1/4}}
\begin{pmatrix}
0 & r_p+1 & 0 & -s_p & 0 & -s_p & 0 & 0 \\
r_p+1 & 0 & s_p & 0 & s_p & 0 & 0 & 0 \\
0 & s_p & 0 & 0 & 0 & r_p+1 & 0 & s_p \\
-s_p & 0 & 0 & 0 & r_p+1 & 0 & -s_p & 0 \\
0 & -s_p & 0 & -r_p-1 & 0 & 0 & 0 & s_p \\
s_p & 0 & -r_p-1 & 0 & 0 & 0 & -s_p & 0 \\
0 & 0 & 0 & s_p & 0 & -s_p & 0 & r_p+1 \\
0 & 0 & -s_p & 0 & s_p & 0 & r_p+1 & 0 \\
\end{pmatrix},\\
\mathbf{B}_{pv_v} = \frac{a}{2^{1/4}}
\begin{pmatrix}
0 & r_p+1 & 0 & s_p & 0 & -s_p & 0 & 0 \\
r_p+1 & 0 & -s_p & 0 & s_p & 0 & 0 & 0 \\
0 & s_p & 0 & -r_p-1 & 0 & 0 & 0 & s_p \\
-s_p & 0 & -r_p-1 & 0 & 0 & 0 & -s_p & 0 \\
0 & -s_p & 0 & 0 & 0 & -r_p-1 & 0 & s_p \\
s_p & 0 & 0 & 0 & -r_p-1 & 0 & -s_p & 0 \\
0 & 0 & 0 & s_p & 0 & s_p & 0 & r_p+1 \\
0 & 0 & -s_p & 0 & -s_p & 0 & r_p+1 & 0 \\
\end{pmatrix}.
	\end{gather}
\end{subequations}
\end{widetext}

\subsubsection{Bloch CM\label{sec:AppendixQuadrupolePEPS_BlochCM}}

We can now compute the Bloch CM $\MCMF_{\ket{\psi}} (\mathbf{k})$ of the physical state on a torus by evaluating Eq.~\eqref{SchurComplementFourier}. Here, the Majorana CM of the local fiducial state of the unit cell with its blocks $\mathbf{A}$, $\mathbf{B}$ and $\mathbf{D}$ is given in the previous subsection. 

In the same basis as $\mathbf{D}$, the Bloch CM $\MCMF _{\omega}(\mathbf{k})$ of the total virtual maximally entangled state of size $16 \times 16$ is
\begin{subequations}
\begin{equation}
\MCMF _{\omega}(\mathbf{k}) = \begin{pmatrix}
\MCMF _{\omega, v_h}(\mathbf{k}) & 0 \\
0 & \MCMF _{\omega, v_v}(\mathbf{k})
\end{pmatrix}
\end{equation}
where the blocks $\MCMF _{\omega, v_h}(\mathbf{k})$ and $\MCMF _{\omega, v_v}(\mathbf{k})$ refer to the horizontal and vertical virtual fermions, respectively. Since the horizontal fermions on the sublattice 1 couple only to the horizontal fermions on sublattice 3 (and similarly for 2 and 4), the CM $\MCMF _{\omega, v_h}(\mathbf{k})$ is given by a sum of two copies of the CM from Eq.~\eqref{CMMajoranaFourierVirtual} for an MPS with $\xi = 1$. Hence,
\begin{equation}
\MCMF _{\omega, v_h}(\mathbf{k}) = \begin{pmatrix}
0 & 0 & -\sigma_1 e^{-ik_x} & 0 \\
0 & 0 & 0 & -\sigma_1 e^{-ik_x}\\
\sigma_1 e^{ik_x} & 0 & 0 & 0\\
0 & \sigma_1 e^{ik_x} & 0 & 0
\end{pmatrix}.
\end{equation}
\end{subequations}
The expression for $\MCMF _{\omega, v_v}(\mathbf{k})$ is analogous with $k_x \rightarrow - k_y$ (where the negative sign indicates that the bonds are oriented from top to bottom and hence point in the direction of negative $y$). Note that $ \MCMF _{\omega, v_h}(\mathbf{k})  = Z^{(2)} (0, e^{-ik_x}, 0)$ and $\MCMF _{\omega, v_v}(\mathbf{k}) = Z^{(2)} (0, e^{ik_y}, 0)$ are of the form of Eq.~\eqref{ZMatrix_2}.

For the quadrupole model pumping PEPS, the evaluation of Eq.~\eqref{SchurComplementFourier} requires the inversion of the matrix $\mathbf{D} + \MCMF _{\omega}(\mathbf{k})$ of size $16 \times 16$. This can be done analytically using the special representation from Eq.~\eqref{ZMatrix_2}. Indeed, one can show that the matrix $Z^{(2)}$ has the properties
\begin{subequations}
	\begin{gather}
	Z^{(2)} (r,s,u) + Z^{(2)} (r',s',u') = Z^{(2)} (r + r', s + s', u + u'),\\
	\det \left[Z^{(2)} (r,s,u)\right] = \big(r^2 + ss^* + uu^*\big)^4,\\
\label{InverseZ2}	[Z^{(2)} (r,s,u)]^{-1} = - \frac{ Z^{(2)} (r,s,u)}{[\det Z^{(2)} (r,s,u)]^{1/4}}.
	\end{gather}
\end{subequations}
Moreover, conjugation with the matrix $\mathbf{B}_{v_hv_v}$ from Eq.~\eqref{Bvirtual}, which gives the off-diagonal block of $\mathbf{D}$, returns a matrix of the form of Eq.~\eqref{ZMatrix_2},
\begin{equation}
\mathbf{B}_{v_hv_v} Z^{(2)} (r,s,u) \mathbf{B}_{v_hv_v}^T = Z^{(2)} (r',s',u')
\end{equation}  
where $r' = r[(1-r_v)^2 - 2s_v^2] + 2(1-r_v)s_v\Re(s-u)$, $s' = -s(1-r_v)^2 + 2rs_v(r_v-1) - 2s_v^2[\Re (u) + i \Im (s)]$ and $u' = u(1-r_v)^2 + 2rs_v(r_v-1) + 2s_v^2[\Re (s) + i \Im (u)]$.
Using these identities, the inverse of $\mathbf{D} + \MCMF _{\omega}(\mathbf{k})$ can be evaluated blockwise, and we compute the determinant in Eq.~\eqref{DeterminantSchurComplement} as
\begin{multline}
q(\mathbf{k}) = \frac{2^8\left( 1 + a^4 + b^4 + a^2 b^2 \cos k_x + a^2 b^2 \cos k_y \right)^4 }{\left( 1 + 2 \sqrt{2} a^2 + 2a^4 + b^4\right)^4}.
\end{multline}
This is strictly positive unless $\gamma = 0$ and $|\alpha| = |\beta|$, such that the PEPS is well-defined everywhere except for these parameter values.

Proceeding thus, we find that the Bloch Majorana CM $\MCMF_{\ket{\psi}} (\mathbf{k})$ of the physical state is also of the form of Eq.~\eqref{ZMatrix_2},
\begin{subequations}\label{FourierCMQuadrupolePEPS}
	\begin{equation}
	\MCMF_{\ket{\psi}} = Z^{(2)} \left(r(\mathbf{k}),s(\mathbf{k}),u(\mathbf{k}) \right)
	\end{equation}
	where the parameters are
	\begin{gather}
	r(\mathbf{k}) = \frac{1 - a^4 - b^4 - a^2 b^2 \cos k_x - a^2 b^2 \cos k_y}{1 + a^4 + b^4 + a^2 b^2 \cos k_x + a^2 b^2 \cos k_y},\\
	s(\mathbf{k}) = \frac{-\sqrt{2} \left( b^2 + a^2 e^{-i k_x} \right)}{1 + a^4 + b^4 + a^2 b^2 \cos k_x + a^2 b^2 \cos k_y},\\
	u(\mathbf{k}) = \frac{\sqrt{2} \left( b^2 + a^2 e^{i k_y} \right)}{1 + a^4 + b^4 + a^2 b^2 \cos k_x + a^2 b^2 \cos k_y}.
	\end{gather}
\end{subequations}
In the denominator, we recognize the fourth root $q(\mathbf{k})^{1/4}$ coming from the matrix inversion according to Eq.~\eqref{InverseZ2}.

\subsubsection{Parent Hamiltonian\label{sec:AppendixQuadrupolePEPS_PH}}

We are now in a position to find a parent Hamiltonian for the physical state on a torus, using Eq.~\eqref{GaussianPH} and the result for the Bloch Majorana CM $\MCMF_{\ket{\psi}}$ from Eq.~\eqref{FourierCMQuadrupolePEPS}. To get an intuitive understanding of the Hamiltonian, we find it useful to express it in terms of the original fermionic modes before the particle-hole transformation of Eq.~\eqref{PHTrafoQuadrupole}. Their FT is related to the FT of the new modes as $\pmode_{\tau, \mathbf{k}} = \pmodenew_{\tau, \mathbf{k}}$ for $\tau = 1,2$ and $\pmode_{\tau, \mathbf{k}} = \pmodenew^{\dagger}_{\tau, - \mathbf{k}}$ for $\tau = 3,4$. With respect to the original fermionic modes, Eq.~\eqref{GaussianPH} is given by
\begin{multline}
H_{\epsilon} = \sum _{\mathbf{k}} \epsilon(\mathbf{k}) \times \\
\pmode_{\tau, \mathbf{k}}^{\dagger}
\begin{pmatrix}
r(\mathbf{k}) & 0 & - s(\mathbf{k})^* & u(\mathbf{k}) \\
0 & r(\mathbf{k}) & - u(\mathbf{k})^* & - s(\mathbf{k})\\
-s(\mathbf{k}) & -u(\mathbf{k}) & -r(\mathbf{k}) & 0\\
u(\mathbf{k})^* & -s(\mathbf{k})^* & 0 & -r(\mathbf{k})\\
\end{pmatrix}_{\tau\tau'}
\pmode_{\tau', \mathbf{k}}
\end{multline}
where a summation over $\tau, \tau'$ is implied in the second line. 

As discussed in Appendix~\ref{sec:AppendixQuadrupolePEPS_PH}, the properties of the parent Hamiltonian are determined by our choice of dispersion relation $\epsilon(\mathbf{k})$. By setting $\epsilon(\mathbf{k}) = q(\mathbf{k})$, we would obtain a gapped parent Hamiltonian with coupling between up to fourth-nearest neighbour unit cells. However, we can find a more short-ranged parent Hamiltonian due to the special form of the fiducial state CM using Eq.~\eqref{ZMatrix_2}. Indeed, by choosing the dispersion function
\begin{equation}
\epsilon (\mathbf{k}) = \frac{1 + a^4+b^4 + a^2 b^2 \cos k_x + a^2 b^2 \cos k_y}{a^4 + b^4 + 1}
\end{equation}
which is proportional to the denominator of $r$, $s$ and $u$ from Eq.~\eqref{FourierCMQuadrupolePEPS}, we obtain a second-nearest neighbour parent Hamiltonian with Bloch representation
\begin{multline}\label{BlochPHQuadrupolePEPS}
H_{\epsilon}(\mathbf{k}) = \frac{1}{a^4 + b^4 + 1} \times \\
\Bigg[ \left(1 - a^4-b^4 -  a^2 b^2 \cos k_x- a^2 b^2 \cos k_y\right) \sigma_3 \otimes \sigma_0 +\\
\sqrt{2} \left(b^2 + a^2 \cos k_x\right) \sigma_1 \otimes \sigma_0 + \sqrt{2} a^2 \sin k_x \left( - \sigma_2 \otimes \sigma_3 \right) \\
+\sqrt{2} \left(b^2 + a^2 \cos k_y\right) \left( - \sigma_2 \otimes \sigma_2 \right) \\
+ \sqrt{2} a^2 \sin k_y \left( - \sigma_2 \otimes \sigma_1 \right)\Bigg],
\end{multline}
where $\sigma_0$ is the identity matrix of dimension two.

\subsection{ES in topological quadrupole phase\label{sec:AppendixQuadrupolePEPS_ES}}

When $b = 0$, the system described by the PEPS of Eq.~\eqref{QuadrupolePEPS_Parametrisation} splits into decoupled four-site plaquettes shifted from the unit cell by one site in both directions, corresponding to the OAI dimerized phase of the quadrupole model if $a= 1$. The Majorana CM $\Gamma_{\mathrm{Plaquette}}$ describing the physical state of one such decoupled plaquette takes the form of Eq.~\eqref{ZMatrix_2} with 
\begin{subequations}
	\begin{gather}
\label{RTopologicalPlaquette}	r = \frac{1-a^4}{1 + a^4},\\
\label{UTopologicalPlaquette}	u = s = - \frac{\sqrt{2} a^2} {1 + a^4}.
	\end{gather}
\end{subequations}
As an application, we will derive the ES contributions from the edges and corners given in Eqs.~\eqref{EntEnDimerizedQuadrupoleEdge} and~\eqref{EntEnDimerizedQuadrupoleCorner}, respectively. 

We begin with the four corners. The Majorana CM $\Gamma_{\mathrm{corner}, \tau}$ for a single corner site on the sublattice $\tau$ is given by the corresponding block of dimension $2$ on the diagonal of $\Gamma_{\mathrm{Plaquette}}$. Specifically,
\begin{equation}
\Gamma_{\mathrm{corner}, \tau} = \begin{pmatrix}
0 & -r \\
r & 0
\end{pmatrix}
\end{equation}
for $\tau = 1, 2, 3, 4$. We now transform $\Gamma_{\mathrm{corner}, \tau}$ to the basis of the original complex fermionic modes before the particle-hole transformation of Eq.~\eqref{PHTrafoQuadrupole}. Then, the CM of the corner site has a vanishing off-diagonal block $\hat{Q}^*_{\mathrm{corner}, \tau} = 0$ and a diagonal block $\hat{R}^*_{\mathrm{corner}, \tau} = \mp \frac{i}{2} \lambda_{\mathrm{corner}}$ with $\lambda_{\mathrm{corner}} = r$, where the negative and positive signs hold for the sublattices $\tau = 1 , 2$ and $\tau = 3,4$, respectively. Using the expression for $r$ from Eq.~\eqref{RTopologicalPlaquette} and $a = \alpha / \gamma$, we obtain the formula for the corner ES level given in Eq.~\eqref{EntEnDimerizedQuadrupoleCorner} with one level per corner.

Similarly, the Majorana CM $\Gamma_{\mathrm{edge}, \tau_1, \tau_2}$ for two decoupled edge sites on the sublattices $\tau_1$ and $\tau_2$ is given by a block of dimension $4$ of $\Gamma_{\mathrm{Plaquette}}$. Concretely, 
\begin{equation}
\Gamma_{\mathrm{edge}, \tau_1\tau_2} = \begin{pmatrix}
0 & -r & 0 & -s \\
r & 0 & -s & 0 \\
0 & s & 0 & -r \\
s & 0 & r & 0
\end{pmatrix}
\end{equation}
for $(\tau_1, \tau_2) \in \{ (3, 1), (4, 2), (4, 1), (2, 3)\}$. In the basis of the original complex fermionic modes before the particle-hole transformation of Eq.~\eqref{PHTrafoQuadrupole}, $\Gamma_{\mathrm{edge}, \tau_1, \tau_2}$ takes the following form: It has a vanishing off-diagonal block $\hat{Q}^*_{\mathrm{edge}, \tau_1\tau_2} = 0$ and a non-zero diagonal block $\hat{R}^*_{\mathrm{edge}, \tau_1\tau_2}$ with doubly degenerate eigenvalues $\pm \frac{i}{2} \lambda_{\mathrm{edge}}$ and 
\begin{equation}
\lambda_{\mathrm{edge}} = \sqrt{r^2 + s^2} = \frac{\sqrt{1 + a^8}}{1 + a^4}.
\end{equation}
This corresponds to the formula given in Eq.~\eqref{EntEnDimerizedQuadrupoleEdge}.

\end{document}